\documentclass[accepted]{article}

\usepackage{mathptmx} 

\usepackage{fancyhdr}
\usepackage[normalem]{ulem}
\usepackage{pifont}
\usepackage{comment}
\usepackage{xspace}
\usepackage[table]{xcolor}
\usepackage{soul}
\usepackage{makecell}
\usepackage{subcaption}
\usepackage{pifont}
\usepackage{multirow}
\usepackage{arydshln}
\usepackage{stmaryrd}
\usepackage{tikz}
\usepackage[english]{babel}
\usepackage{colortbl}
\usepackage{microtype}
\usepackage{graphicx}
\usepackage{booktabs} 
\usepackage{xurl}
\usepackage{hyperref}
\hypersetup{breaklinks=true}
\usepackage{amsthm}

\usepackage[accepted]{mlsys2025}
\usepackage{amsmath}
\usepackage{bm}
\usepackage{amsfonts}
\usepackage{amssymb}

\newcommand{\squishlist}{
 \begin{list}{$\bullet$}
  { \setlength{\itemsep}{0pt}
     \setlength{\parsep}{0pt}
     \setlength{\topsep}{3pt}
     \setlength{\partopsep}{0pt}
     \setlength{\leftmargin}{1.5em}
     \setlength{\labelwidth}{1em}
     \setlength{\labelsep}{0.5em} } }

\newcommand{\squishnums}{
 \begin{list}{$\bullets$}
  { \setlength{\itemsep}{0pt}
     \setlength{\parsep}{3pt}
     \setlength{\topsep}{3pt}
     \setlength{\partopsep}{0pt}
     \setlength{\leftmargin}{1.5em}
     \setlength{\labelwidth}{1em}
     \setlength{\labelsep}{0.5em} } }

\newcommand{\squishlisttwo}{
 \begin{list}{$\bullet$}
  { \setlength{\itemsep}{0pt}
     \setlength{\parsep}{0pt}
    \setlength{\topsep}{0pt}
    \setlength{\partopsep}{0pt}
    \setlength{\leftmargin}{2em}
    \setlength{\labelwidth}{1.5em}
    \setlength{\labelsep}{0.5em} } }

\newcommand{\squishnobulletlist}{
 \begin{list}{}
  { \setlength{\itemsep}{0pt}
     \setlength{\parsep}{0pt}
     \setlength{\topsep}{0pt}
     \setlength{\partopsep}{0pt}
     \setlength{\leftmargin}{0em}
     \setlength{\labelwidth}{0em}
     \setlength{\labelsep}{0em} } }
     
\newcommand{\squishend}{
  \end{list}  }

\newcommand{\squishnobullet}{
 \begin{list}{}
  { \setlength{\itemsep}{0pt}
     \setlength{\parsep}{0pt}
     \setlength{\topsep}{3pt}
     \setlength{\partopsep}{0pt}
     \setlength{\leftmargin}{0.5em}
     \setlength{\labelwidth}{1em}
     \setlength{\labelsep}{0.5em} } }

\newcommand{\+}{{\text{+}}}

\newcommand{\secref}[1]{\S\ref{#1}}
\newcommand{\figref}[1]{Fig.~\ref{#1}}
\newcommand{\tabref}[1]{Tab.~\ref{#1}}

\definecolor{Gray}{gray}{0.85}
\definecolor{LightCyan}{rgb}{0.88,1,1}

\newcolumntype{a}{>{\columncolor{Gray}}c}
\newcolumntype{b}{>{\columncolor{white}}c}
\usepackage{makecell}

\newcommand{\framework}{{\textsc{Privatar}}\xspace}
\newcommand{\frameworknospace}{{\textsc{Privatar}}}
\newcommand{\artifactbadges}{%
  \vspace{-5em}%
  \noindent\makebox[\textwidth][l]{
    \includegraphics[height=65pt]{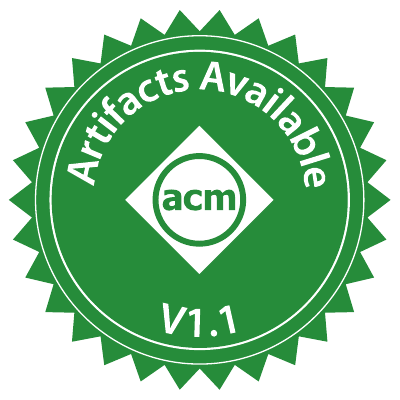}\hspace{0.2em}%
    \includegraphics[height=65pt]{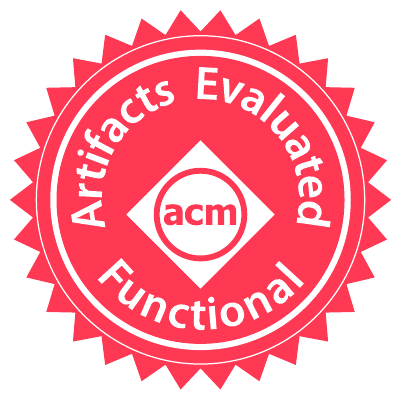}\hspace{0.2em}%
    \includegraphics[height=65pt]{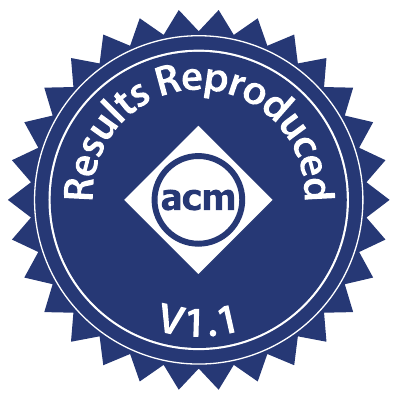}%
  }\par\vspace{-0.3em}%
}
\mlsystitlerunning{\framework: Enabling Privacy-preserving Real-time Multi-user VR through Secure Offloading}

\newif\ifcommenton
\commentontrue

\ifcommenton
\newcommand{\TODO}[1]{\textcolor{red}{[TODO] #1}}
\newcommand{\JT}[1]{{\color{brown}\bfseries [Jianming: #1]}}
\newcommand{\HS}[1]{{\color{blue}\bfseries [Hanshen: #1]}}
\newcommand{\AI}[1]{{\color{blue}\bfseries [Anirudh: #1]}}
\newcommand{\TK}[1]{{\color{violet}\bfseries [TK: #1]}}

\newcommand{\ES}[1]{{\color{blue}#1}}
\newcommand{\fixme}[1]{{{\color{blue} #1}}}
\newcommand{\leonote}[1]{{\color{red}[\textbf{Leo:} #1 ]}}
\else
\newcommand{\TODO}[1]{}
\newcommand{\ES}[1]{}
\newcommand{\AI}[1]{}
\newcommand{\JT}[1]{}
\newcommand{\TK}[1]{}
\newcommand{\HS}[1]{}
\newcommand{\fixme}[1]{}
\newcommand{\leonote}[1]{}

\fi

\newif\ifrevisionon
\revisionontrue

\ifrevisionon
\newcommand{\rvs}[1]{{\color{blue}#1}}
\newcommand{\LC}[1]{{\color{blue}#1}}
\else
\newcommand{\rvs}[1]{}
\newcommand{\LC}[1]{}
\fi

\definecolor{colorRevA}{rgb}{0.97, 0.80, 0.796}
\definecolor{colorRevB}{rgb}{0.85, 0.90, 0.98}
\definecolor{colorRevC}{rgb}{0.87, 0.83, 0.90}
\definecolor{colorRevD}{rgb}{0.83, 0.91, 0.83}
\definecolor{colorRevE}{rgb}{1, 0.90, 0.80}
\definecolor{colorRevF}{rgb}{0.77, 0.78, 0.83}
\definecolor{blond}{rgb}{0.98, 0.94, 0.75}

\newif\ifrevisionon
\revisionontrue

\begin{document}

\twocolumn[
\artifactbadges
\mlsystitle{\framework: Enabling Privacy-preserving Real-time Multi-user VR through Secure Offloading}




\mlsyssetsymbol{equal}{*}

\begin{mlsysauthorlist}
\mlsysauthor{Jianming Tong}{to,mit,google}
\mlsysauthor{Hanshen Xiao}{go}
\mlsysauthor{Krishnakumar Nair}{google}
\mlsysauthor{Hao Kang}{to}
\mlsysauthor{Ziqi Zhang}{uiuc}
\mlsysauthor{Ashish Sirasao}{amd}
\mlsysauthor{G. Edward Suh}{ed}
\mlsysauthor{Tushar Krishna}{to}
\end{mlsysauthorlist}

\mlsysaffiliation{to}{Georgia Institute of Technology, Atlanta, Georgia, USA}
\mlsysaffiliation{mit}{Massachusetts Institute of Technology, Cambridge, Massachusetts, USA}
\mlsysaffiliation{go}{Purdue University/NVIDIA, West Lafayette, Indiana, USA}
\mlsysaffiliation{google}{Google, Mountain View, California, USA}
\mlsysaffiliation{amd}{AMD, Santa Clara, California, USA}
\mlsysaffiliation{uiuc}{University of Illinois Urbana-Champaign, Champaign, Illinois, USA}
\mlsysaffiliation{ed}{Cornell University/NVIDIA, Westford, Massachusetts, USA}

\mlsyscorrespondingauthor{Jianming Tong}{jianming.tong@gatech.edu}

\mlsyskeywords{Privacy-preserving AR/VR, Secure Workload Offloading}

\begin{abstract}
\label{sec:abstract}
Multi-user virtual reality (VR) applications such as football and concert experiences rely on real-time avatar reconstruction to enable immersive interaction. However, rendering avatars for numerous participants on each headset incurs prohibitive computational overhead, fundamentally limiting scalability. This work introduces a framework, Privatar, to offload avatar reconstruction from headset to untrusted devices within the same local network while safeguarding sensitive facial features against adversaries capable of intercepting offloaded data.

Privatar builds on the insight that ``domain-specific knowledge of avatar reconstruction enables provably private offloading at minimal cost". (1) \textit{System level}. We observe avatar reconstruction is frequency-domain decomposable via block-wise DCT with negligible quality drop, and propose Horizontal Partitioning (HP) to keep high-energy frequency components on-device and offloads only low-energy components. HP offloads local computation while reducing information leakage to low-energy subsets only. (2) \textit{Privacy level}. For \textit{individually} offloaded, \textit{multi-dimensional} signals without aggregation, worst-case local Differential Privacy requires prohibitive noise, ruining utility. We observe users’ expression statistical distribution are \textit{slowly changing over time and trackable online}, and hence propose Distribution-Aware Minimal Perturbation (DAMP). DAMP minimizes noise based on each user’s expression distribution to significantly reduce its effects on utility and accuracy, retaining formal privacy guarantee. Combined, HP provides empirical privacy protection against expression identification attack. And DAMP further augments it to offer a formal guarantee against arbitrary adversaries.

On a Meta Quest Pro, Privatar supports up to 2.37$\times$ more concurrent users at 5.7$\sim$6.5\% higher reconstruction loss and $\sim$9\% energy overhead, providing a better throughout-loss Pareto frontier over SotA quantization, sparsity, and local reconstruction baseline. \framework further provides both provable privacy guarantee and stays robust against both empirical attack and NN-based Expression Identification Attack, proving its resilience in practice. Our code is open-sourced at \url{https://github.com/georgia-tech-synergy-lab/Privatar}.
\end{abstract}

]



\printAffiliationsAndNotice{}  

\vspace{-3mm}
\section{Introduction}
\label{sec:introduction}
Virtual Reality (VR) is rapidly evolving to support shared, immersive 3D environments for applications ranging from concerts~\cite{virtual_concert}, sports games~\cite{VR_sports}, cinemas~\cite{Kim_2024} to collaborative design\cite{zhou2025surveymethodologicalapproachescollaborative}. Central to this evolution is the ability to render photorealistic avatars of \textit{multiple users} in real-time, enhancing social interaction. 

An ideal multi-user VR system must achieve four competing goals: i) high fidelity, ii) strong user privacy, iii) power and heat efficiency for better user experience, and iv) scalability to multiple users. 
In current paradigm, avatar reconstructions of all users happen locally within a single VR headset, which satisfies i)-iii) but fails to scale as the rendering workload from multiple users quickly overwhelms the limited computational resources of mobile devices. 
This paper explores an alternative of offloading local computation to a powerful but untrusted external device for throughput improvement.
However, offloading introduces privacy leakage, as it exposes private user facial input individually to the local network, potentially leading to confidentiality and integrity risks. An adversary on the same local network (communication)  or residing at other devices (compute) can eavesdrop on, infer from, or tamper with offloaded data.

\vspace{-3mm}
\subsection{Technical Challenges}
Broadly speaking, there are three lines of provable privacy protection for offloading in untrusted servers: cryptographically based encryption, trusted hardware based isolation, and information-theoretical based perturbation. But they all fall short in either efficiency or utility (quality of avatar).

Cryptographic solutions enable a user to encrypt their personal VR data without leakage, and the server to perform computation over ciphertexts without decryption, which perfectly satisfies i) accuracy and ii) privacy. However, SotA solutions~\cite{tong2025CROSS} cannot satisfy iii) efficiency and iv) scalability requirements listed above. For example, Homomorphic Encryption (HE) encrypts offloaded data and its computation, leading to 3 magnitude higher computation latency~\cite{tong2025CROSS}, hardly supporting more users. Multiple Party Computation (MPC) leverages more than one non-colluding servers to collaboratively compute the private data, but requiring up-to Giga-Bytes of communication overhead to forbid real-time processing~\cite{keller2022securequantizedtrainingdeep, NEURIPS2021_27545182}. 

Confidential computing, using technologies like Trusted Execution Environments (TEEs) in CPU Intel SGX~\cite{intel_SGX} and AMD SEV-SNP or SME~\cite{amd_sev_snp}, offers sufficient privacy guarantee and somewhat throughput improvement, (1.79$\times$ in our experiments) because of slow throughput on CPU and extra encryption overhead.

Information-theoretical privacy preservation largely relies on adding noise to obfuscate release. Compared to cryptographic encryption or confidential computing, noise does not cause any efficiency loss but it hurts  utility and introduces privacy risk. In this line, Differential Privacy (DP)~\cite{differential_privacy} is one of the most representative frameworks. DP aims to ensure that, in the worst case for an adversary with arbitrary prior belief, they cannot gain additional knowledge from the release to distinguish an individual's participation in a data processing. Operationally, DP requires first determining an \emph{upper bound} of the \emph{sensitivity} of the processing function \cite{clipping_bias}, i.e., the worst-case change on the release when one arbitrarily replaces one datapoint. Afterwards, noise is calibrated according to both sensitivity and security parameters \cite{sensitivity_geometry}. However, depending on the application, finding a usable noise solution without significantly hurts utility is non-trivial~\cite{local-DP, near2025guidelines}.

\vspace{-3mm}
\subsection{Our Contributions}

We introduce \framework, the first framework to enable privacy-preserving offloading without sever utility degradation, ensuring private, high-fidelity, and real-time multi-user avatar reconstruction at scale. We systemically protect privacy of individually released facial data in two levels.

On the \textbf{System Level}, \framework's core innovation is a strategic partitioning of the avatar reconstruction pipeline. It separates the frequency representation of an avatar's texture, assigning different components to either the local VR headset or an untrusted external device based on variance. The facial mesh and the high-variance frequency components of the texture are processed securely on the local VR headset. Conversely, the reconstruction of the low-variance frequency components is offloaded to untrusted third-party devices to minimize the noise required by provable privacy protection. \textit{This physical isolation reduces the sensitivity of offloaded data} by only exposing a subset of components, providing empirical protection against expression identification attack~\cite{kopalidis2024advances,zhang_fer}.


On the \textbf{Privacy Level}, \framework roots in using exploitation of data entropy for privacy protection. Compared to DP, which fully relies on algorithmic randomness to ensure indistinguishability, we further consider incorporating the data entropy for privacy enhancement. That is to say, we consider formalizing the adversarial inference hardness based on the randomness from both the data distribution and the additional noise perturbation instead of purely on noise perturbation in DP. Based on this, we propose Distribution-Aware Minimal Perturbation to reduce the amount of noise required by local DP by up-to 17.6$\times$ for the same level of provable privacy guarantee, reducing utility degradation.



Intuitively, \framework leverages the temporal invariant nature of users' facial expression to reduce the required noise, i.e. users turn to make the similar expressions as they did recently (analogue to user's quirk) and all expressions a user could make cannot change drastically over time. Using users’ historical expression statistics and PAC Privacy \cite{xiao2023pac,pac_survery,pac_thesis}, we compute per-dimension, provably minimal perturbations that meet a target privacy guarantee. Our contributions are: 

\begin{figure*}[t]
    \centering
    \begin{minipage}[t]{0.59\textwidth}
        \centering
        \vspace{0pt} 
        \subfloat{{\includegraphics[width=0.15\textwidth]{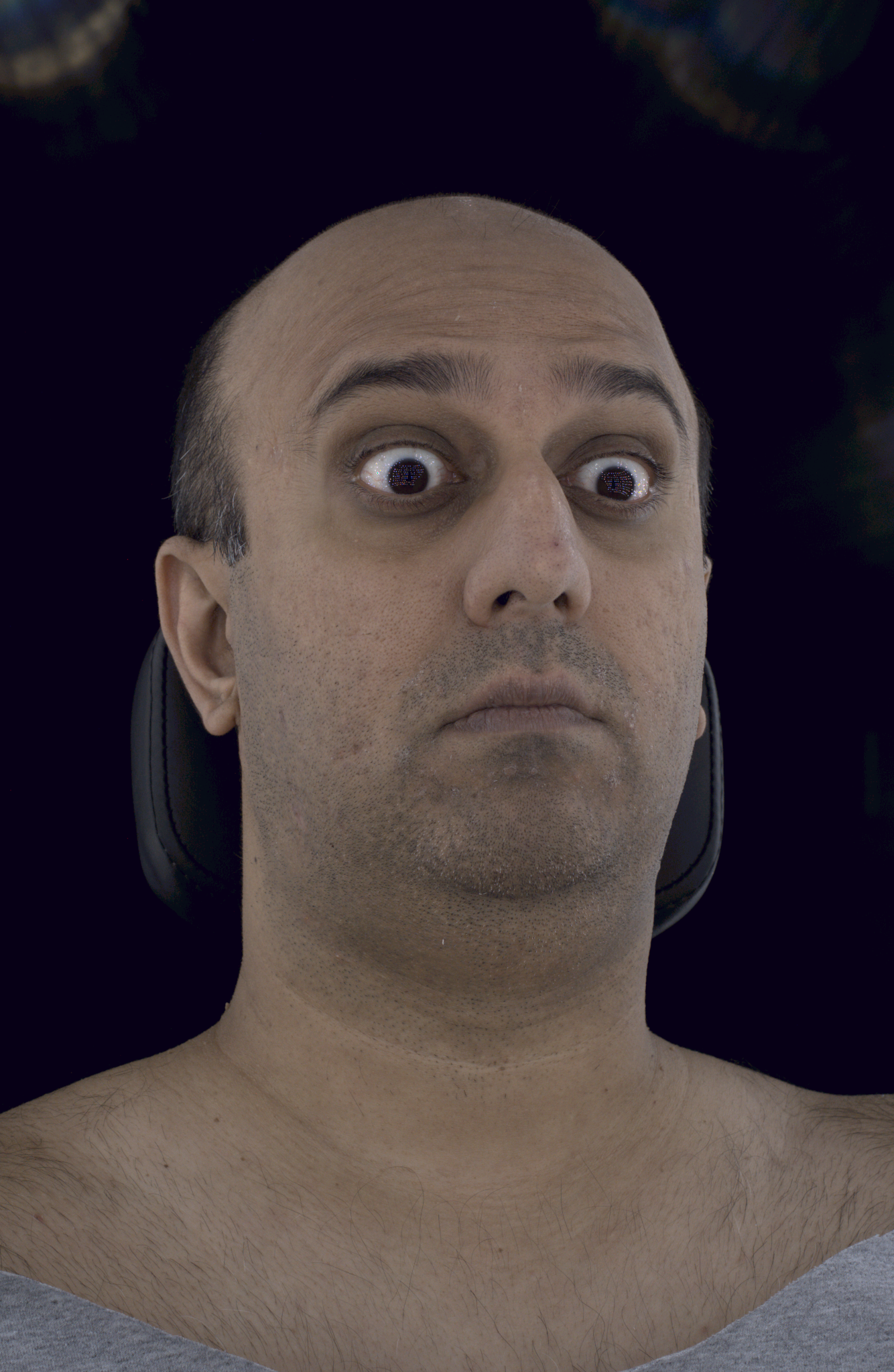}}} \ 
        \subfloat{{\includegraphics[width=0.15\textwidth]{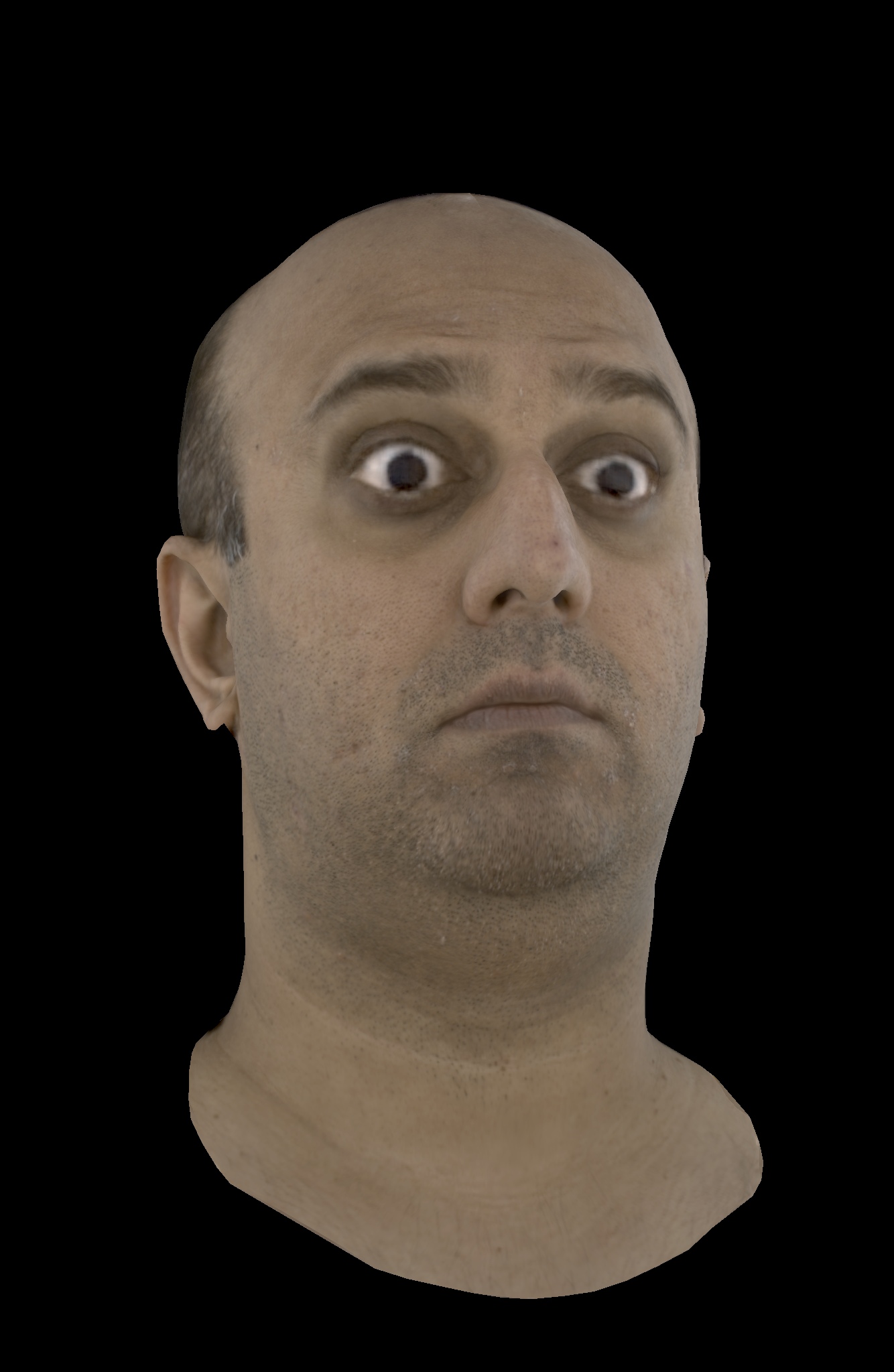}}} 
        \ 
        \subfloat{{\includegraphics[width=0.15\textwidth]{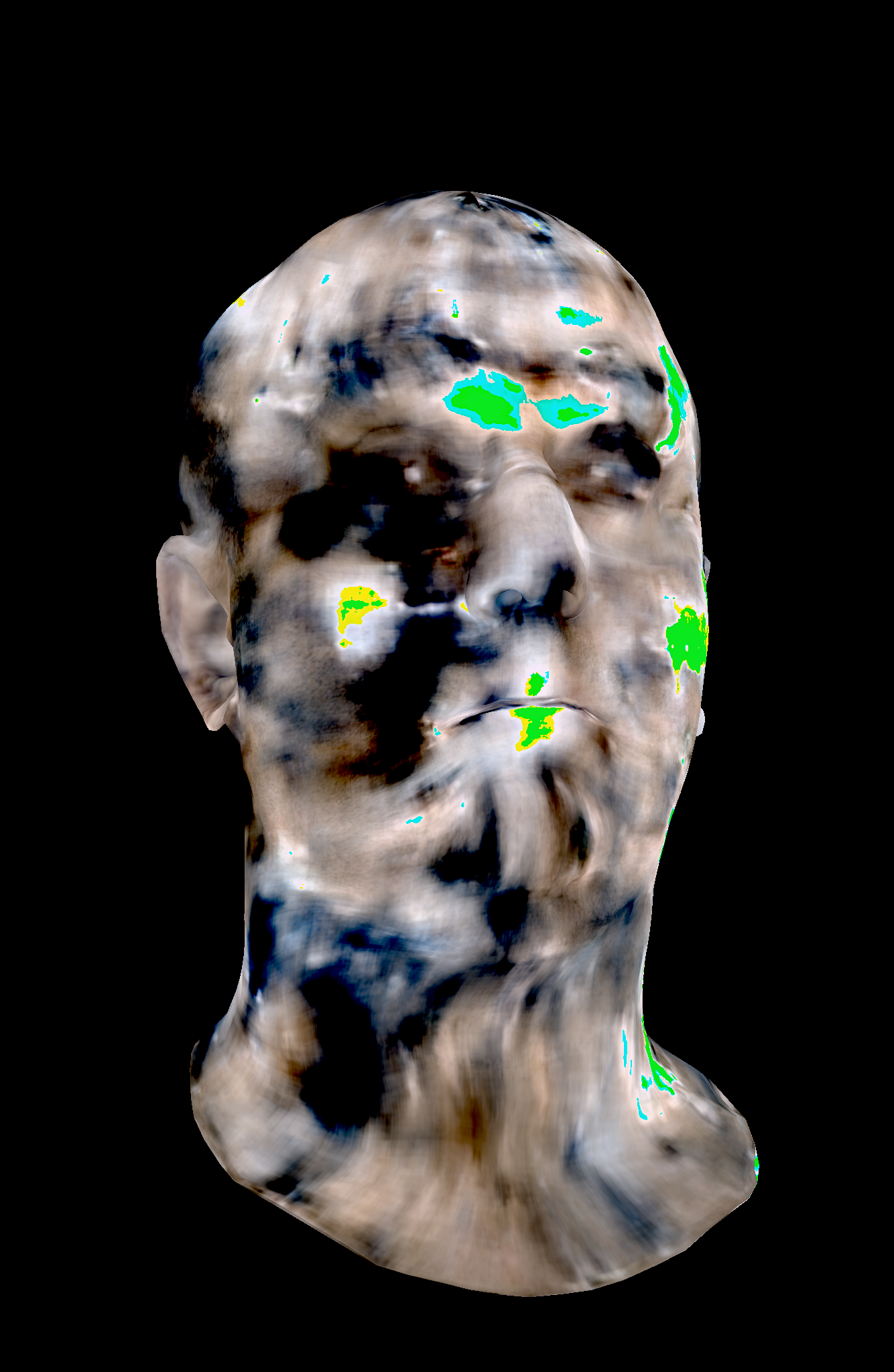}}} 
        \ 
        \subfloat{{\includegraphics[width=0.15\textwidth]{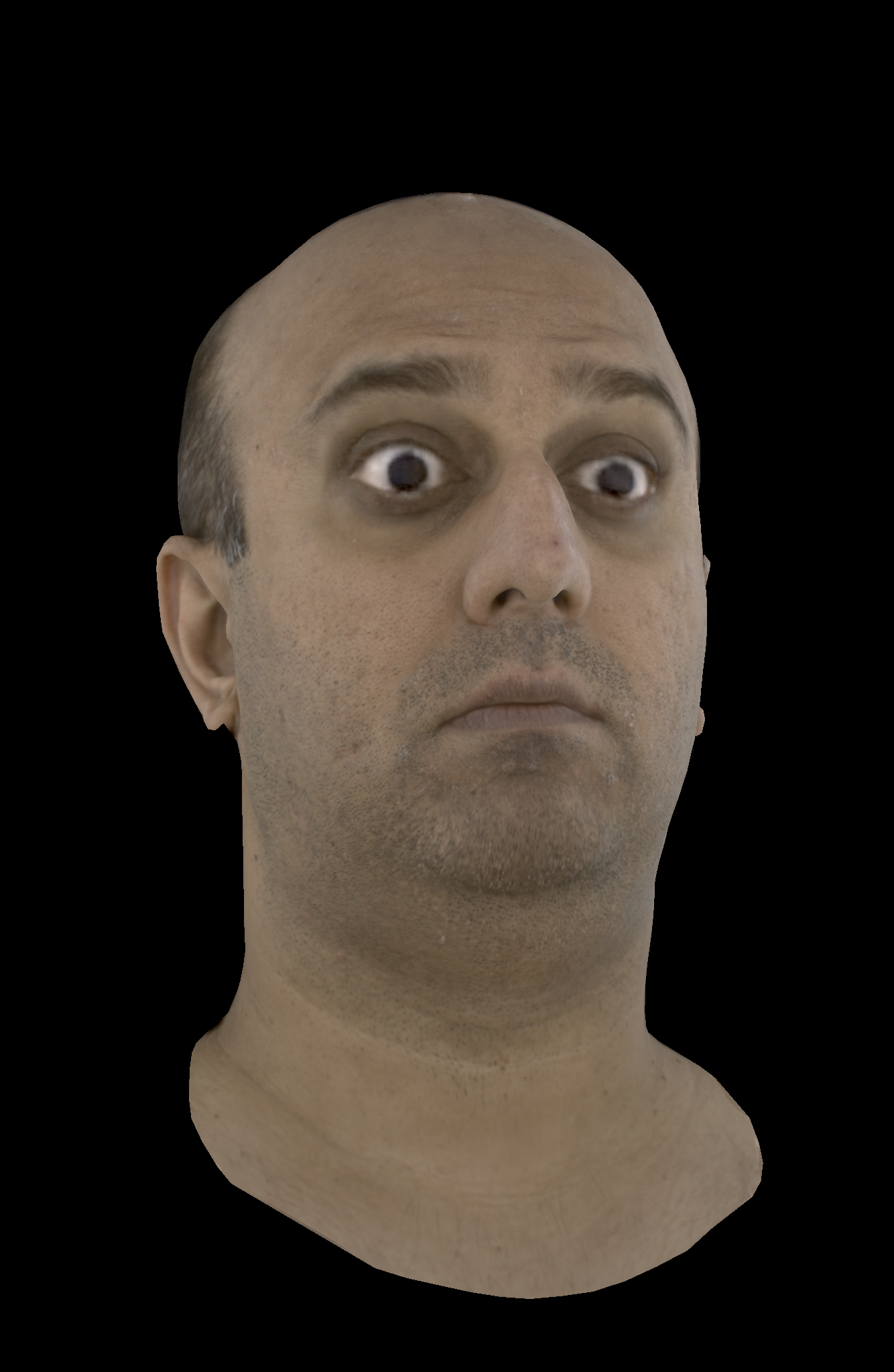}}} 
        \ 
        \subfloat{{\includegraphics[width=0.15\textwidth]{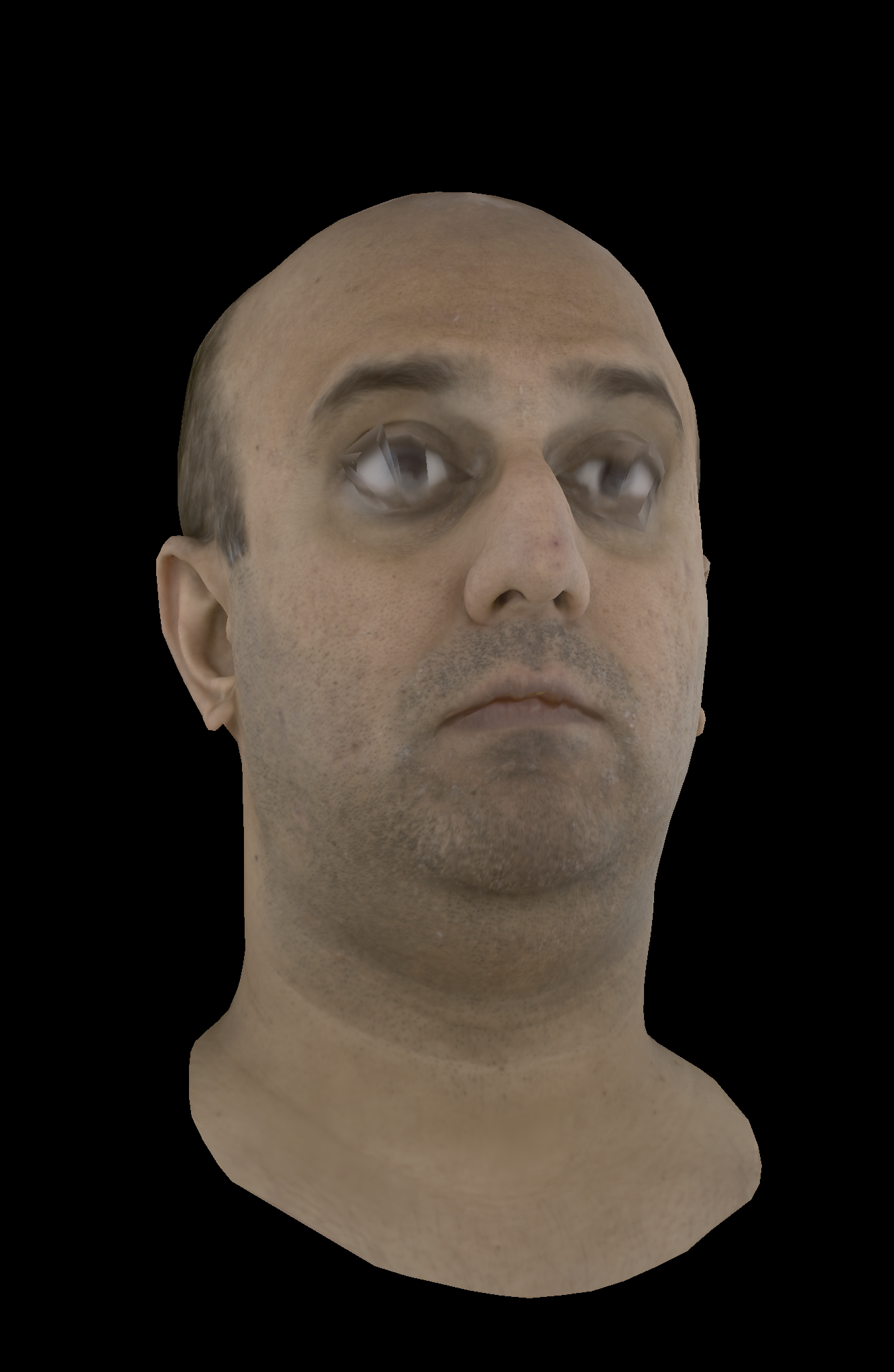}}} 
        \ 
        \subfloat{{\includegraphics[width=0.15\textwidth]{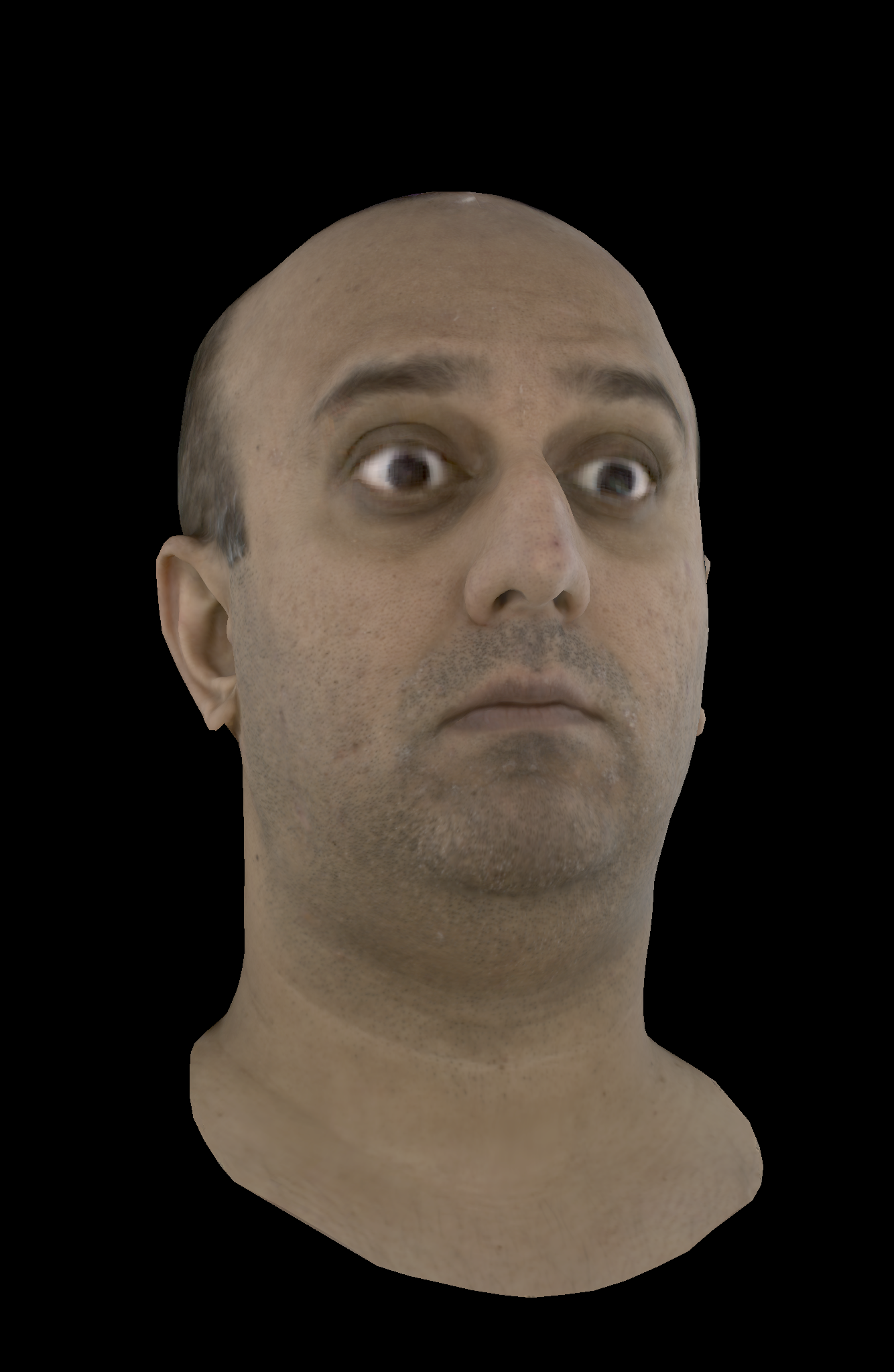}}} \\
        \subfloat{{\includegraphics[width=0.15\textwidth]{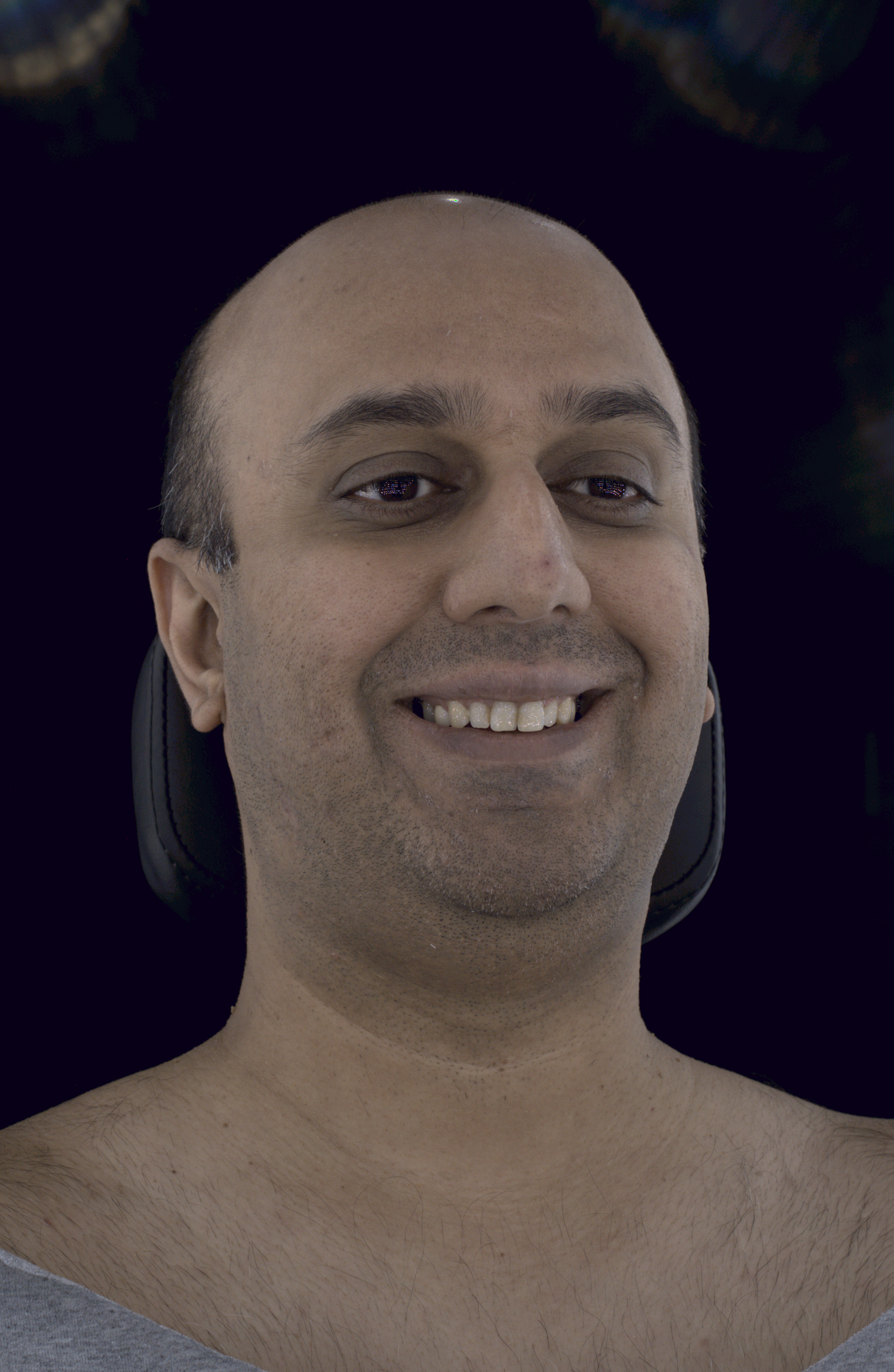}}} \ 
        \subfloat{{\includegraphics[width=0.15\textwidth]{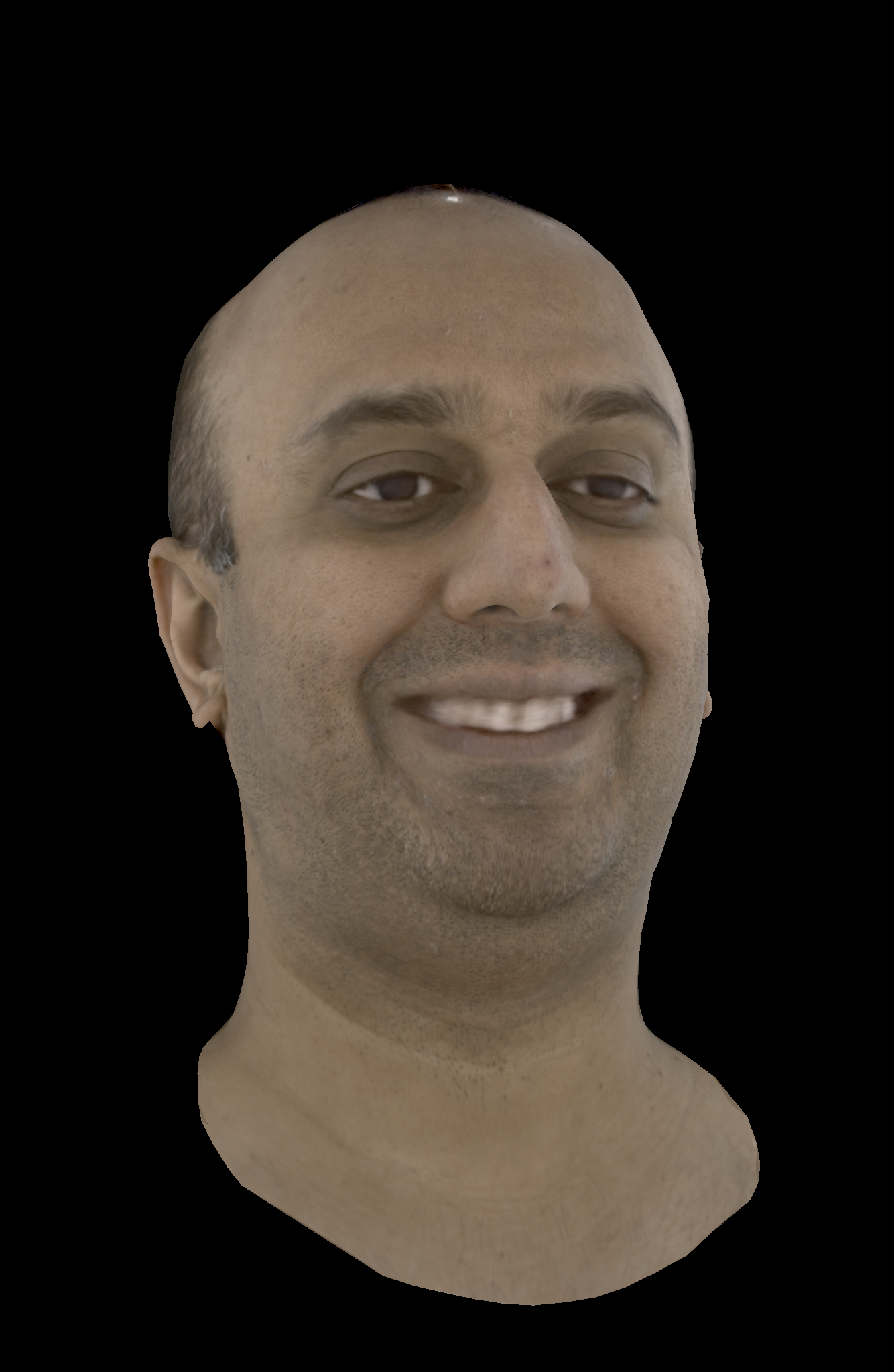}}} 
        \ 
        \subfloat{{\includegraphics[width=0.15\textwidth]{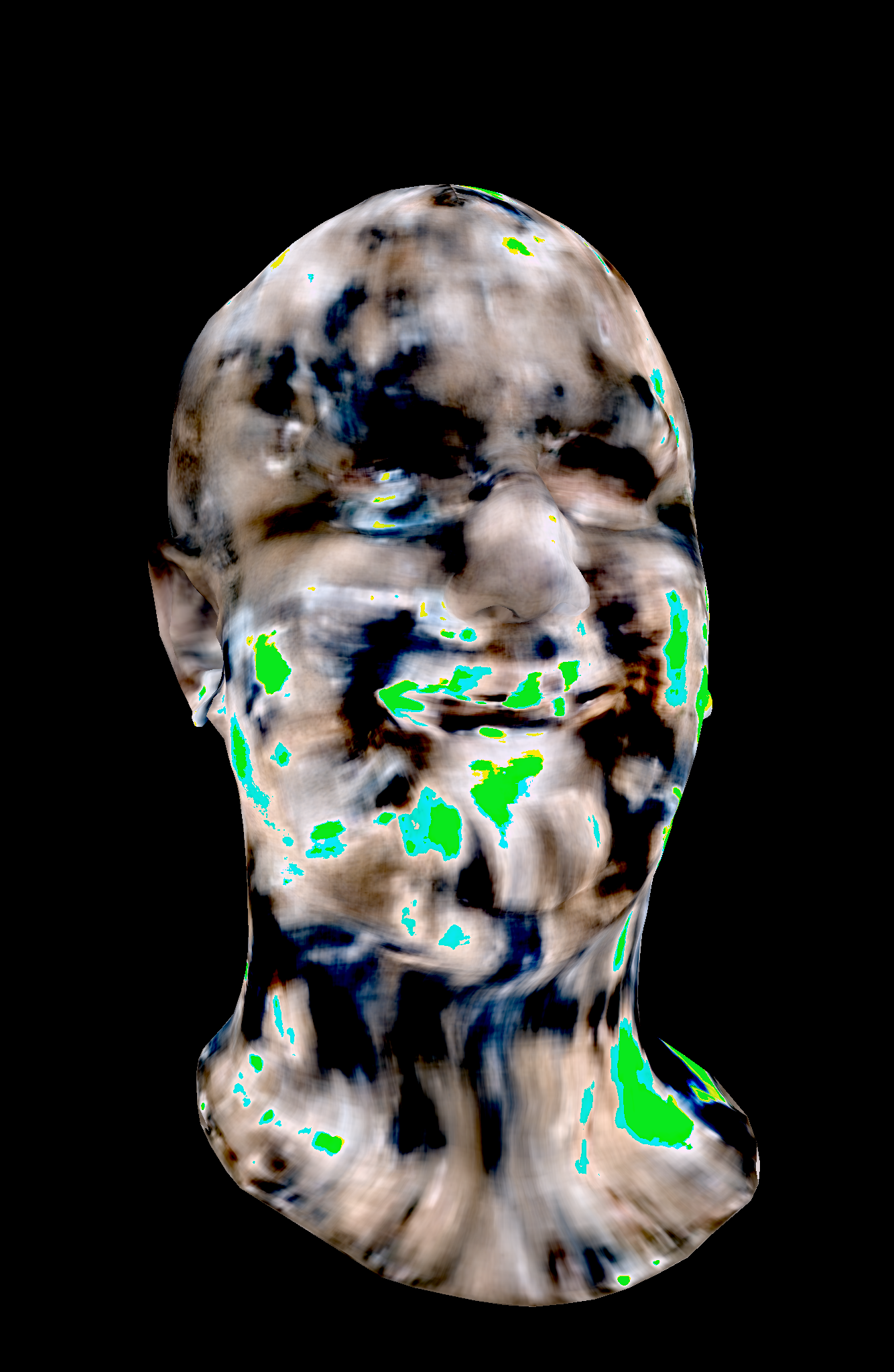}}} 
        \ 
        \subfloat{{\includegraphics[width=0.15\textwidth]{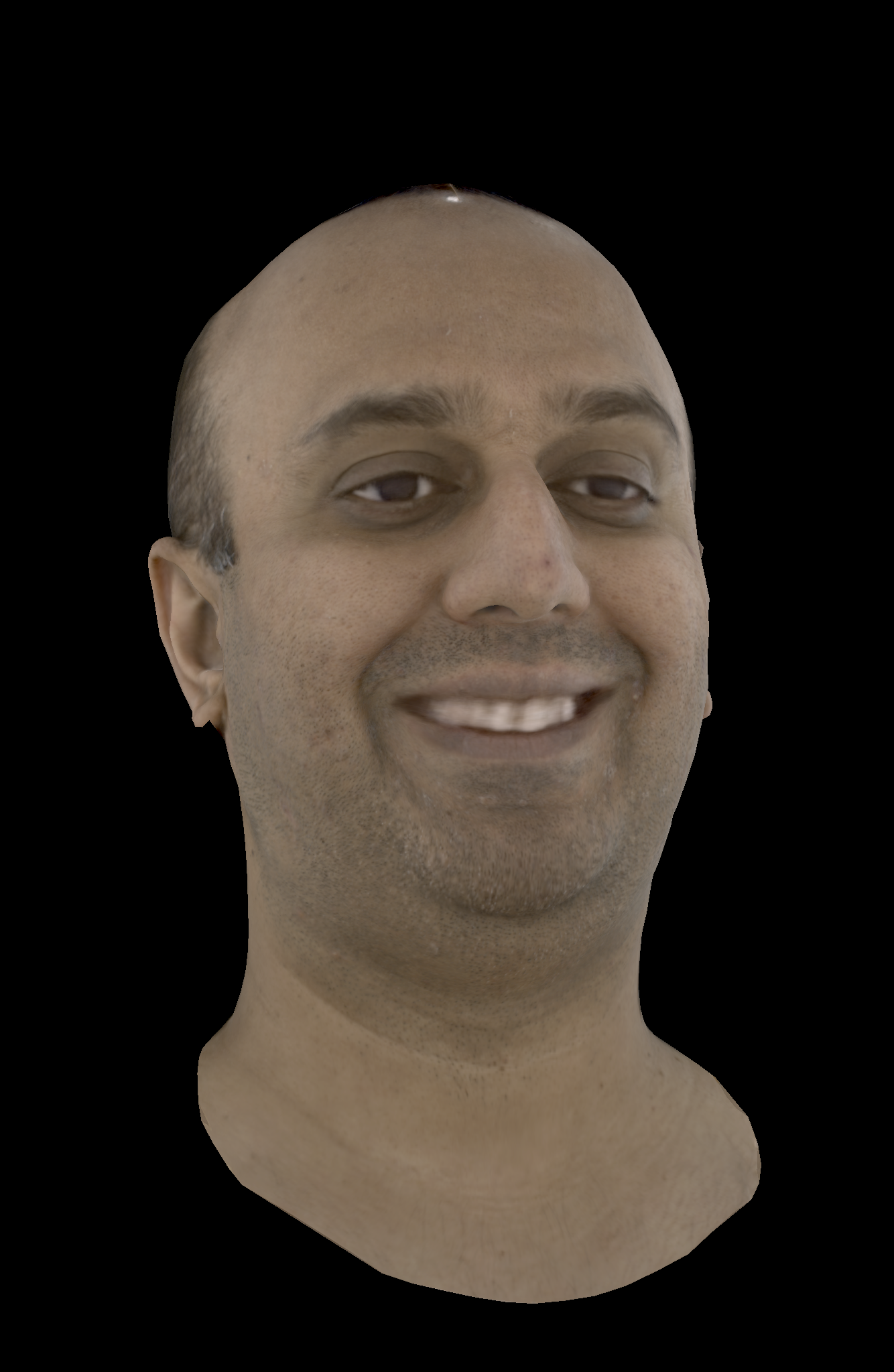}}} 
        \ 
        \subfloat{{\includegraphics[width=0.15\textwidth]{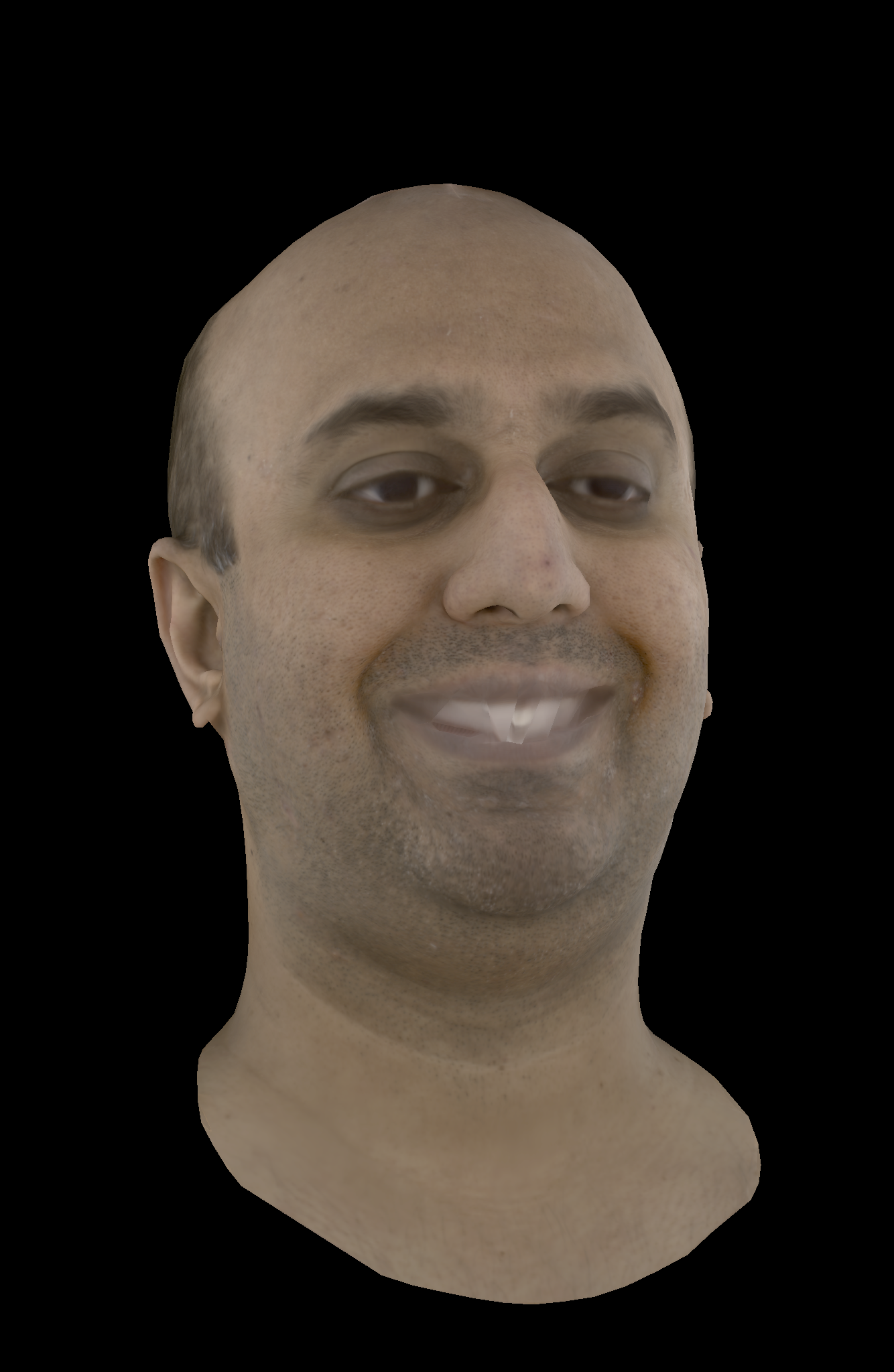}}} 
        \ 
        \subfloat{{\includegraphics[width=0.15\textwidth]{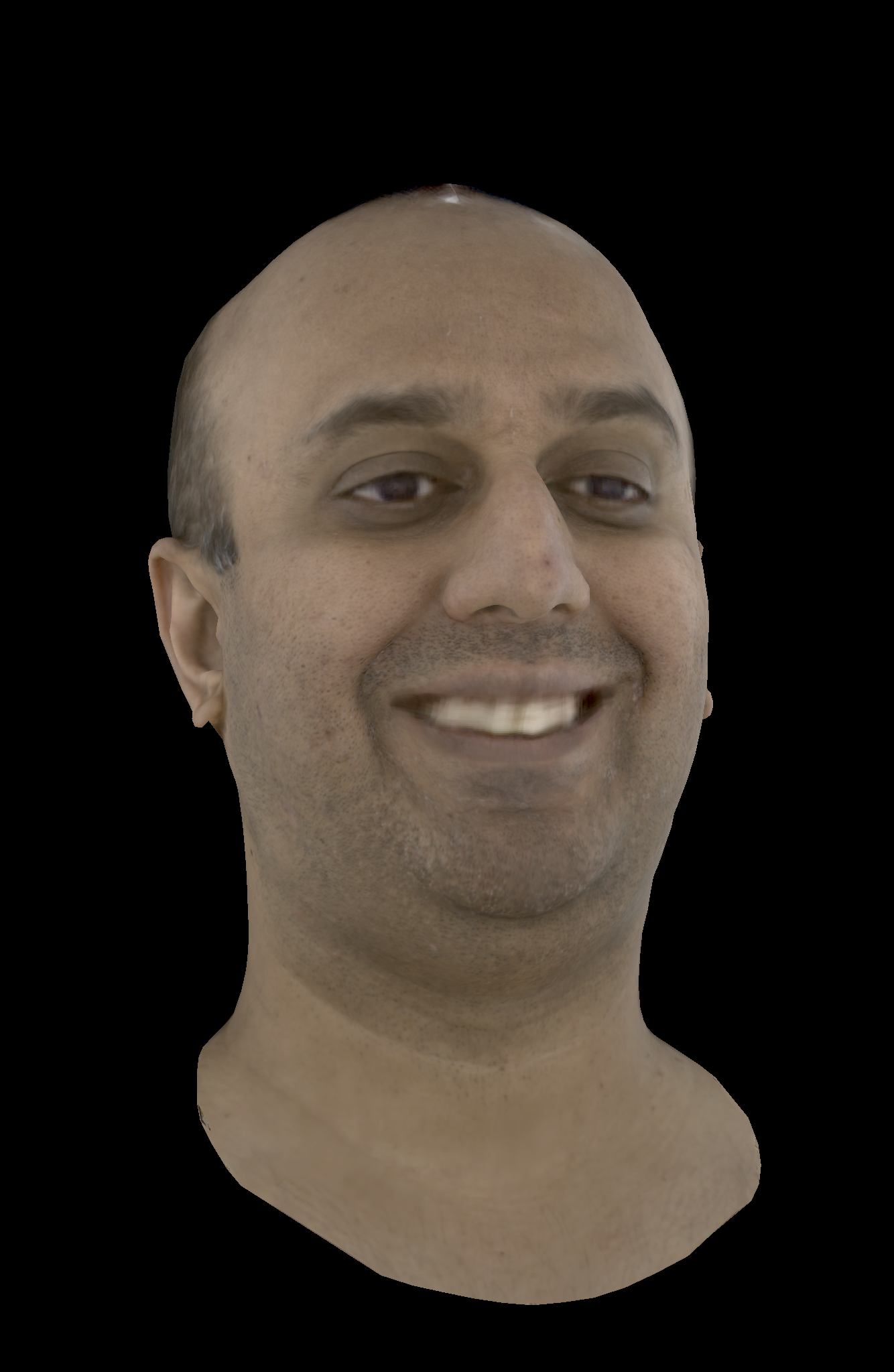}}}  \\ 
        \hfil \footnotesize (a) Truth \quad (b) Baseline \quad (c) FO \quad (d) Quantize\quad (e) Sparsity \quad (f) \framework
    \end{minipage}%
    \hfill
    \begin{minipage}[t]{0.4\textwidth}
        \centering
        \vspace{0pt} 
        \subfloat{\includegraphics[width=0.955\textwidth]{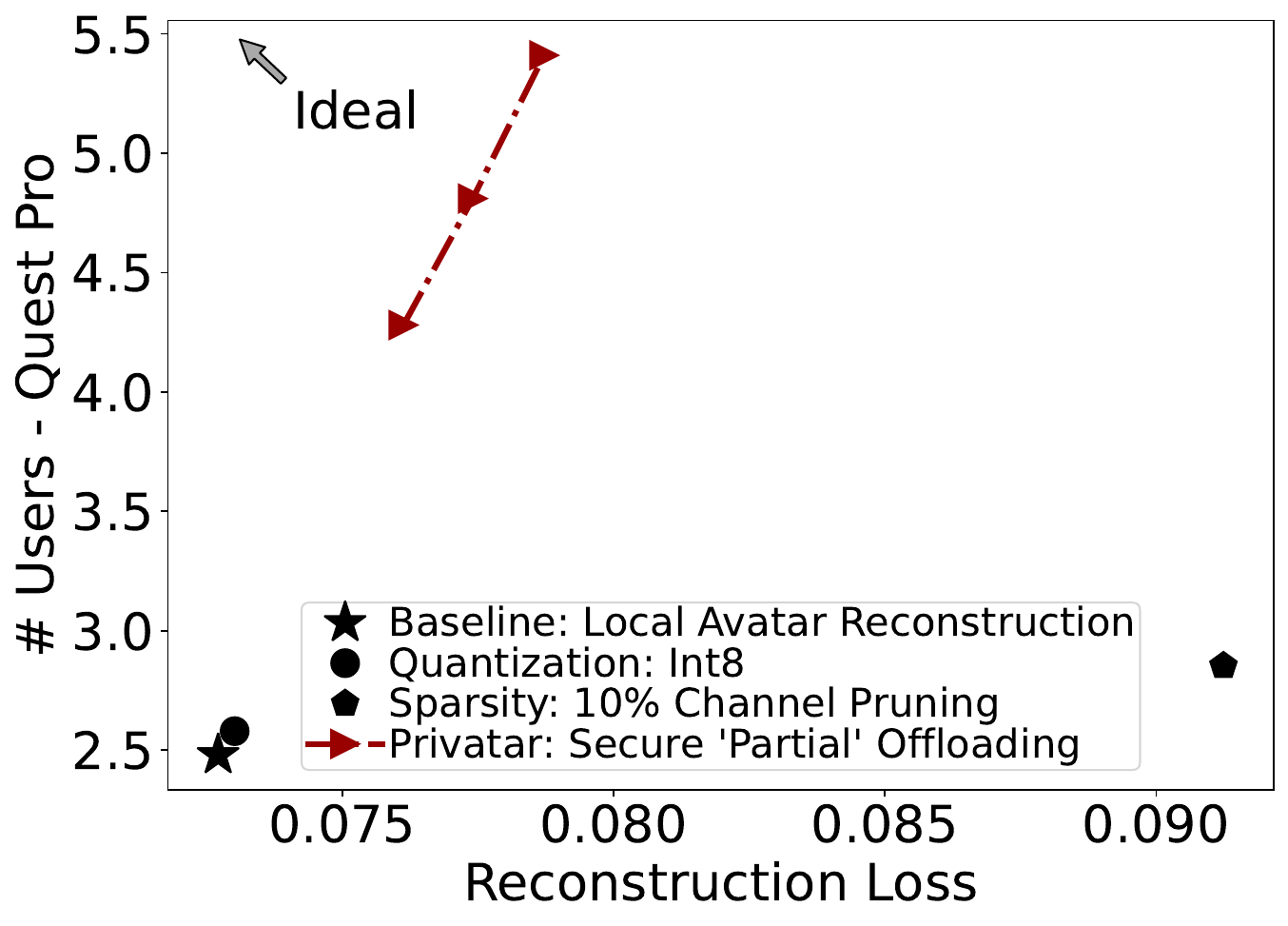}}
        \\ \footnotesize  (g) Latency–accuracy tradeoff in all settings.
    \end{minipage}
    \vspace{-3mm}
    \caption{\textbf{A visual and quantitative comparison among efficient avatar reconstruction techniques}. 
    \textbf{\framework (Secure Partial Offloading, 1f)} achieves 2.37$\times$ throughput (\#users per second) than SotAs (1b$\sim$1e) under similar quality (loss). FO: Fully Offloading.}
    \label{fig:reconstruction_quality_complete_outsource}
    \vspace{-2mm}
\end{figure*}


$\bullet$ Horizontal Partitioning (HP): Horizontally split of the avatar reconstruction into two independent paths for locally processing on device and offloading to untrusted devices, and a frequency-domain split of unwrapped facial textures that enables fine-grained data/compute relocation among two paths. This reduces local computation while only restricts the untrusted devices to see an incomplete view of the private data for privacy protection, improving scalability while maintaining privacy. 

$\bullet$ Distribution-Aware Minimal Perturbation (DAMP): A dimension-wise noise determination mechanism that exploits the entropy of private data (statistical distributions) into minimizing perturbation (noise) needed to achieve provable privacy protection. DAMP calibrates multi-dimensional noises for individually released data via PAC privacy, yielding markedly better reconstruction quality (less reconstruction loss) than local DP at the same privacy level. 

$\bullet$ \framework: Combined, HP reduce offloaded information to an incompleted subset, reducing noises needed by DAMP, and DAMP further augments the empirical privacy protection of HP to a formal guarantee against arbitrary adversaries. Our evaluation on commercial VR hardware (Meta Quest Pro) shows that \framework supports up-to 3 more users (2.37$\times$) with negligible 5.7\% accuracy degradation and only a minor (9\%) increase in energy consumption, as shown in \figref{fig:reconstruction_quality_complete_outsource}. \framework enables a better throughput-loss Pareto frontier over quantization,
sparsity, and local processing baseline, and is empirically robust against NN-based expression identification attacks, and provable robust against arbitrary block-box attacks.

\section{Background}
\label{sec:BackgroundMotivation}
\begin{figure}[h]
    \vspace{-3mm}
    \centering
    \includegraphics[width=0.5\textwidth]{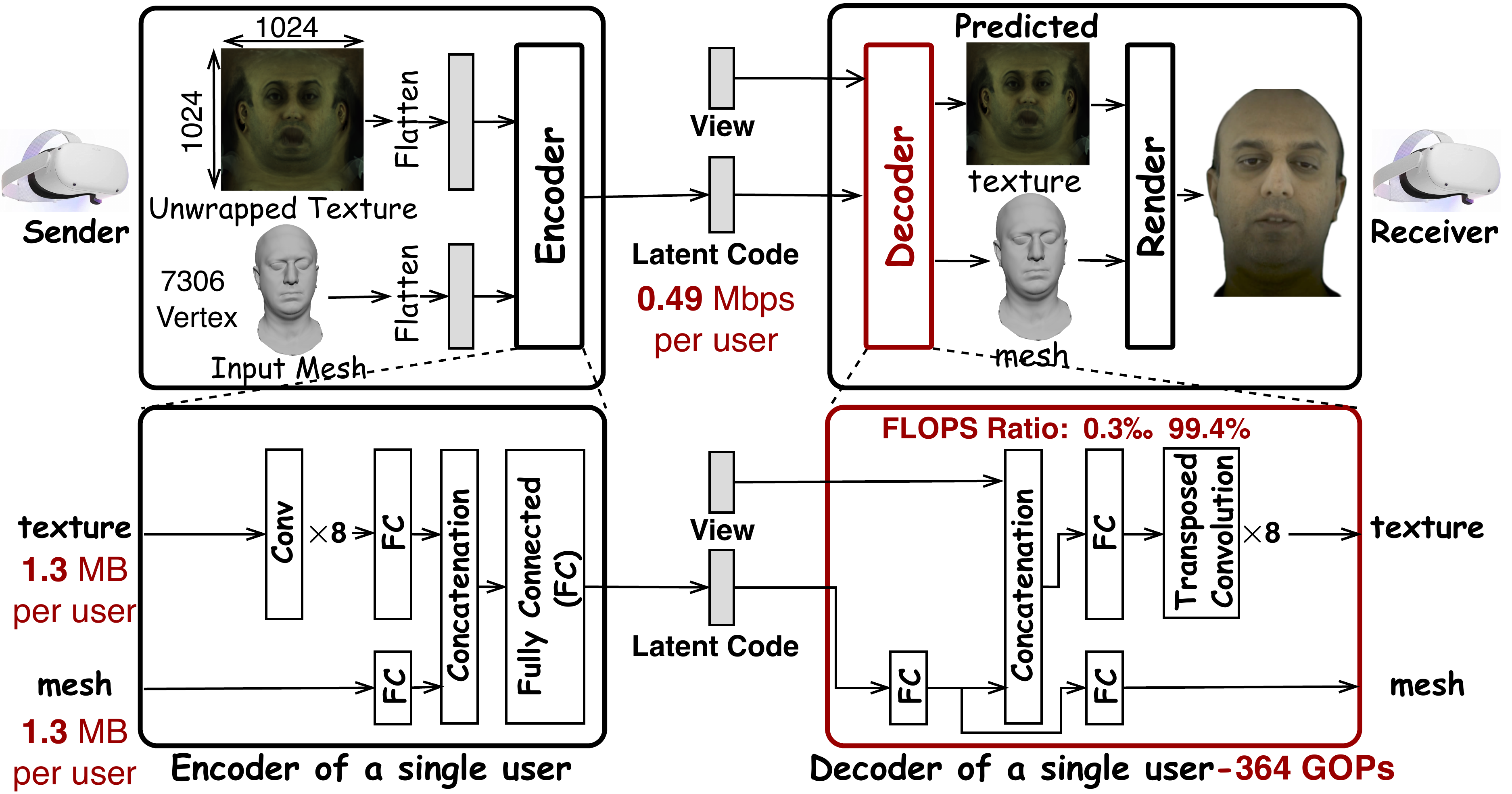}
    \vspace{-8mm}
    \caption{\textbf{Overview of the Variational Auto-Encoder (VAE) architecture used for avatar reconstruction.} \textbf{\textit{Takeaway:}} The texture reconstruction stage (transposed convolution) accounts for 99.4\% of the decoder’s FLOPs, forming the primary computational bottleneck that \framework mitigates through partial offloading.}
    \vspace{-4mm}
    \label{fig:avatar_reconstruction}
\end{figure}

\begin{figure*}
\subfloat[Setup and Threat Model \label{fig:threat_model}]{{\includegraphics[width=0.35\textwidth]{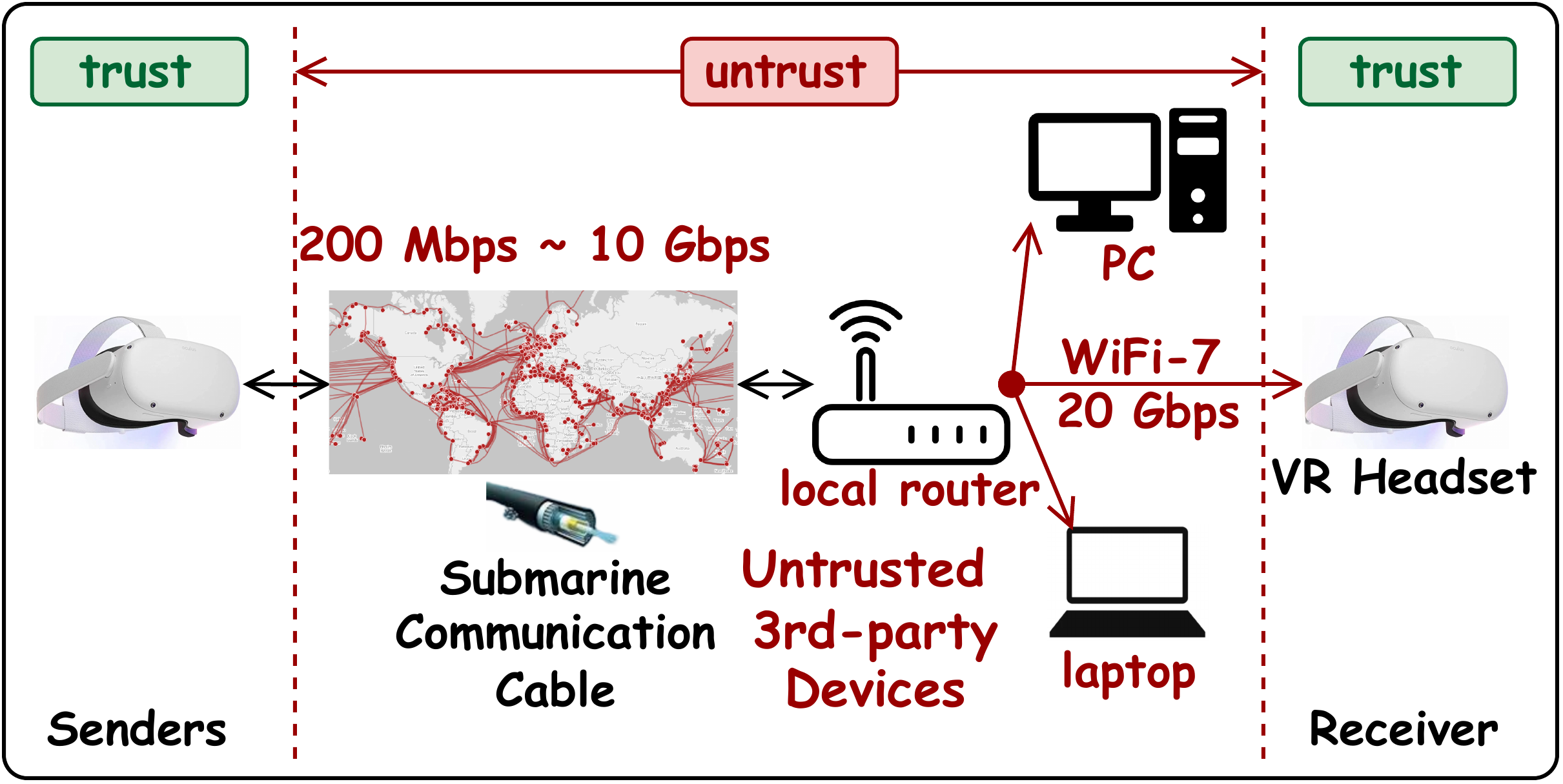}}} \ 
\subfloat[Baseline: Local Reconstruction \label{fig:all_local}]{{\includegraphics[width=0.285\textwidth]{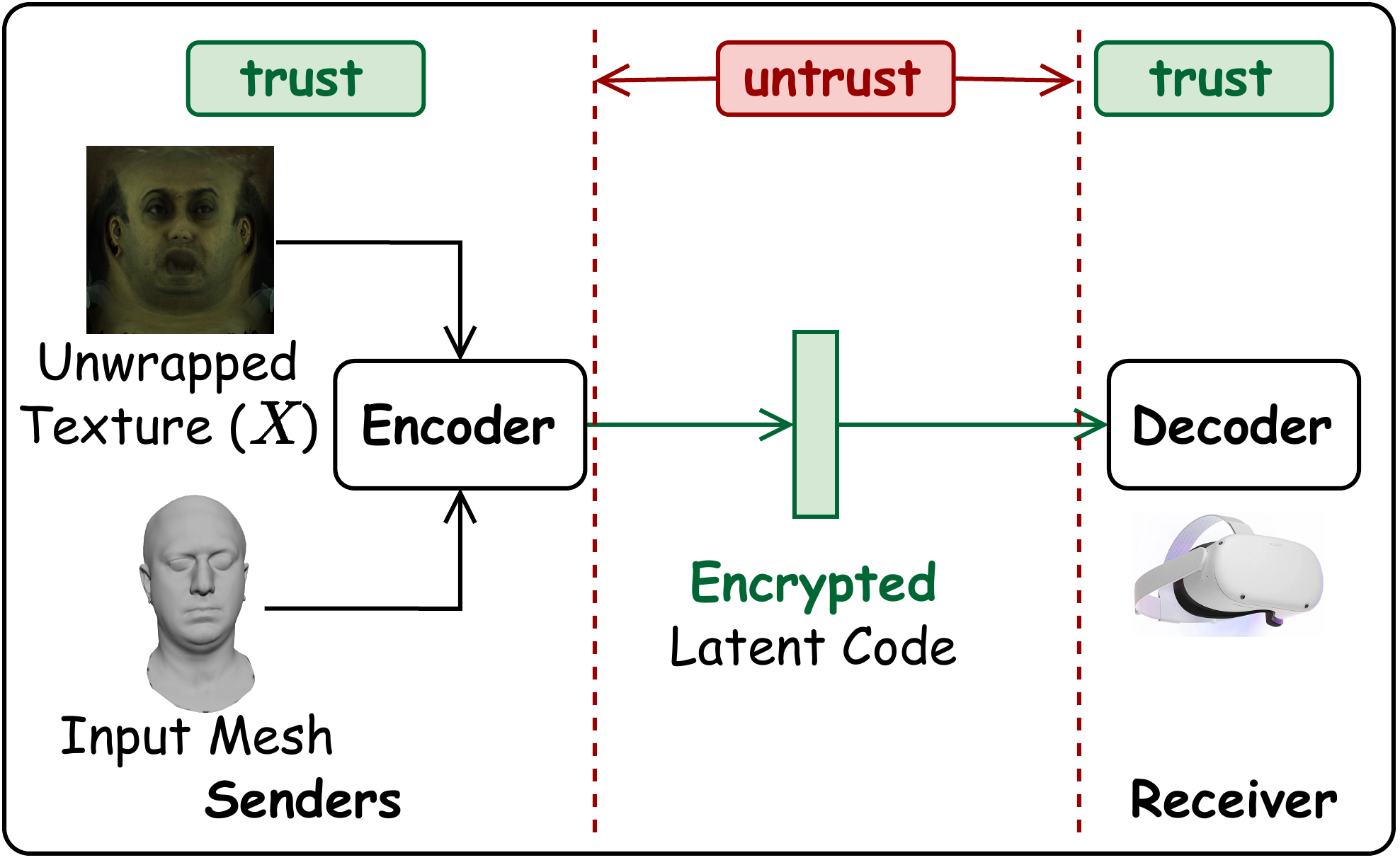}}} \ 
\subfloat[Privatar (Local + Offloading) \label{fig:privatar_offload}]{{\includegraphics[width=0.355\textwidth]{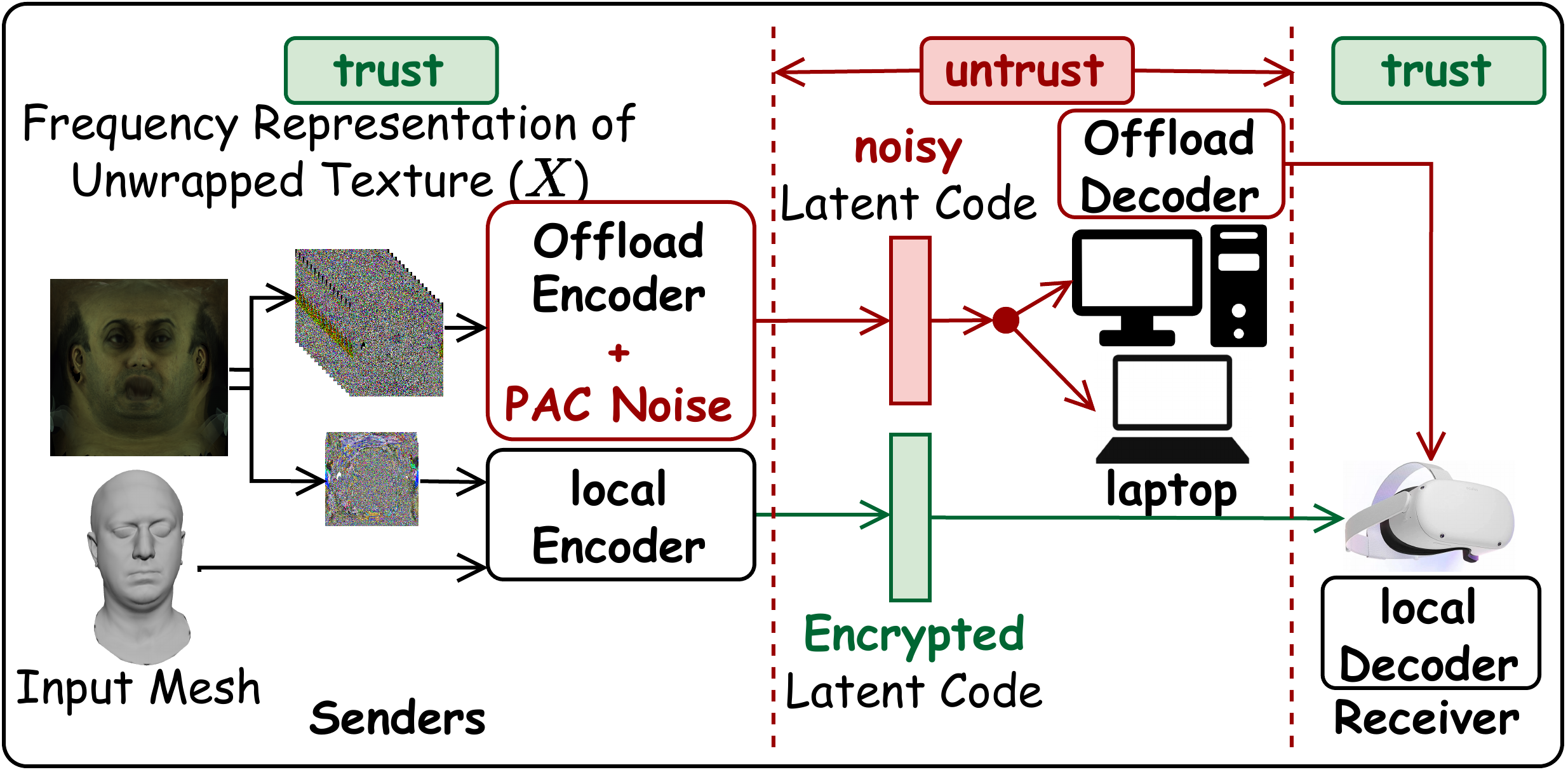}}} 
\vspace{-3mm}
\caption{\textbf{Overview of \framework against baseline under realistic threat model.} \textbf{(\ref{fig:threat_model}) Setup and Threat Model:} In a multi-user VR session, each user only trust its own VR headset. \textbf{(\ref{fig:all_local}) Baseline} processes \textbf{entire decoder} locally on receiver headset. It ensures privacy but the limited compute on device is not sufficient to support many users. 
 \textbf{(\ref{fig:privatar_offload}) Privatar (This work)}: Our framework introduces Horizontal Partitioning, which splits the reconstruction into two paths and offloaded one path to untrusted PC to reduce local computation. Local Path (Green): The relatively privacy-sensitive data, such as the facial mesh and the highly identifying frequency components of the texture, are processed securely on the trusted receiver headset. Offloaded Path (Red): Less-sensitive, but computationally intensive, frequency components are perturbed with minimal, precisely calibrated noise and offloaded to untrusted devices. \textbf{\textit{Key Innovation:}} Privatar prevents the untrusted device from ever accessing the complete private facial data. By only processing a partial and obfuscated frequency components, it cannot reconstruct user's sensitive expression, achieving both scalability and privacy without sacrificing quality.}
\label{fig:setup}
\vspace{-2.5mm}
\end{figure*}

\subsection{Facial Avatar Reconstruction Pipeline}

An avatar needs two key data: a high-resolution unwrapped texture (the ``facial skin", $\sim$1.3 MB) and a detailed facial mesh (the 3D model structure, $\sim$1.3 MB) in \figref{fig:avatar_reconstruction}. Transmitting raw data for every user at a smooth 60 frames per second (FPS) is impractical. It would require $\sim$1.25 Gbps per user, which far exceeds the limited and unstable bandwidth of internet connections, which can be as low as 200 Mbps, as shown in \figref{fig:threat_model}. To reduce bandwidth, modern VR systems use a Variational Auto-Encoder (VAE)~\cite{Lombardi_2018, ma2021pixelcodecavatars}.

$\bullet$ On the sender's end, an \textbf{encoder} compresses texture and mesh into a compact ``\textit{latent code}". This reduces required bandwidth by over 99\%, to just 0.49 Mbps per user, making transmission over limited-bandwidth submarine cable feasible, as shown in \figref{fig:threat_model}.

$\bullet$ On the receiver's end, a \textbf{decoder} receives this compact latent code and decompresses it to reconstruct the original high-resolution texture and mesh for avatar rendering.
\subsection{Compute Bottleneck in Multi-user Reconstruction}
While VAE solves the bandwidth problem, it introduces a new one: a severe computational bottleneck on the receiver's headset. In a multi-user session, the headset must decode the latent code from all senders simultaneously. 

The computation demand of decoder quickly overwhelms the limited computational budget of a standalone VR headset, restricting the number of users that can be supported. For instance, a commercial headset like the Meta Quest Pro has a computational capacity of 902 GFLOPS, which supports a maximum of \textit{two} users at 60 FPS before losing FPS. The bottleneck is the decoder of the pipeline. Within the decoder, a series of eight transposed convolution layers responsible for reconstructing the unwrapped texture accounts for 99.4\% of the total computation, as shown in \figref{fig:avatar_reconstruction}. In contrast, reconstructing the facial mesh is computationally trivial, contributing only 0.6\% overall FLOPS of a decoder.

\subsection{Vulnerability of Facial Avatars}
Facial avatars, built from textures and expression-driven meshes, introduce additional risks of identity leakage (via facial recognition or soft-biometric profiling) and expression leakage (revealing emotion or cognitive state).

\textbf{Identity Leakage:}
The facial texture carries rich biometric information that uniquely identifies individuals, similar to a face image in traditional vision datasets. Leveraging facial textures, adversaries can recover identity through facial recognition models or soft-biometric profiling, linking avatars back to real-world users~\cite{nair2023uniqueid}. In large-scale multi-user environments, this enables cross-session tracking and deanonymization attacks.

\textbf{Expression and Emotion Leakage:}
Facial meshes encode temporally dynamic expression parameters, such as smiles, frowns, and subtle muscle movements, which can reveal a user’s mood, cognitive state, or reactions to specific stimuli. Studies show that such motion telemetry can leak private emotional or health-related information, even when identity is obfuscated~\cite{nguyen2024penetrationvision}.



In summary, reducing local computational overhead of decoder is the critical challenge for enabling VR experiences with a higher number of simultaneous users. This work proposes a \textit{secure partial offloading mechanism to achieve this without compromising user privacy or visual quality.}

\section{Challenges and Key Insights}
This section presents multi-user VR setup in \figref{fig:threat_model}, defines corresponding threat model, outlines two key challenges in secure offloading, and highlights two observations that guide Privatar’s innovations on \textit{what to offload} and \textit{how to add noise to offloaded data for privacy protection}.

\label{sec:problem_formulation}

\subsection{Threat Model of Reconstruction Offloading}

\textbf{Setup:} In multi-user VR session, each participant wears a headset located at a different physical site (e.g., user's homes or offices). Headsets communicate with one another through the Internet via a local router, which represents the household or institutional network gateway connecting the headset to external infrastructure and nearby personal devices (e.g., PC, laptop). All traffic leaving the headset traverses this local router and wide-area links (e.g., submarine communication cables), exposing the data to untrusted local and network-side devices that can intercept or analyze offloaded information, as shown in \figref{fig:threat_model}.

\textbf{Trust Boundary:}  We aim to provide provable and strong privacy definitions similar to local DP~\cite{local-DP}, where a user only trusts its own local headset, and may not trust any other devices even under the same local network, and will randomize his private data locally to ensure privacy before releasing, which leads to:
\squishlist
\item \textbf{Untrusted network} and hence any third-party device connected to network, including the router, laptops and PCs. Only VR headsets involved are trusted.
\item \textbf{Untrusted communication channel} between any device and receiver VR headsets, as shown in \figref{fig:threat_model}.
\squishend

\textbf{Protected Asset:} facial expressions are primary assets requiring protection. A leakage can reveal personal identification~\cite{linkage_attack} and motion~\cite{yang2024vrprotect}.

\textbf{Adversary:} We consider a computationally-unbounded adversary who can observe users' release and has {\em full knowledge} on the following: 1) encoding and noise mechanism applied by the user, 2) the underlying distribution of each user's facial data. That is to say, the only thing the adversary does not know is the randomness in both user's facial data generation and encoding and perturbation mechanisms. For the formal, provable privacy analysis (\secref{sec:formal_privacy_guarantee}), \framework ensures protection of any selected private asset (e.g., identity, expression etc.) from user’s release against an arbitrary adversary attempting to extract that asset.
To complement this formal guarantee with concrete empirical validations, we instantiate two adversaries as ``expression identification attacker", whose goal is to correctly infer the user’s facial expression (\figref{fig:empirical_attack}, detailed in \secref{sec:formal_guarantee_empirical_attack}). This represents real-world attack to demonstrate the effectiveness of \framework. 

\subsection{Challenges in Reconstruction Offloading}
\label{sec:fully_offloading}
Offloading reconstruction requires sending latent codes of users and decoder from the receiver VR headset to untrusted devices. This involves exposing sensitive user facial data and introduces privacy concerns. Privacy protection requires increasing the ambiguity of offloaded data, i.e. force different offloaded data to appear similar, thus making it difficult for adversaries to distinguish between them. 

Arguably, noise is the most popular way of randomizing data to increase ambiguity. For individually released latent code, the SotA local Differential Privacy (DP)-based randomization solutions \cite{croft2021obfuscation, practical_image_obfuscation_provable_privacy, survey_dp} mostly apply uniform noise to all dimensions of offloaded data, also called isotropic noise. However, adding DP-based noise perturbation into offloaded latent codes either suffers from privacy leakage or ruins reconstruction quality. Our empirical results show:

\squishlist
\item With low noise, an ML attacker can identify expressions with 86.15\% accuracy (\tabref{tab:attack_cmp}), a clear privacy breach.
\item With enough noise for privacy, the reconstruction quality is ruined, as shown in Fully Offloading (FO in \figref{fig:reconstruction_quality_complete_outsource}), with a 105$\times$ increase in reconstruction loss compared to the baseline (local avatar reconstruction).
\squishend

Such a flaw of imbalance between privacy and utility (quality) boils down to two key reasons.

$\bullet$ \textbf{Offloading the Entire Asset:} It is contradictory to make \textbf{\textit{entire}} offloaded data to be \textit{\textbf{ambiguous}} for high privacy protection and \textbf{\textit{accurate}} for high  reconstruction quality at the same time. High ambiguity requires high noises while high accuracy calls for low noises.

$\bullet$ \textbf{Prohibitive Isotropic Noise:} In avatar reconstruction, data are released individually without being aggregated. This invalidates typical DP for lack of a aggregation of users to hide the private data of a single one. The local-DP~\cite{local-DP} is the only viable solution, which typically uses isotropic noise, applying uniform level of randomization across all dimensions of the data. This noise is conservatively calibrated to worst-case sensitivity (the maximal possible change in data). As a result, dimensions with small values receive the same large amount of noise as the max-value dimension, causing the noise to overwhelm the actual signal and severely degrade reconstruction quality.

These two challenges make secure offloading a non-trivial open question, motivating \framework (\figref{fig:privatar_offload}).

\subsection{Two Key \framework Observations and Insights}
\framework balances the privacy and accuracy by addressing two failures with two core insights.

\subsubsection{Insight 1: Imbalanced Frequency Distribution}
\begin{figure}
\centering
\subfloat[\(L_2\) Norm of Components \label{fig:frequency_info_distribution}]{{\includegraphics[width=0.21\textwidth]{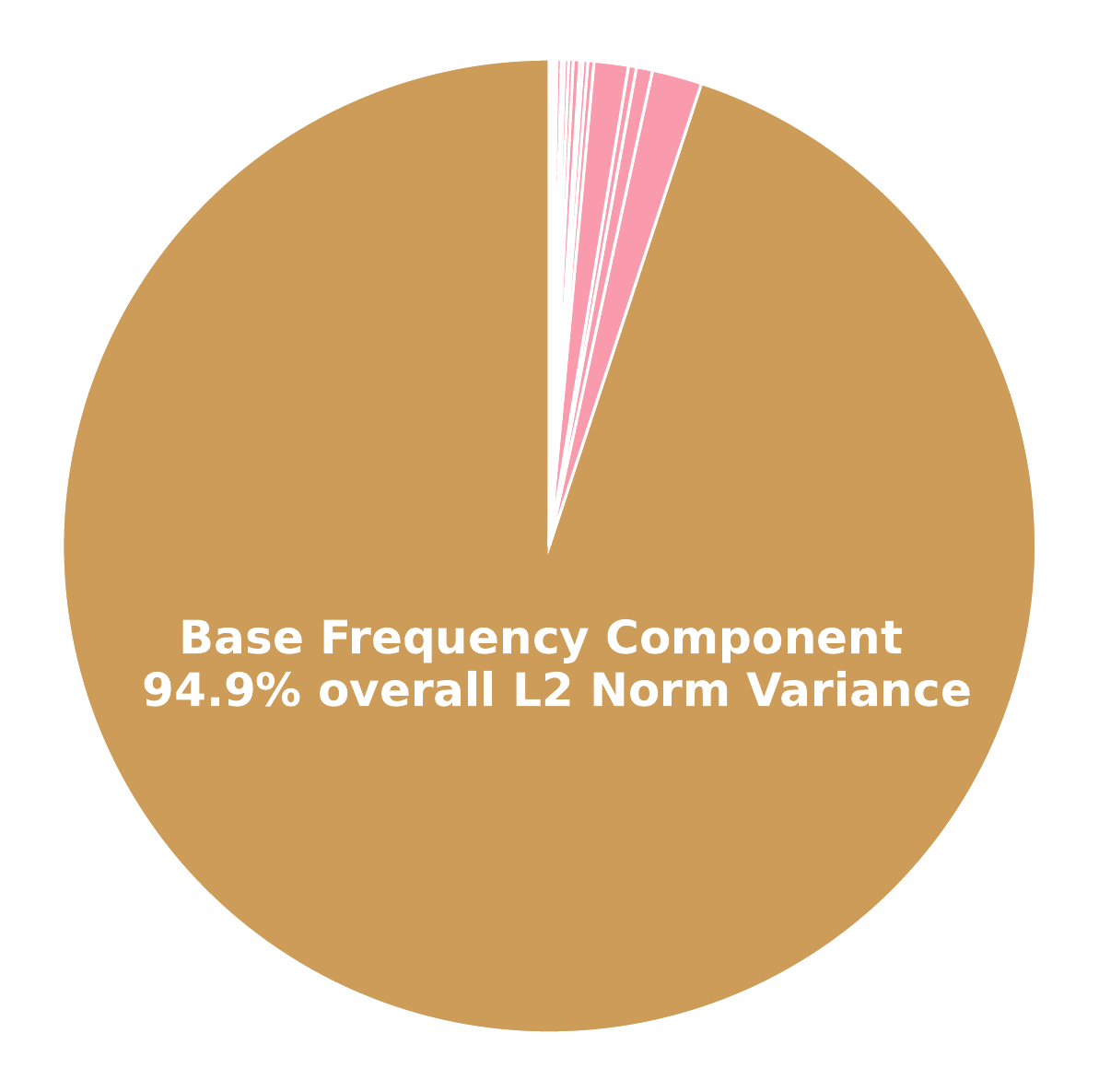}}}
\subfloat[Distribution Comparison\label{fig:l2_norm_comparison}]{{\includegraphics[width=0.24\textwidth]{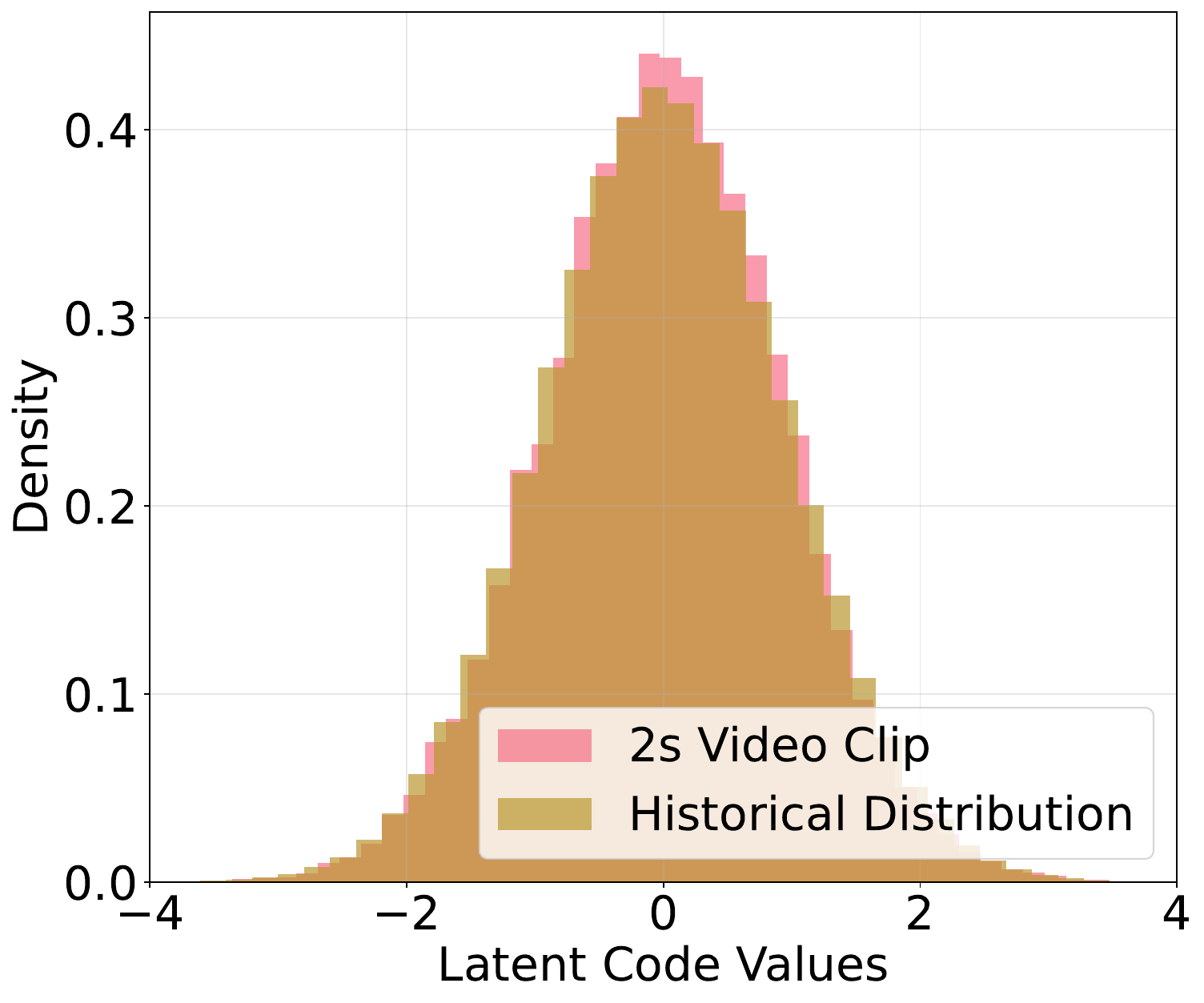}}} 
\vspace{-3mm}
\caption{\textbf{Illustration of two key observations}. \textbf{\textit{Takeaway:}} (\ref{fig:frequency_info_distribution}) The energy ($L_2$ norm) has extremely imbalanced distribution over frequency. Base frequency component contains 94.9\% energy of entire data. (\ref{fig:l2_norm_comparison}) Historical distribution (training dataset, \secref{sec:Experiments}) are similar to distribution of random 2 second clip of avatar trace.}
\vspace{-5.5mm}
\end{figure}

We observe that entropy of private facial information has extreme imbalanced distribution over frequency, as illustrated in \figref{fig:frequency_info_distribution}. Specifically, after decomposing the facial unwrapped texture into a spectrum of sixteen frequency components, the \(L_2\) norm\footnote{\(L_2\) norm (squared Euclidean length) often serves as a proxy for ``energy" as it measures energy in an orthonormal basis.} of the base frequency component contributes to 94.9\% of that in the original unwrapped facial texture while these other 15 frequency components overall contribute to $\sim 5$\%, as shown in \figref{fig:frequency_info_distribution}. This indicates the base frequency component contains higher energy than the union of all remaining frequency components. Therefore, instead of offloading the entire unwrapped texture, Privatar only \textit{\textbf{offloads a partial set of non-base components with less \(L_2\) norm}}, keeping the high-energy components exclusively on the trusted VR headset. Privacy is therefore achieved through the inherent difficulty of reconstructing a complete face from incomplete data, offering empirical privacy protection for the expression identification attack, detailed in \secref{sec:formal_guarantee_empirical_attack}.

\begin{figure*}[t!]
    \centering
    \includegraphics[width=\textwidth]{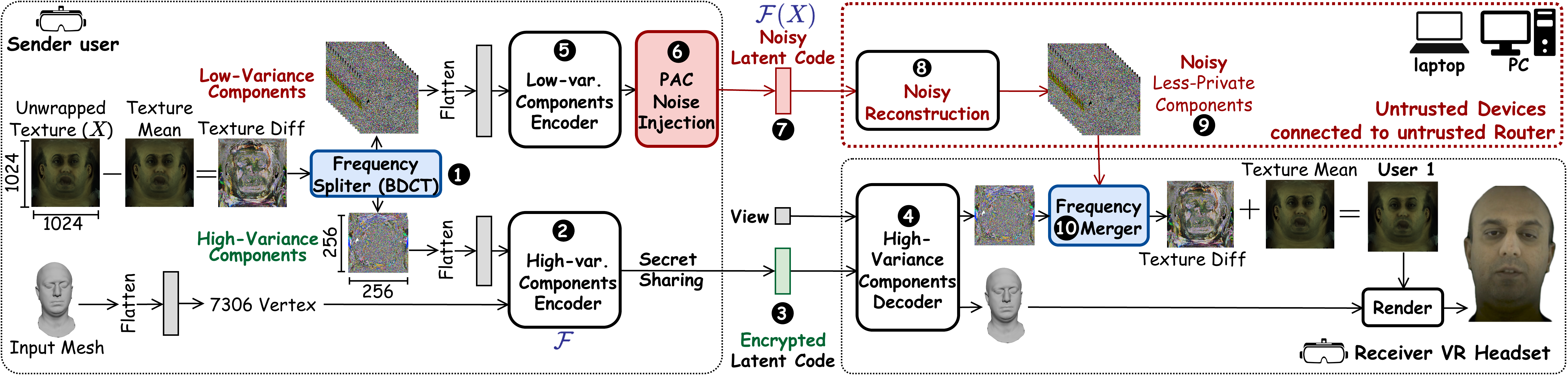}
    \vspace{-8mm}
    \caption{\textbf{Overview of \framework avatar reconstruction flow.} VR reconstruction is \textit{horizontally partitioned} into two paths, running on the local device (bottom right) and untrusted devices (upper right red dashed box), separately. Noisy reconstruction is highlighted in red.}
    \label{fig:HP_overview}
    \vspace{-2mm}
\end{figure*}

\subsubsection{Insight 2: User-Unique Slowly Drifting Distribution}

To further provide a proved guarantee for arbitrary attacks, \framework adds noise to offloaded frequency components. Albeit the reduction of information from the offloaded \textit{partial} frequency components, the noise required for provable privacy protection could still be prohibitive when following local DP-based noise determination. It is because noise of all dimensions has to be the same large to hide potential highest possible dimensional value to ensure privacy guarantee for arbitrary cases, i.e. \textit{isotropic noise without consideration of the data distribution.} We observe that such worst-case protection is over conservative for avatar reconstruction, as the distribution of each user's expression is slowly changing over time, i.e. being relatively static within a short duration, which could be used to reduce noises for dimensions with less entropy. 

Intuitively, the extreme dynamite of expressions a person could make cannot drastically change from what he could make before, e.g. a person could not suddenly make his mouth to be 2$\times$ wider than the largest mouth that he could make before.  Quantitatively, this is reflected by a slowly changing distribution. Specifically, for a randomly selected user, we show in \figref{fig:l2_norm_comparison} that the distribution of latent code for expressions in training dataset is \emph{almost identical} to that for expressions of randomly selected 2 seconds (120 frames). This motivates the \framework's \textit{Distribution-Aware Minimal Perturbation (DAMP)}, which \textit{\textbf{leverages the latest statistical distribution of each user's expressions}} to apply \textbf{\textit{non-uniform, minimum noise}}, i.e. more for high-variance dimensions, less for low-variance ones. And DAMP keeps updating the statistical distribution with new expressions online to track distribution in the long run. This tailored approach provides a formal privacy guarantee while reducing required noise by up-to $17.6\times$ compared to DP, preserving high reconstruction accuracy, as quantified later in \figref{fig:privacy_vs_num_offloaded_components}.

\section{\framework}
\label{sec:Methodology}

\begin{figure*}[t!]
\subfloat{{\includegraphics[width=0.057\textwidth]{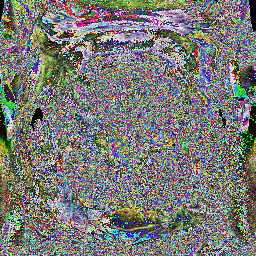}}} \ 
\subfloat{{\includegraphics[width=0.057\textwidth]{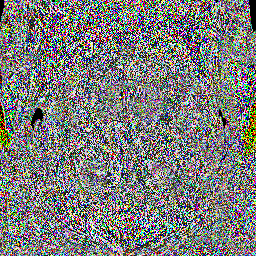}}} \ 
\subfloat{{\includegraphics[width=0.057\textwidth]{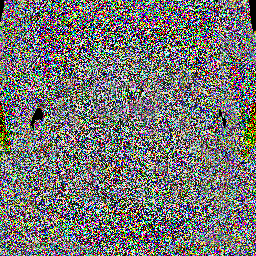}}} \ 
\subfloat{{\includegraphics[width=0.057\textwidth]{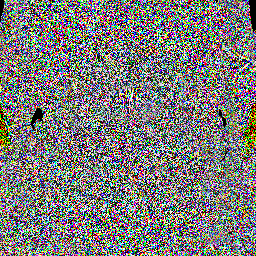}}} \ 
\subfloat{{\includegraphics[width=0.057\textwidth]{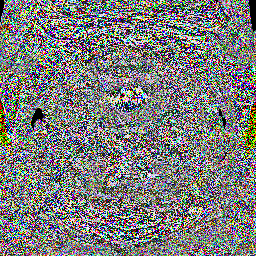}}} \ 
\subfloat{{\includegraphics[width=0.057\textwidth]{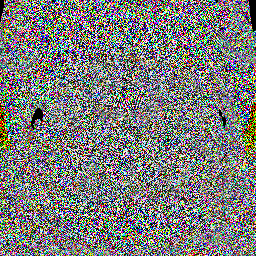}}} \ 
\subfloat{{\includegraphics[width=0.057\textwidth]{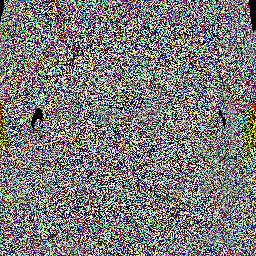}}} \
\subfloat{{\includegraphics[width=0.057\textwidth]{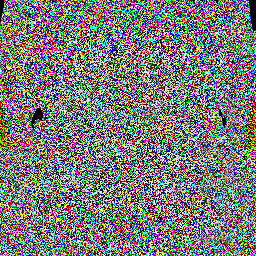}}} \
\subfloat{{\includegraphics[width=0.057\textwidth]{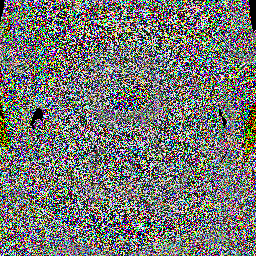}}} \ 
\subfloat{{\includegraphics[width=0.057\textwidth]{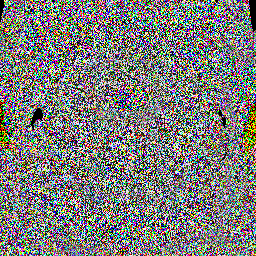}}} \ 
\subfloat{{\includegraphics[width=0.057\textwidth]{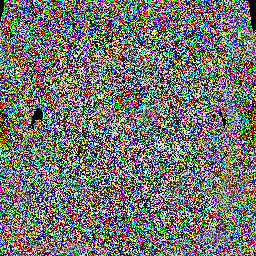}}} \ 
\subfloat{{\includegraphics[width=0.057\textwidth]{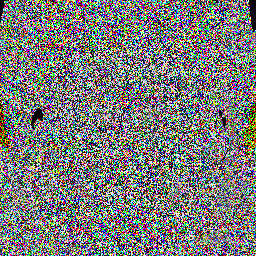}}} \ 
\subfloat{{\includegraphics[width=0.057\textwidth]{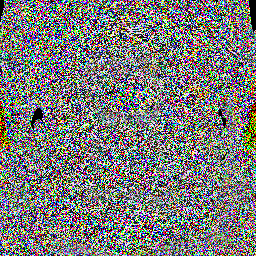}}} \ 
\subfloat{{\includegraphics[width=0.057\textwidth]{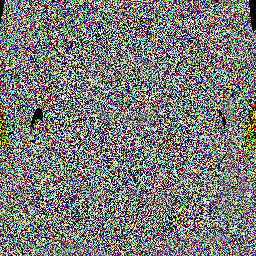}}} \ 
\subfloat{{\includegraphics[width=0.057\textwidth]{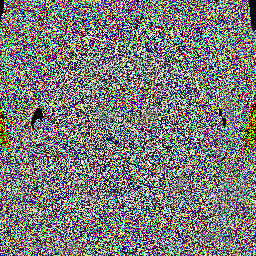}}} \ 
\subfloat{{\includegraphics[width=0.057\textwidth]{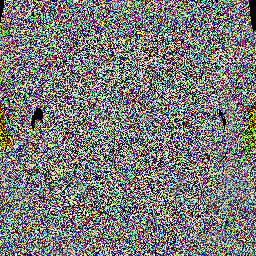}}} \ \\
\subfloat{{\includegraphics[width=\textwidth]{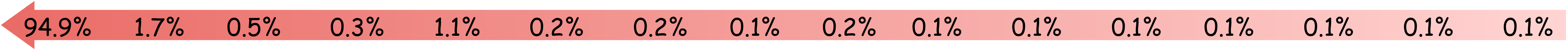}}}
\vspace{-5mm}
\caption{\textbf{Illustration of 16 frequency components and ratio of energy of each in overall components} after applying block DCT with $B=4$ to the difference between input unwrapped texture and its average value in \figref{fig:HP_overview}. Frequency increases from left to the right. Visually, all components look like random data, protecting the privacy of user expression information. The bottom bar shows profiled \(L_2\) norm variance of each frequency component. The base frequency component has the highest \(L_2\) norm ($94.9\%$ of overall \(L_2\) norm), indicating a highest energy. Other frequency components have similarly small \(L_2\) norm, much lower than that of the base frequency component. \textit{\textbf{Takeaway:} the base frequency component carries the most energy. HP keeps it local and only offloads low-energy components to reduce noises needed for obfuscation, improving utility.} HP gives empirical protection as only subset instead of all components are offloaded, exposing a partial view of information, while formal privacy guarantee is obtained from DAMP through noise perturbation.}
\label{fig:Compare_to_SoTA}
\vspace{-4mm}
\end{figure*}

\textbf{Goal.} Enable secure offloading a subset of frequency components in private facial unwrapped texture to reduce computation overhead in the receiver VR headset, so that (i) the high-energy frequency components of texture and facial mesh never leave the trusted headset, (ii) low-energy components are safely offloaded with minimal noise perturbation.

\subsection{Horizontal Partitioning (HP)}
\label{sec:hp}

\subsubsection{HP Overview}
HP splits the encoder–decoder pair of VAE (\figref{fig:avatar_reconstruction}) into two independent parallel paths: a \emph{local path} runs on the trusted receiver headset and an \emph{offloaded path} runs on untrusted devices under the same local network with receiver headset (\figref{fig:HP_overview}). The private facial texture is split in frequency, reconstructed independently, then merged for rendering.

\noindent \textbf{Local Path (\ding{183}$\rightarrow$\ding{184}$\rightarrow$\ding{185})}
The encoder and decoder for selected frequency components run on trusted sender and receiver VR headset, separately, with the latent code to be encrypted during the communication for privacy protection. Encryption/decryption key are shared once ahead of actual transmission~\cite{secret_key_sharing}, and runtime encryption and authentication lead to negligible overhead~\cite{dworkin2007recommendation}.

\noindent \textbf{Offloaded Path (\ding{186}$\rightarrow$\ding{187}$\rightarrow$\ding{188}$\rightarrow$\ding{189}$\rightarrow$\ding{190})} The sender VR headset encodes the remaining components locally, injects distribution-aware noise (\secref{sec:DAMP}), and then offloads noisy latent codes for decoding to untrusted devices. Only \emph{partial and obfuscated} view is ever exposed to reduce leakage.

\subsubsection{Frequency Partitioning of Unwrapped Texture}
\label{sec:ps}

\textbf{Frequency Decomposition of Unwrapped Texture}: Unwrapped facial texture is a 3 dimensional image in $\mathbb{R}^{H\times W \times 3}$. HP first subtracts it by the average texture in training dataset and applies a block Discrete Cosine Transform (DCT)~\cite{ji2022privacypreserving, strang1999discrete} of size $B\times B$ to each non-overlapping block, yielding $B^2$ frequency components, each of shape $\frac{H}{B}\!\times\!\frac{W}{B}\!\times\!3$.\, Each component is $\tfrac{1}{B^2}$ the size of unwrapped texture, so splitting across paths preserves the total size of data while enabling compute relocation. 

\textbf{Variance-Based Partitioning of Freq. Components}: \framework uses the \(L_2\) norm as a proxy of the energy contained in each frequency component to partition all components among local and offloaded paths.

\textbf{Step 1:} Given the goal of local computation reduction, \framework computes the number of components to offload.

\textbf{Step 2:} \framework selects required number of components with lower \(L_2\) norm (variance) for offloading. Note that facial mesh is always kept local for privacy protection.

\noindent \textbf{Differences to original VAE}: Compared to \figref{fig:avatar_reconstruction}, \framework reduces local compute and memory by: (1) downsampling the texture input to $\frac{1}{B}$ spatial resolution, which removes $log_2(B)$ layers from both the local encoder (\ding{183}) and decoder (\ding{185}). This lowers the per-component reconstruction cost to $\approx \frac{1}{B^2}$ of the original VAE, so reconstructing all $B^2$ components has comparable total cost; (2) offloading $X$ of the $B^2$ components leaves only $\frac{B^2-X}{B^2}$ locally ($X!\in![2, B^2!-!2]$), reducing both local memory footprint and computation.

\subsubsection{Utility Effects of Horizontal Partitioning}
Partitioning has negligible utility (quality) effects. By partitioning $B^2$ components across two paths, \framework gives robust privacy protection against expression identification attack (\figref{fig:empirical_attack}). As a result, different choices of partitioning create the design space to trade off privacy, local compute latency, and communication latency. 


\subsubsection{Privacy Protection of Horizontal Partitioning}
HP only provides \emph{empirical} protection. Achieving \emph{provable} privacy for the \textit{individually released user latent codes} still requires noise perturbation; however, HP reduces the required noise by exposing only a subset of texture information.

As shown in \figref{fig:Compare_to_SoTA}, each post-transform component appears noise-like, which suppresses direct expression cues from any single component or any incomplete subset. \emph{Empirically}, HP splits private information across two paths so the offloaded path observes only an incomplete, low-energy view, mitigating expression identification attacks (\figref{fig:empirical_attack}). By limiting exposed content, HP also reduces identity leakage. 

\begin{figure}[t]
    \centering
    \subfloat[Statistical Visualization\label{fig:pac_noise_visualization_value_N_noise}]{{\includegraphics[width=0.247\textwidth]{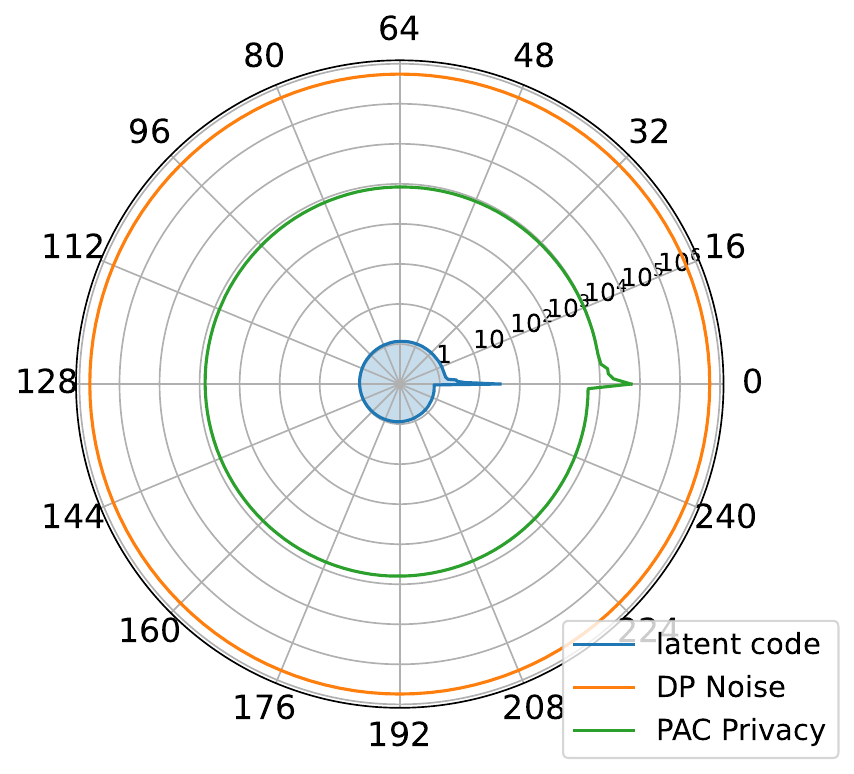}}} \hfil 
    \subfloat[Single Entry Visualization \label{fig:noisy_value_visualization_dp}]
    {{\includegraphics[width=0.227\textwidth]{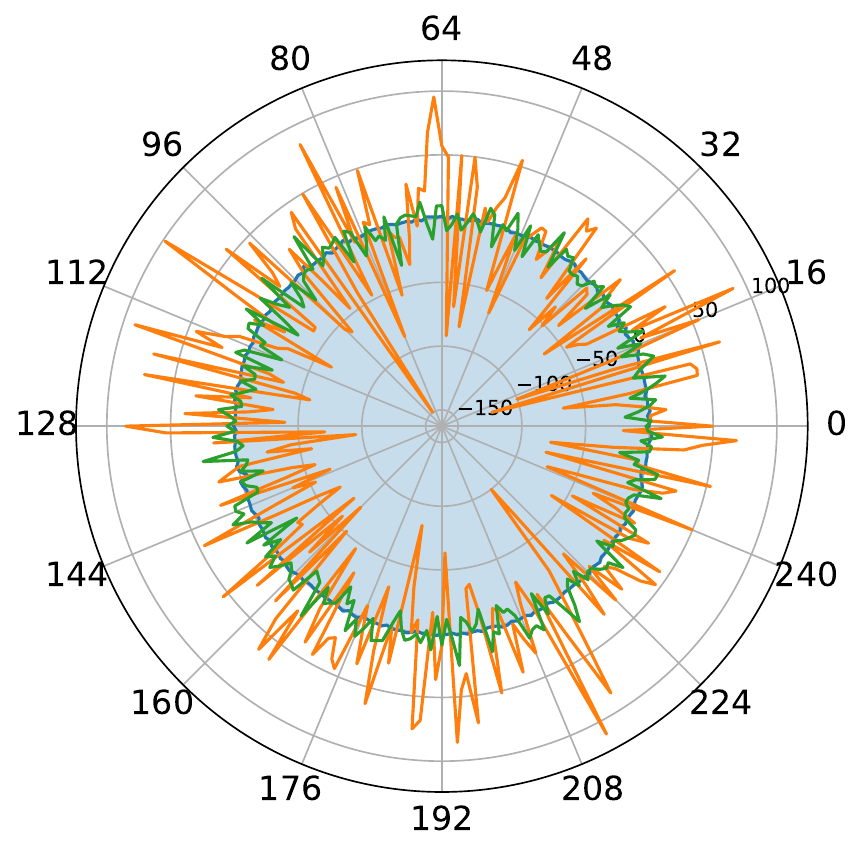}}}
    \vspace{-3mm}
    \caption{\textbf{Statistical noise comparison between Differential Privacy based noise or PAC Privacy based noise in DAMP.} Each direction of the chart shows one dimension of the 256-dimensional latent code. \textbf{\textit{Takeaway:}} Statistically, DAMP leverages the actual distribution of the latent code, blue in (\ref{fig:pac_noise_visualization_value_N_noise}), to reduce the trace of covariance (reflecting the amount of noise statistically) of DP based noise, orange in (\ref{fig:pac_noise_visualization_value_N_noise}), by $10^3$ into PAC based noise, green in (\ref{fig:pac_noise_visualization_value_N_noise}). For each individual released latent code, PAC Privacy reduces the overall noisy value of DP (orange) by 17.6$\times$ into green curve in (\ref{fig:noisy_value_visualization_dp}), reflected by less fluctuation in the wave, significantly reducing accuracy degradation caused by noise.}
    \label{fig:noise_visualization}
    \vspace{-2mm}
\end{figure}
\subsection{Distribution-Aware Min. Perturbation (DAMP)}
\label{sec:DAMP}

Even with HP, providing \emph{provable} privacy for \textit{individually released user latent codes} can require prohibitive noise. We introduce DAMP, which reduces the required noise by leveraging the \textit{statistical distribution} of the offloaded data, thereby minimizing utility loss while maintaining the same provable privacy level for any selected sensitive asset (e.g., identity, expression etc.).

\subsubsection{Key Idea}
\figref{fig:noise_visualization} highlights the core problem: a fundamental mismatch between standard Differential Privacy (DP) and the data's distribution. DP typically injects \textit{isotropic} noise (uniform variance in all dimensions), as shown by the orange ring in \figref{fig:pac_noise_visualization_value_N_noise}. However, the actual distribution of the latent codes is highly \textit{anisotropic} (non-uniform, concentrated in specific dimensions), shown in blue. This mismatch forces DP to add excessive, unnecessary noise in low-variance dimensions, inflating the total ``noise energy" (covariance of generated noise) and severely degrading data utility. DAMP solves this by being distribution-aware. It is the first technique to leverage \textit{PAC privacy}~\cite{xiao2023pac} for time-series facial data, enabling the generation of calibrated, \textit{\textbf{anisotropic}} noise. As shown by the green ring in \figref{fig:pac_noise_visualization_value_N_noise}, DAMP aligns the noise's dimensional distribution with the data's statistical distribution, allocating minimal perturbation to low-variance dimensions. This reduces the covariance of noise by three orders of magnitude. Consequently, in \figref{fig:noisy_value_visualization_dp}, DAMP shrinks the per-sample perturbation error by $17.6\times$ (green vs. orange wave). DAMP thus preserves data utility for the same provable privacy guarantee. 

\subsubsection{Initial Distribution and Runtime Update}
\label{sec:formal_privacy_guarantee}
DAMP initializes with user's historical distribution, and then updates this distribution continuously as the user generates new expressions. During runtime, DAMP tracks the latest distribution, computes the minimum noise needed to meet the target privacy guarantee, and adds this noise to the offloaded latent code. By recalibrating noise as the distribution drifts slowly over time, DAMP preserves provable privacy protection throughout long-term use.

Crucially, latent code distribution is computed and stored entirely on user's local device (e.g., VR headset). This distribution is \textbf{never shared} and only used to calibrate the noise injected into the latent codes before they are offloaded, ensuring the distribution statistics themselves remain private.

\subsubsection{Provable Formal Privacy Guarantee}
\label{sec:formal_privacy_guarantee}

To model privacy leakage and provide formal privacy guarantee, DAMP adopts the \textit{PAC Privacy} framework~\cite{xiao2023pac,formal_proof_pac,pac_thesis,one-sided-pac}, which characterizes privacy through adversary’s inference hardness. The privacy risk is quantified by \textit{posterior success rate}, i.e. the probability that an adversary can correctly infer private information after observing the leakage.

\textbf{Definition} [PAC Privacy \cite{xiao2023pac}, \cite{pac_survery}] 
A processing function $\mathcal{F}$ satisfies $(\delta_{\rho}, \rho, \mathsf{D})$-\textit{PAC Privacy} (terminology listed in \tabref{tab:Terminology}) if the following experiment is {\em impossible}:  

A user samples $X$ from the distribution $\mathsf{D}$ and sends $\mathcal{F}(X)$ to an adversary who knows $\mathsf{D}$ and $\mathcal{F}$. The adversary returns an estimate $\hat{X}$ such that
\[
\Pr\big[\rho(\hat{X}, X) = 1\big] \geq 1 - \delta_{\rho},
\]
where $\rho(\cdot,\cdot)$ is a correctness criterion (e.g., correctly guessing the expression in expression identification attack). 

A smaller $\delta_{\rho}$, i.e., an impossibility of a stronger adversarial reconstruction, intuitively implies a higher privacy risk.

\label{sec:attack_case_introduction}
In multi-user avatar reconstruction, each user’s private texture $X$ (e.g., facial expression) is drawn from a distribution $\mathsf{D}$. When the encoded latent code $\mathcal{F}(X)$ is offloaded to the cloud, we must ensure that any adversary can NOT reconstruct the exact sensitive expression $\hat{X}$ from $\mathcal{F}(X)$, i.e.  $\rho(\hat{X}, X)\neq1$.  
For instance, under an \textit{Expression Identification Attack}, $\rho(\hat{X}, X)=1$ if the adversary correctly classifies the user’s expression (e.g. \textit{laughing} or \textit{surprised}).

The posterior success rate $(1 - \delta_{\rho})$ quantifies this risk. To bound it below a desired threshold for arbitrary adversary, DAMP introduces minimal perturbation noise $\bm{e}$ to the latent code $\mathcal{F}(X)$, as illustrated in \ding{187} of \figref{fig:HP_overview}.

To calibrate $\bm{e}$, we first define the \textit{optimal prior success rate}, the adversary’s best guess \textbf{before} seeing the leakage:
\[
\delta_{\rho,o} = \min_{X' \in \mathcal{X}^*} \Pr_{X \sim \mathsf{D}}\big[\rho(X', X) \neq 1\big]
\]
where $\mathcal{X}^*$ denotes a uniform distribution of private samples. Once $\mathsf{D}$ and $\rho$ are fixed, $(1 - \delta_{\rho,o})$ is determined.
According to~\cite{xiao2023pac}, the posterior success rate $(1-\delta_{\rho})$ is bounded by the mutual information between $X$ and its noisy representation $\mathcal{F}(X) + \bm{e}$:
\begin{equation}
\small
\delta_{\rho}\ln\!\left(\frac{\delta_{\rho}}{\delta_{\rho,o}}\right)
+ (1-\delta_{\rho})\ln\!\left(\frac{1-\delta_{\rho}}{1-\delta_{\rho,o}}\right)
\leq \mathsf{MI}\!\left(X; \mathcal{F}(X) + \bm{e}\right)
\label{main_inequality}
\end{equation}
where $\mathsf{MI}(\cdot;\cdot)$ denotes mutual information, noted as $v$.  
\begin{table}[!t]\centering
\vspace{-2mm}
\caption{\textbf{Formal Privacy Analysis Terminology}}
\label{tab:Terminology}
\scriptsize
\resizebox{\linewidth}{!}{
\begin{tabular}{rl}\hline
Term (Symbol) & Definition \\\hline
Private Data ($X$) & Unwrapped Texture in avatar reconstruction.\\
Data Distribution ($\mathsf{D}$) & Distribution of Private Data ($X$).\\
Processing Function ($\mathcal{F}$) & Encoder (\ding{184}) in avatar reconstruction.\\
Observation ($O$) & The offloaded noisy latent code.\\
Estimation of Secret ($\hat{X}$) & The guess of adversary after observing $O$.\\
Prior Successful Rate ($1-\delta_{\rho,o}$) & The \underline{S}uccessful \underline{R}ate of correct guess \textbf{before} observing $O$.\\
 Posterior SR (PSR, $1-\delta_{\rho}$) & The \underline{S}uccessful \underline{R}ate of correct guess \textbf{after} observing $O$.\\
noise $\bm{e}$ & The multi-dimensional noise $\bm{e}$, same shape as $\mathcal{F}(X)$.\\
mutual information bound $v$ & The mutual information between $X$ and $\mathcal{F}(x)+\bm{e}$.\\
\hline
\end{tabular}}
\vspace{-3mm}
\end{table}

This inequality directly links the added noise $\bm{e}$ to an upper bound on adversarial inference capability, forming the theoretical foundation for DAMP noise calibration. We provide detailed step-by-step noise calculation in \secref{sec:theory_step_noise_calculation}.

\section{Experiments}
\label{sec:Experiments}
In this section, we evaluate whether \framework maintains utility (quality of avatar) and privacy while improving throughput under realistic VR hardware constraints.

\subsection{Setup}

\textbf{Dataset.}
We use Multiface \cite{wuu2023multiface}. The training set contains 13 identities, 65 or 143 labeled expressions per identity, 40 views per expression from various angles, in total $\sim$ 1.7 million frames and $\sim$ 15 TB data. The synthetic test set has 1,172 consecutive frames.

\noindent \textbf{Hardware}  We evaluate on a Meta Quest Pro (902 GFLOPS GPU). The untrusted offload host is a PC with RTX 5090 (non-TEE GPU) and AMD Threadripper 7985WX (secure memory encryption, SME) \cite{amd_sev_snp} as CPU TEE. Additional results for Quest 3 are in \secref{sec:quest3_result}. Headset and PC are connected by WiFi-7 ($\leq$20 Gbps).


\noindent \textbf{\framework configuration.}
 We adopt a VAE \cite{Lombardi_2018,wuu2023multiface} and set B=4, yielding 16 frequency components. We rank components by \(L_2\) norm variance and offload the 14 lowest-variance components. The base component is always local. The offloaded branch is trained for 2 epochs, batch size 10. Noise is calibrated by MI budget 
$v$ following \secref{sec:theory_pac} (default $v=0.1$), and injected into the offloaded latent code. The choices of partitioning should be selected to maximize on-device compute utilization. For reconfigurable hardware \cite{tong2026MINISA, tong2024FEATHER} supporting diverse workload shapes with high compute utilization, finer-grained partitioning exposes a richer tradeoff space.

\begin{figure}[t!]
    \centering
    \includegraphics[width=\linewidth]{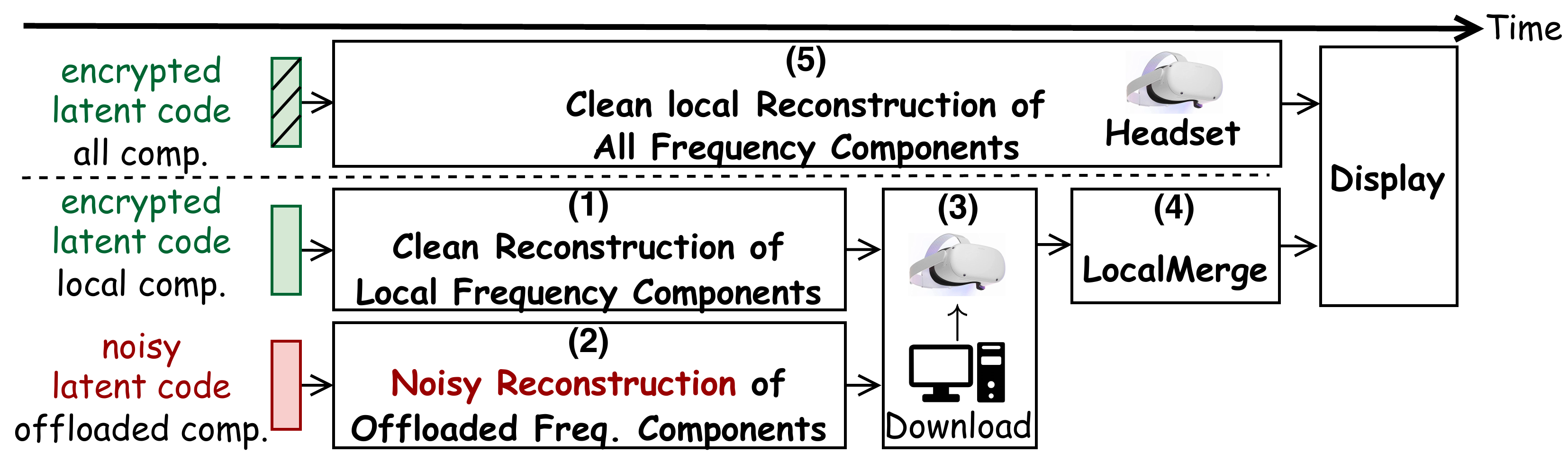}
    \vspace{-8mm}
    \caption{\textbf{Latency breakdown of avatar reconstruction in \framework.} (1–4) show collaborative headset–PC reconstruction. (5) represents full local reconstruction for highest throughput.}
    \label{fig:latency_breakdown_illustration}
    \vspace{-3mm}
\end{figure}
\begin{figure}[t!]
    \centering
    \subfloat[Empirical Att.\label{fig:empirical_attacker}]{{\includegraphics[width=0.17\textwidth]{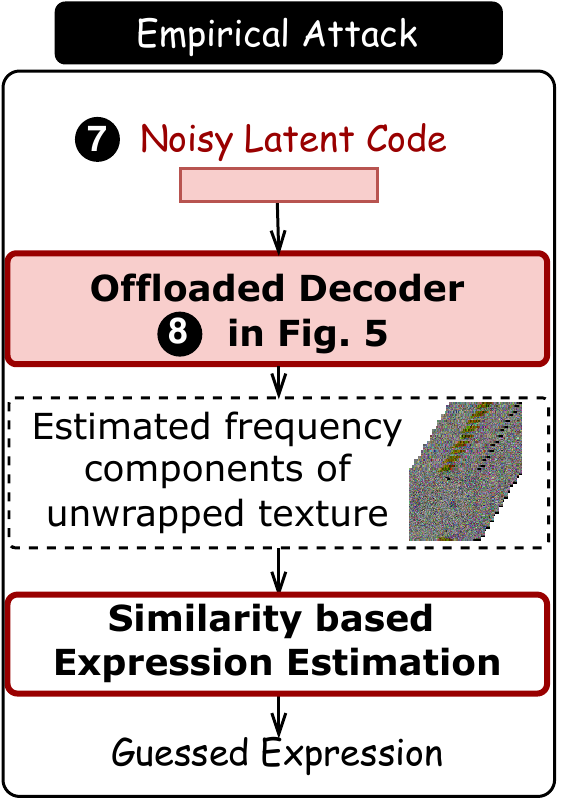}}} \hfil 
    \subfloat[NN-based Attacker \label{fig:nn_attacker}]
    {{\includegraphics[width=0.3\textwidth]{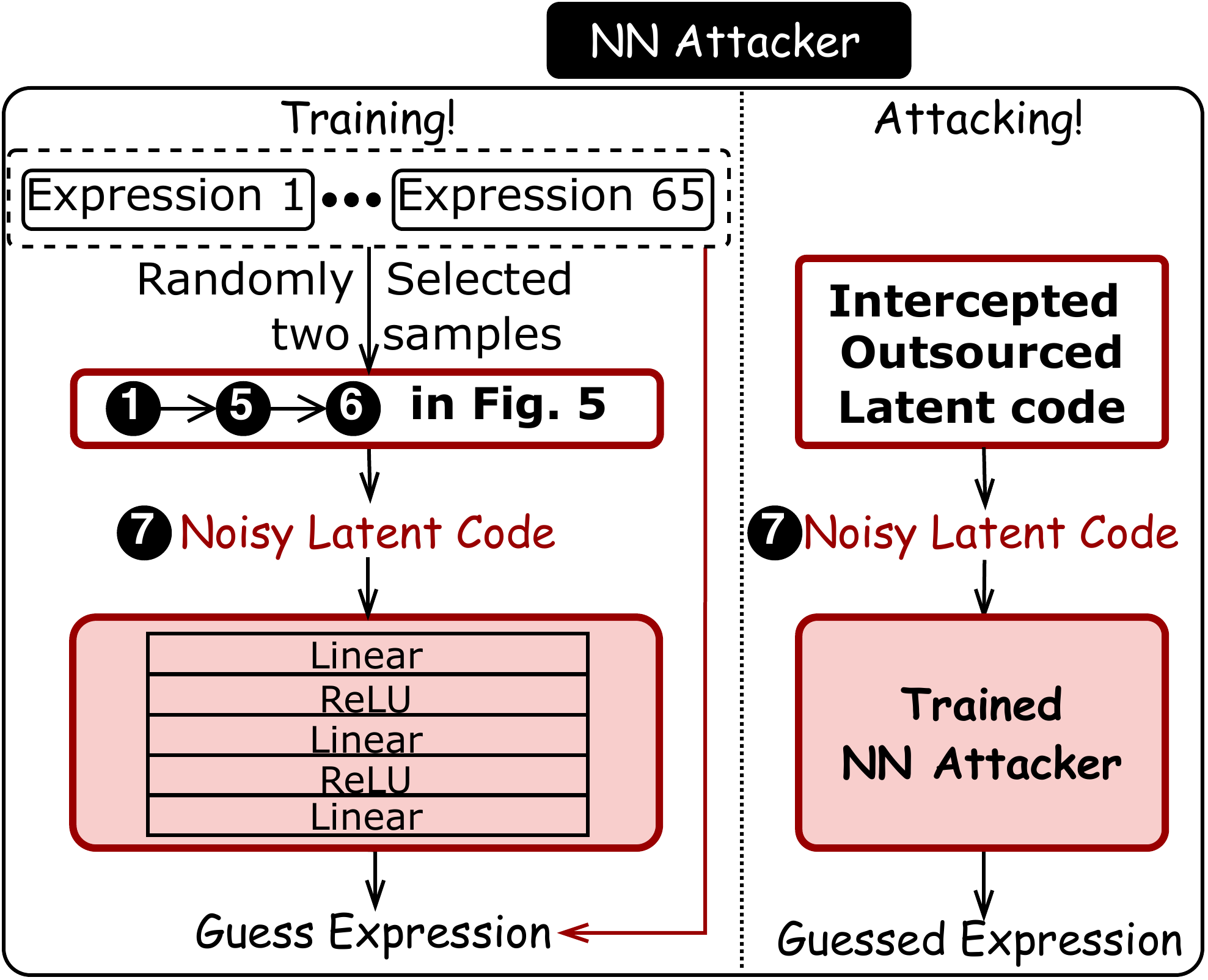}}}
    \vspace{-3mm}
    \caption{\textbf{Illustration of empirical attackers.} (\ref{fig:empirical_attacker}) Empirical attacker takes in reconstructed offloaded components and estimate the expression based on its similarity to all frequency components of two random samples from all expressions. It achieves 86.5\% PSR when offloading all components. (\ref{fig:nn_attacker}) NN attacker has three fully connected layers, and guesses expression from offloaded noisy latent code. Constructions are detailed in \secref{sec:formal_guarantee_empirical_attack}.}
    \vspace{-6mm}
    \label{fig:empirical_attack}
\end{figure}

\noindent \textbf{Baseline.} The entire decoder runs locally on the headset (de-facto VR deployment). On Quest Pro at 60 FPS, the baseline supports 2.48 users with loss as $L_0=0.072$.

\noindent \textbf{Metrics for Loss, Throughput, Privacy, Energy.} \ \hfill \\
\noindent $\bullet$ \emph{\textbf{Loss:}} Mean-squared error (MSE) between reconstructed texture and ground truth. We report normalized loss relative to $L_0=0.072$ of the unpartitioned VAE baseline.\newline
\noindent $\bullet$ \emph{\textbf{Latency:}} We breakdown latency as in \figref{fig:latency_breakdown_illustration}:
(1/4/5) Local compute (measured by roofline model~\cite{roofline}),
(2) Offloaded compute (measured on RTX 5090 or AMD 7985WX),
(3) Communication (download offloaded frequency components from PC to headset via WiFi-7).\newline
\noindent $\bullet$ \emph{\textbf{Throughput:}} We report max concurrent users at 60 FPS, i.e. throughput = \#users/(critical-path-latency$\times$60).\newline
\noindent $\bullet$ \emph{\textbf{Privacy:}} the maximal Posterior Success Rate (PSR) of guessing an expression correctly. A PSR bound $\pi$ means that for \emph{any} adversary, the probability of correctly guessing the expression after observing the release is $\le \pi$. We provide
(a) t-PSR: a theoretical upper bound on posterior success rate following PAC Privacy (\secref{sec:theory_pac}).
(b) e-PSR: the empirical PSR, i.e., the maximum success rate across both (a) empirical attacker and (b) NN-based attacker in \figref{fig:empirical_attack}. \newline
\noindent $\bullet$ \emph{\textbf{Energy:}} compute and communication use 32 GOP/s/W \cite{nerual_comp} and 13.94 nJ/bit \cite{sun2014modeling}.

\vspace{-2mm}
\subsection{\framework vs SotAs}

We compare normalized loss, \#users and empirical PSR of \framework against SotAs in \tabref{tab:baseline_setup}.

\noindent $\bullet$  \textbf{\framework vs Quantization:} Post-training \emph{weight-only} 8-bit quantization of the decoder primarily reduces weight storage and memory loads of it. Total MAC count, i.e. local compute, is unchanged as activations (facial unwrapped texture) require floating-point data type to preserve fidelity. Hence \framework achieves 2.27$\times$ more throughput improvement over compute-bounded quantization at the same privacy protection level of random guess. \newline
\noindent $\bullet$ \textbf{\framework vs Sparsity:} Channel pruning~\cite{he2017channel} (remove 10\% channels and associated kernels with smallest \(L_2\) norm). It does effectively reduce local computation but makes the shape of workload irregular, which requires dedicated hardware support that is not available on Qualcomm Snapdragon XR2+ Gen 1 on Quest Pro. Hence \framework provides 2.05$\times$ higher throughput over sparsity. Higher sparsity ratio introduces unbearable loss increase, and being impractical, and thus we did not compare against it. \newline
\noindent $\bullet$ \textbf{{\framework vs CPU TEE:}} We offload the same component set as \framework to a CPU (AMD 9950X3D with SME). We use CPU because GPU TEEs are unavailable on RTX 5090. While secure memory encryption offers stronger privacy guarantee than \framework, the CPU with SME runs slower than GPU without TEE. Therefore \framework achieves 1.32$\times$ higher throughput at the same level of privacy guarantee (random guess). \newline
\noindent $\bullet$  \textbf{{Fully Offload (FO):}} FO applies isotropic Gaussian DP noise to the latent code of entire facial texture, and offloads it to the GPU. This is distribution-agnostic and thus conservative for anisotropic latent code. Consequently, it introduces prohibitive 105$\times$ loss increase and becomes impractical. 

In summary, with 14 offloaded components, \framework reaches 2.37$\times$ users and 2.17$\times$ users/watt, and achieves 2.06$\times$, 2.27$\times$, 1.32$\times$ higher throughput than sparsity (10\% channel pruning), quantization (8-bit weights) and TEE-like (CPU with secure memory encryption) under the same privacy guarantee of e-PSR bound of random guess (1.54\%).

\noindent
\textbf{Key Takeaway:} 
At equal e-PSR (random-guess, 1.54\%), \framework lies on the throughput–loss Pareto frontier: partial offloading reduces headset decode cost, while DAMP achieves the target e-PSR with minimal noise perturbation.

\begin{table}[!t]\centering
\vspace{-2mm}
\caption{\textbf{\framework vs. SotAs.} Note: Loss is MSE ratio to local baseline ($\downarrow$: lower is better); \#Users is a ratio to baseline ($\uparrow$: higher is better). e-PSR: max PSR over two empirical attackers. \textbf{\textit{Takeaway:}} \framework offers highest \#users and invisible loss increase.}
\label{tab:baseline_setup}
\resizebox{0.48\textwidth}{!}{
\begin{tabular}{lcccc}
\hline
\textbf{Technique} & \textbf{Loss ($\downarrow$)} & \textbf{\# Users ($\uparrow$)} & \textbf{e-PSR} \\ \hline
\rowcolor[gray]{0.9}
(\textbf{\textit{Baseline}}) Local Reconstruction & 1$\times$ & 1$\times$ & 1.54\% \\\hdashline
(\textit{\textbf{Sparsity}}) 10\% Channel Pruning & 1.25$\times$ & 1.15$\times$ & 1.54\% \\
(\textit{\textbf{Quantization}}) 8-bit Weights & 1.004$\times$ & 1.04$\times$ & 1.54\% \\
(\textbf{\textit{TEE}}) CPU w/ Secure Memory Encryption & 1$\times$ & 1.79$\times$ & 1.54\% \\
(\textbf{\textit{FO}}) Fully Offloading + Isotropic Noise & 105$\times$ & 2.3$\times$ & 85.6\% \\ \hdashline
\rowcolor[HTML]{E6EFDB}
(\textbf{\textit{\frameworknospace}}) Partial Offloading + DAMP Noise & \textbf{1.08}$\times$ & \textbf{2.37$\times$} & \textbf{1.54}\% \\
\hline
\end{tabular}}
\vspace{-4mm}
\end{table}

\subsection{\framework Ablation Study}

\begin{figure*}[t]
    \centering
    \subfloat[Loss -- $m$ \label{fig:reconstruction_loss_vs_num_offloaded_components}]{{\includegraphics[width=0.245\textwidth]{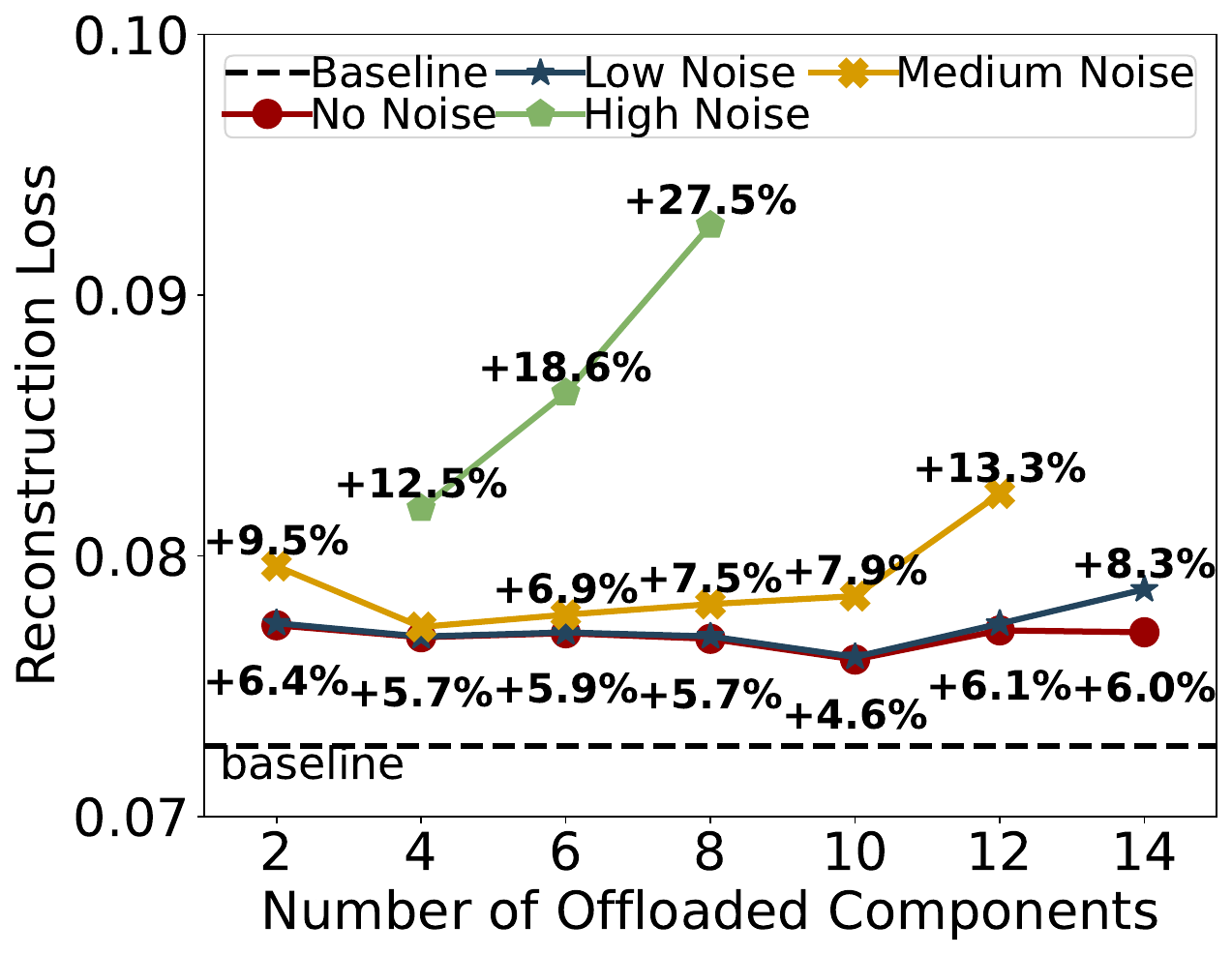}}} \hfil 
    \subfloat[Latency/Throughput -- $m$ \label{fig:total_users_quest_pro}]{{\includegraphics[width=0.255\textwidth]{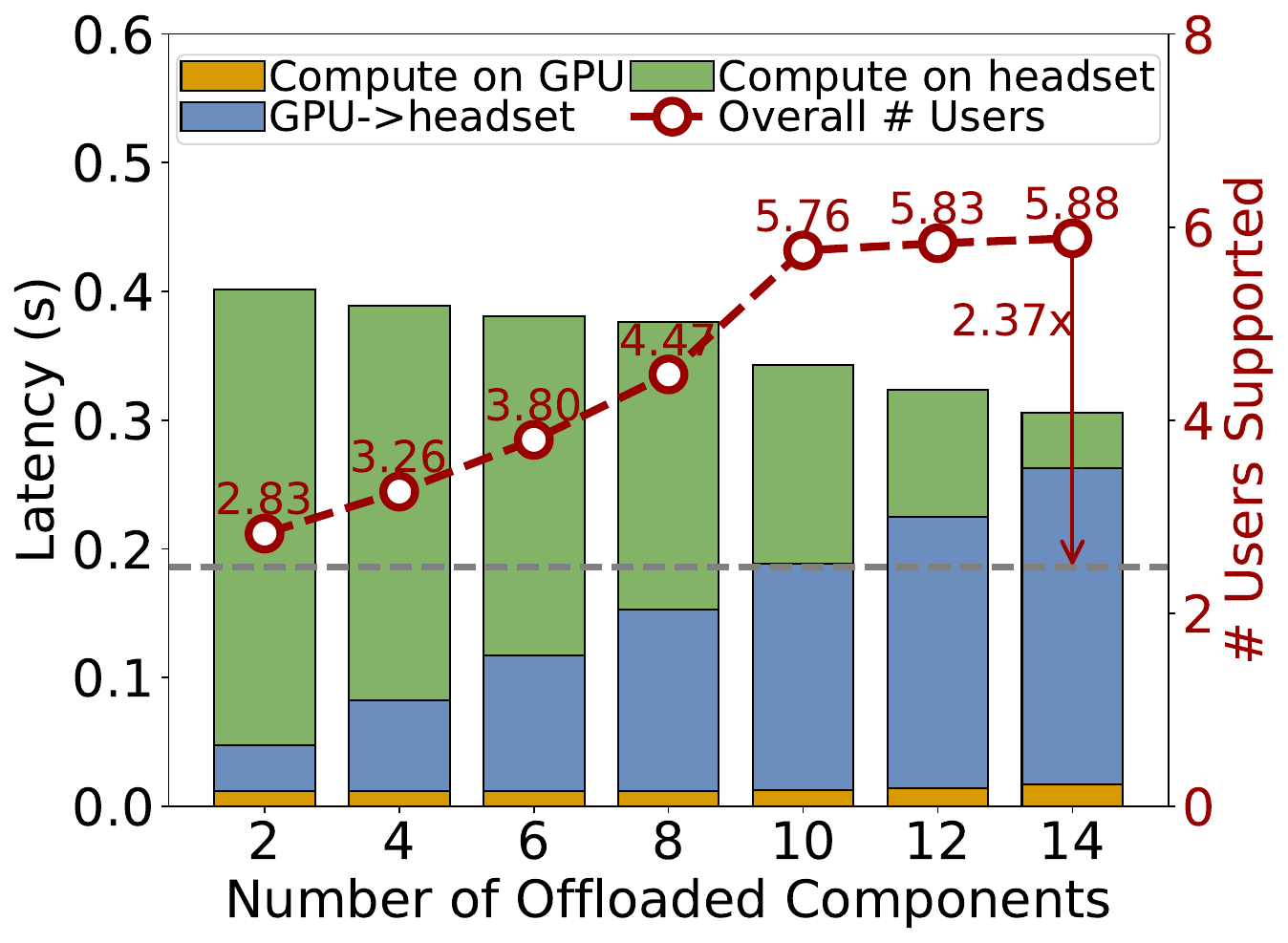}}} \hfil 
    \subfloat[Noise L2 Norm -- $m$\label{fig:privacy_vs_num_offloaded_components}]
    {{\includegraphics[width=0.22\textwidth]{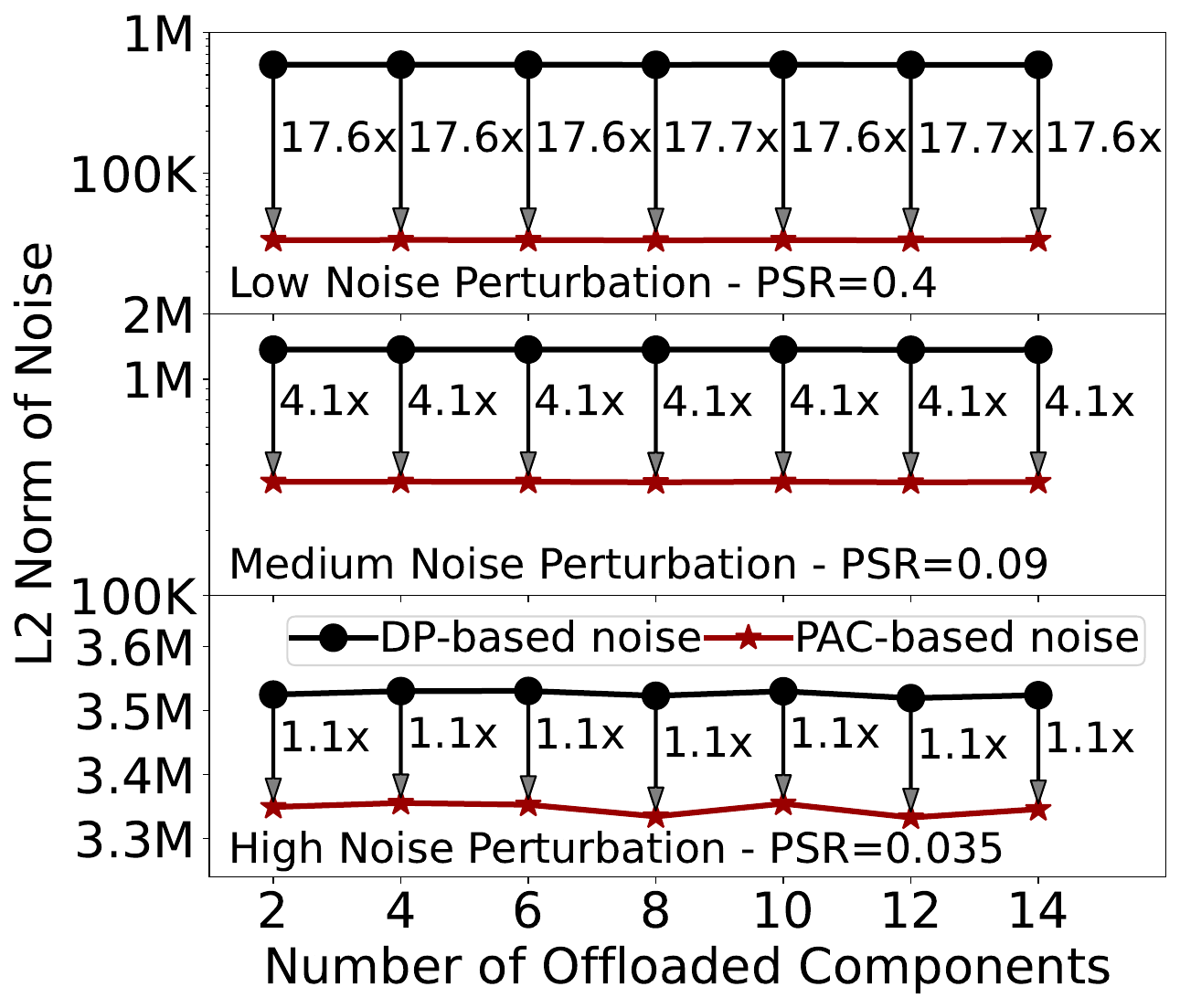}}}
    \subfloat[e-PSR -- $m$ \label{fig:posterior_successful_rate_vs_num_outsourced_components}]
    {{\includegraphics[width=0.25\textwidth]{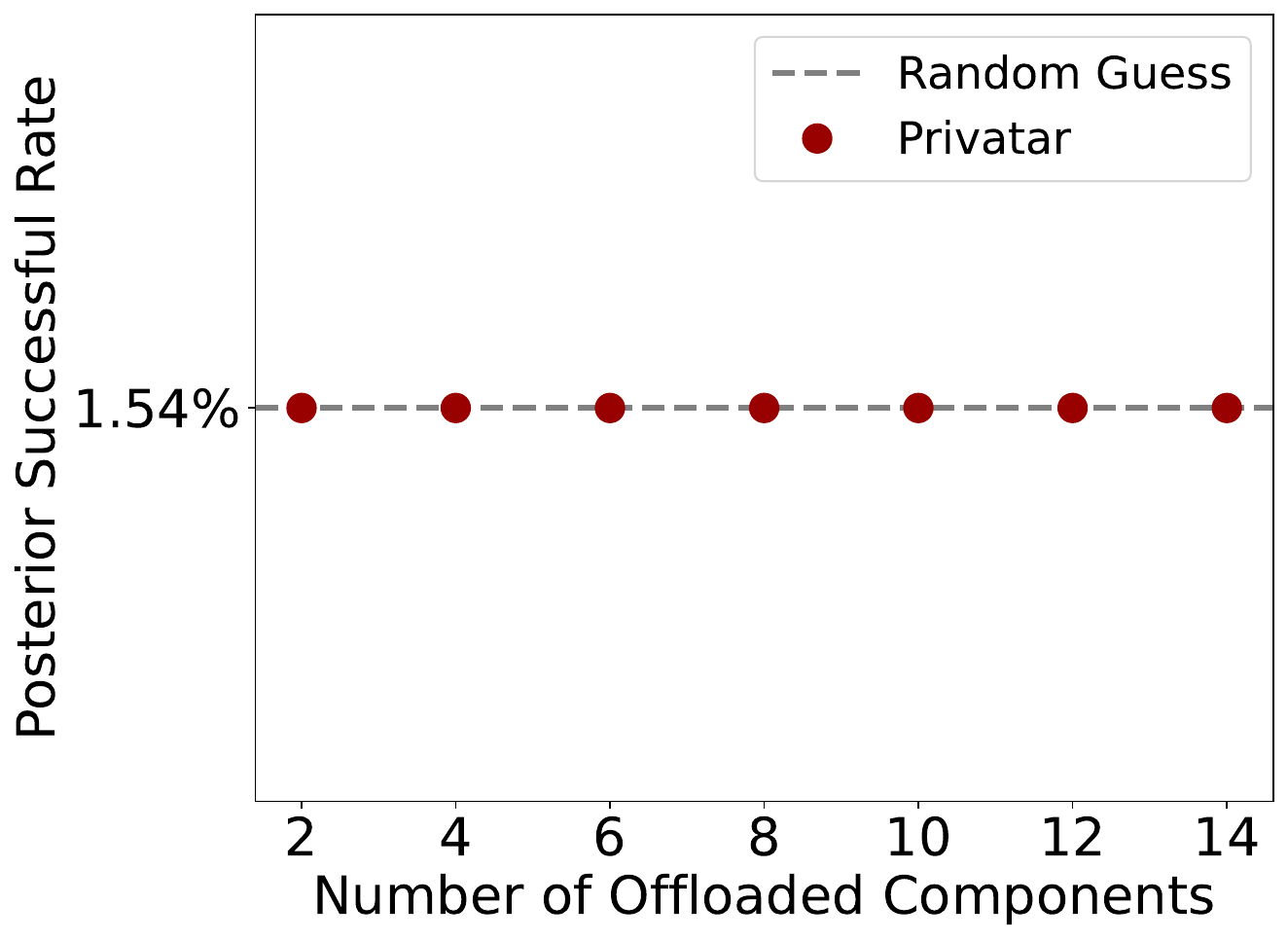}}}
    \vspace{-3mm}
    \caption{\textbf{Effects of number of offloaded components ($m$) on (a) Reconstruction Loss, (b) latency (left bar) and throughput (right red line) of avatar reconstruction, (c) the amount of noises needed for given privacy guarantee, and (d) Robustness against empirical attack.} \textbf{(\ref{fig:reconstruction_loss_vs_num_offloaded_components})} Curves show normalized reconstruction loss relative to the unpartitioned VAE used in baseline ($L_0{=}0.072$) for \framework with no noise and with DAMP at low/medium/high noise. HP alone increases loss by only $5.7\!\sim\!6.4\%$ across $m$. Low and medium DAMP settings keep degradation modest (often visually indistinguishable), while high-noise settings trade accuracy for tighter PSR bounds. \textbf{(\ref{fig:total_users_quest_pro})} The latency breakdown of reconstructing a single user while the overall throughput (number of users, \# user per second) are shown on left and right axis.
    \textbf{(\ref{fig:privacy_vs_num_offloaded_components}) Noise efficiency}: For the same PSR bound, DAMP leverages per-user latent anisotropy to shrink the required $L_2$ norm of noise by up-to $17.6\times$ (PSR$=0.4$), $4.1\times$ (0.09), and $1.1\times$ (0.035) relative to isotropic DP-based noise.  \textbf{(\ref{fig:posterior_successful_rate_vs_num_outsourced_components}) Empirical PSR}: The e-PSR remains at random-guess ($\approx$1.54\%) for \framework across all $m$ (with or without DAMP), since the base frequency never leaves the headset and the offloaded view is low information.}
    \label{fig:effect_num_freq_component}
    \vspace{-2mm}
\end{figure*}

\framework has two design choices \textit{(1) number ($m$) of frequency components to be offloaded}, and \textit{(2) the amount of noises to be added into the offloaded components}.
We sweep $m \in \{2,4,6,8,10,12,14\}$ and three t-PSR targets \{40\%, 9\%, 3.5\%\}, corresponding to mutual information bound as $v=\{1, 0.1, 0.01\}$, plus a no-noise reference and a fully-offloading all components reference in \tabref{tab:attack_cmp}.

\subsubsection{Knob 1: Offloaded Frequency Components}
\paragraph{Loss:}
With \emph{no noise}, partitioning in the frequency domain increases reconstruction loss by $\approx 6\%$ because local/offload path only sees partial frequency components. Qualitatively, this increase is visually subtle in our examples (see \figref{fig:reconstruction_quality_complete_outsource}). We further conduct perceptual study using LPIPS~\cite{zhang2018unreasonable}, where various partitioning choice only lead to negligible 0.72\% LPIPS increase. This proves the negligible utility effects of different partition choices. At \emph{low noise} (e.g., \(v=1\), t-PSR \(=40\%\)), the loss is similar to the no-noise case since the \emph{local merger} (\ding{191} in \figref{fig:HP_overview}) can suppress small perturbations of offloaded low-energy components using local high-energy components. As \(v\) decreases (tighter privacy, larger effective noise), loss grows and the loss increase is more pronounced at larger \(m\) because a higher fraction of components are reconstructed from the noisy latent code (yellow/green curves in \figref{fig:reconstruction_loss_vs_num_offloaded_components}).

\noindent \textbf{Latency / Throughput:} Increasing \(m\) shifts more decoding work off the headset, reducing local compute latency and improving overall throughput (red curve in \figref{fig:total_users_quest_pro}). 
The latency breakdown for one user (bars in \figref{fig:total_users_quest_pro}) shows: (i) \textbf{local} latency decreases with \(m\) (green), (ii) \textbf{communication} latency increases with \(m\) due to downloading more noisy reconstructed frequency components from PC to the headset (blue), and (iii) \textbf{offloaded compute} on RTX 5090 is small relative to the (i) and (ii). Takeaway: higher \(m\) improves throughput until communication dominates.

\noindent \textbf{Noise:} Noise is injected \emph{only} in the offloaded latent (\ding{188} in \figref{fig:HP_overview}). The PC decodes noisy frequency components which are sent back to the headset. Thus, changing \(m\) alters \emph{how many} components are affected by noise, not the per-latent noise calibration. Total noise energy per frame scales with the count of offloaded components rather than changing the noise distribution itself. 

\begin{table}[t]
\centering
\vspace{-2mm}
\caption{\textbf{Ablation setup.} We sweep the number of offloaded components \(m\) and the MI budget \(v\) that calibrates DAMP. t-PSR is the certified PSR bound; e-PSR is the maximum observed empirical PSR bound across the empirical and NN attackers.}
\label{tab:attack_cmp}
\scriptsize
\resizebox{0.47\textwidth}{!}{
\begin{tabular}{c|c|cc}\hline
Privacy Level &  \# offloaded components ($m$)  & e-PSR &  t-PSR \\ \hline
Fully Offload + No Noise      & 16  & 85.6\% & None \\ \hdashline
HP + No Noise      & \{2,4,6,8,10,12,14\}  & 1.54\% & None \\
HP + Low Noise ($v=1$)    & \{2,4,6,8,10,12,14\}  & 1.54\% & 40\% \\ 
HP + Medium Noise ($v=0.1$) & \{2,4,6,8,10,12,14\}  & 1.54\% & 9\% \\
HP + High Noise ($v=0.01$)   & \{2,4,6,8,10,12,14\}  & 1.54\% & 3.5\% \\ \hline
\end{tabular}}
\vspace{-2mm}
\end{table}

\subsubsection{Knob 2: Amount of DAMP Noises}
For a fixed \(m\), decreasing the MI budget \(v\) (stronger privacy) increases the required noise. Across all \(m\), DAMP requires noise with up to \(17.6\times\) less \(L_2\) norm than isotropic DP at t-PSR \(=40\%\), \(4.1\times\) less at \(9\%\), and \(1.1\times\) less at \(3.5\%\) (\figref{fig:privacy_vs_num_offloaded_components}). Empirically, as long as the \emph{base component stays local}, the attacker’s e-PSR remains at random-guess \(\approx 1.54\%\) for all tested \(m\) and \(v\) (\figref{fig:posterior_successful_rate_vs_num_outsourced_components}). DAMP preserves this empirical behavior while certifying tighter t-PSR. The noise reduction gets smaller with increase of the noise, because distribution matters less with larger noise.

In short, \framework picks configuration of \(v{=}0.1\) (t-PSR\(=9\%\)) and \(m{=}14\), and increases loss by \(8.3\%\) over the local baseline (in \figref{fig:all_local}) while ensuring e-PSR at random-guess.

\vspace{-2mm}
\section{Related Work}
\label{sec:RelatedWork}

\subsection{Security and Privacy of Avatars}
Multi-user avatar reconstruction in AR/VR exposes rich motion and interaction data that inherently reveal user identity and behavior. Prior work shows that even without visual access to the face, users can be re-identified from head and hand trajectories~\cite{nair2023uniqueid}, private activities can be inferred from motion traces~\cite{nguyen2024penetrationvision}, and typed text can be reconstructed through subtle movements~\cite{slocum2023motions, luo2024heimdall, ni2025vreckey}. These findings highlight the privacy risks in motion-driven avatar reconstruction.

\subsection{Secure Offloading Techniques}
Secure offloading strategies fall into three categories: \textbf{encryption}, \textbf{partitioning}, and \textbf{perturbation}.

\textbf{Encryption:} Homomorphic Encryption (HE)~\cite{Faster_CryptoNets, cheetah_hpca21, HE_inference_lloret, falcon} offers strong privacy by encrypting both data and computation, but incurs orders-of-magnitude overhead in compute and memory~\cite{tong2025CROSS,tong_smartpaf}. Multi-Party Computation (MPC)~\cite{goldreich1998secure} reduces the computational cost but requires multiple trusted parties, making both impractical for real-time VR workloads.

\textbf{Partitioning:} Input partitioning separates sensitive and less-sensitive features for selective offloading. \textit{Block partitioning}~\cite{vishwamitra2017blur, yuan2024bicryptonets} masks facial regions, while \textit{frequency partitioning}~\cite{wang2020high, ji2022privacypreserving} transmits high-frequency components. However, avatar reconstruction depends on complete facial recovery, lacking formal privacy guarantees. 

\textbf{Perturbation:} Offloading noisy data and related computation to untrusted devices causes utility degradation. The minimal noise required for a given privacy level is preferred. Privacy protection of individually released data would require local Differential Privacy (LDP)~\cite{local-DP}. LDP adds noise before offloading to protect user data without trusting the collector~\cite{usenix22-apple-cms, sp21-ldp-manipulation}. Yet, LDP noise grows with data dimensionality, significantly degrading utility for 256-dimensional latent code used in Variational Auto-Encoder. To address this, \framework optimizes multi-dimensional noise allocation and applies PAC Privacy~\cite{xiao2023pac} to achieve minimal noise with provable privacy bounds.

\vspace{-2mm}
\section{Conclusion}
\label{sec:Conclusion}
This work presents \framework, the first framework enabling secure partial offloading of avatar reconstruction to untrusted devices, reducing headset computation and scaling multi-user VR. \framework reframes privacy as a \emph{representation} decision by introducing two key components: (1) \emph{Horizontal Partitioning (HP)}, which splits the private data so untrusted devices never see a complete, identifying view, and (2) \emph{Distribution-Aware Minimal Perturbation (DAMP)}, which minimizes injected noise while providing formal privacy guarantees. Combining, Privatar enables VR headset to support 2.37$\times$ users with 9\% more energy, and being robust against both empirical attacker and NN-based attacker. 
\subsection{Acknowledgments}
This work was supported in part by ACE, one of the seven centers in JUMP 2.0, a Semiconductor Research Corporation (SRC) program sponsored by DARPA. We thank reviewers for their feedbacks.

\nocite{langley00}

\bibliography{refs}
\bibliographystyle{mlsys2025}

\appendix

\appendix
\section{Artifact Appendix}

\subsection{Abstract}

This artifact contains the full implementation and evaluation pipeline for Privatar, a privacy-preserving real-time multi-user VR avatar reconstruction system. It horizontally partitions a frequency-decomposed VAE decoder, keeping privacy-sensitive low-frequency components on the local VR headset while offloading high-frequency components to an untrusted cloud with calibrated noise injection. The artifact includes: (1)~training and testing scripts for the baseline VAE model and five design-choice variants (direct split, quantization, sparsity, frequency decomposition, and horizontally partitioned frequency decomposition), (2)~latency profiling and FLOPs calculation, (3)~differential privacy (DP) and PAC privacy noise generation, (4)~noisy inference under various mutual information budgets, and (5)~empirical and NN-based expression identification attacks. All experiments are reproducible within an NVIDIA GPU Docker container. Link to code: \url{https://github.com/georgia-tech-synergy-lab/Privatar}.

\subsection{Artifact check-list (meta-information)}

{\small
\noindent $\bullet$ {\bf Algorithm:} Variational Autoencoder (VAE) with Block DCT frequency decomposition, horizontal partitioning, DP and PAC privacy noise injection



\noindent $\bullet$ {\bf Data set:} Multiface dataset (subject 6795937), downloaded via provided script ($\sim$1\,TB)


\noindent $\bullet$ {\bf Hardware:} NVIDIA GPU (validated on RTX 5090/4090/3090), 16\,GB+ GPU memory, 52\,GB+ system RAM.


\noindent $\bullet$ {\bf Metrics:} MSE (screen, texture, vertex), LPIPS, latency (ms), FLOPs, posterior success rate (PSR)

\noindent $\bullet$ {\bf Output:} Trained model weights (\texttt{best\_model.pth}), test metrics, latent codes (\texttt{z\_*.pth}, \texttt{z\_offload\_*.pth}), noise covariance matrices (\texttt{.npy}), attack accuracy (PSR)

\noindent $\bullet$ {\bf Experiments:} Training (6 variants), testing (6 variants), latency profiling (5 variants + FLOPs), DP noise generation (80 files), PAC noise generation (75 files), noisy inference, empirical attack, NN-based attack, frequency covariance analysis


\noindent $\bullet$ {\bf How much time is needed to prepare workflow?:} $\sim$30 minutes (Docker setup, dependency and data setup)

\noindent $\bullet$ {\bf How much time is needed to complete experiments?:} Full training: $\sim$48 hours per variant (100K iterations); functional test (1K iterations): $\sim$16 minutes per variant. Testing, noise calculation, and attacks: $\sim$2 hours total.

\noindent $\bullet$ \textbf{DOI:} \url{10.5281/zenodo.19443137}




}

\subsection{Description}


The artifact is delivered as a GitHub repository containing all source code, configuration files, and experiment scripts. The repository is structured into six model variant directories and a shared \texttt{experiment\_scripts} directory:

\squishlist
{\small
\item \texttt{multiface/} --- Baseline VAE (DeepAppearanceVAE)

\item \texttt{multiface\_direct\_split/} --- Direct architecture split into local + cloud paths

\item \texttt{multiface\_quantization/} --- quantized decoder (8-16 bit)

\item \texttt{multiface\_sparse/} --- Channel-pruned decoder (20--80\% sparsity)

\item \texttt{multiface\_frequency\_decompose/} --- BDCT frequency decomposition (no offloading)

\item \texttt{multiface\_partition\_frequency\_decompose/} --- BDCT + horizontal partitioning (\framework)

\item \texttt{experiment\_scripts/} --- DP/PAC noise analysis, empirical attack configs, BDCT visualization, figure drawing, rendering utilities, and dataset download scripts
}
\squishend

Each variant directory contains 
\squishlist
{\small
\item core logic \texttt{train.py}/\texttt{test.py}.

\item Training/Testing launchers with configurable parameters \texttt{launch\_train\_job\_serial.py} and test script

\item latency measurement on VR headset, CPU or GPU \texttt{latency\_profiling\_script*.py}.

\item architecture definitions: \texttt{models.py}.
}
\squishend

\subsubsection{Hardware dependencies}

An NVIDIA GPU with CUDA support is required. The artifact has been validated on:

\squishlist
{
\item NVIDIA RTX 5090 (primary evaluation platform)

\item NVIDIA RTX 3090/4090 (supported via existing profiling scripts). Other generations like GH200 prefers different launcher like \texttt{torchrun} instead of Python.
}
\squishend

At least 16\,GB GPU memory and 52\,GB system RAM are recommended. The baseline latency profiling script runs on CPU (modeling a VR headset without a discrete GPU); when the device is not a VR headset, it defaults to CPU. All other variant profiling scripts run on GPU (modeling cloud execution).

\subsubsection{Software dependencies}

\squishlist
\item NVIDIA Docker: use based on your system, e.g. \texttt{pytorch:24.01-py3}.

\item OS packages: \texttt{mesa-utils}, \texttt{mesa-common-dev}, \texttt{libegl1-mesa-dev}, \texttt{libgles2-mesa-dev}.

\item Python packages: {\texttt{torch}, \texttt{Pillow}, \texttt{ninja}, \texttt{imageio}, \texttt{opencv-python}, \texttt{torchjpeg}, \texttt{lpips}}

\item \texttt{nvdiffrast}: cloned from GitHub and installed via \texttt{setup.py} inside Python virtual environment.


\squishend

\subsubsection{Data sets}

The Multiface dataset is used, containing facial images, tracked meshes, and unwrapped UV textures across 65+ expressions and 40 camera views. The dataset is downloaded via the provided script at \texttt{experiment\_scripts/dataset\_config/} \texttt{download\_dataset.py} ($\sim$1\,TB for identity 6795937). We also need pretrained model weights (\texttt{6795937\_best\_model.pth}, 97\,MB). 

\subsection{Installation}

Detailed step-by-step commands are provided in the repository's \texttt{README.md}. The installation involves five steps:

\squishlist
\item  Clone the repository and launch the Docker container (\texttt{nvcr.io/nvidia/pytorch:24.01-py3}) with GPU access, mounting the repository to \texttt{/work}.

\item  Install OS-level packages and Python dependencies.

\item  Install \texttt{nvdiffrast} from source. For RTX 5090, apply the provided \texttt{nvdiffrast\_patch.sh} and install the nightly PyTorch build with CUDA 13.0 support.

\item  Download the Multiface dataset ($\sim$1\,TB).

\item  Download the pretrained model checkpoint (\texttt{6795937\_best\_model.pth}, 97\,MB).

\item  All subsequent commands assume \texttt{/work} is the mount point inside Docker.

\squishend 

\subsection{Experiment workflow}

\begin{figure}[t!]
    \centering
    \includegraphics[width=\linewidth]{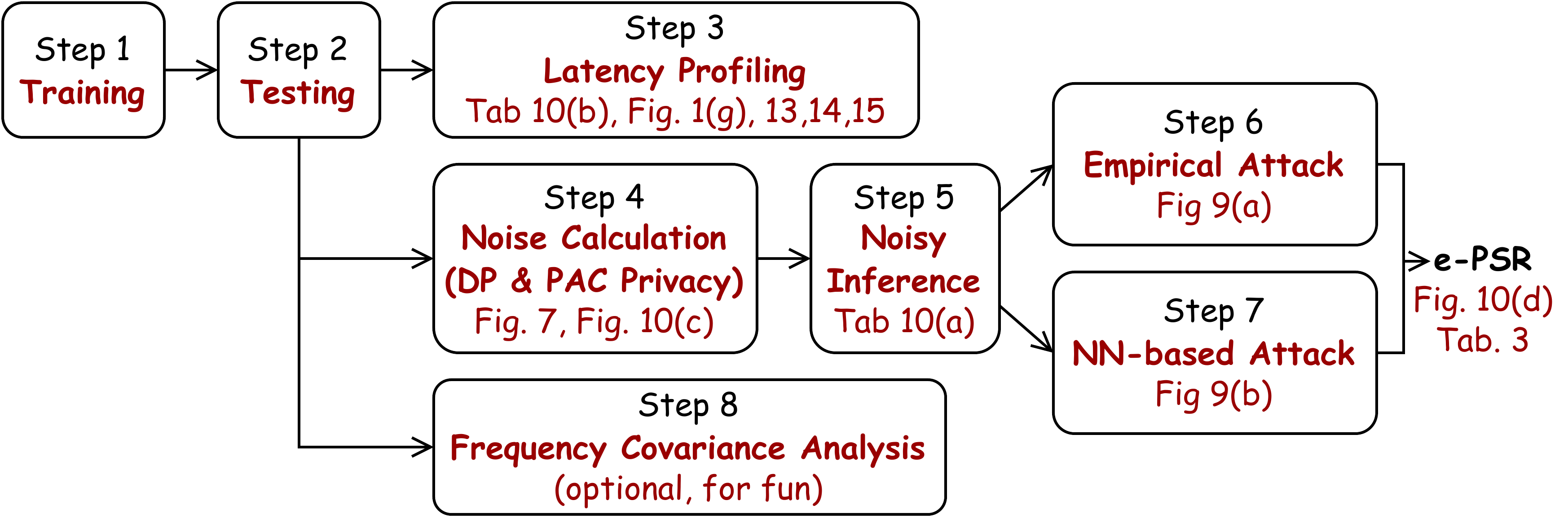}
    \vspace{-8mm}
    \caption{\textbf{Evaluation Flow}}
    \label{fig:evaluation_flow}
    \vspace{-7mm}
\end{figure}

The experiments follow an eight-step pipeline. Each step depends on the outputs of previous steps. The overview of all steps and how to obtain results of the papers are illustrated in \figref{fig:evaluation_flow}.

\textbf{Functional test vs.\ full reproduction.}
For functional testing, set \texttt{val\_num=30} and \texttt{max\_iter=1000} ($\sim$16 minutes per variant). For full reproduction, use \texttt{val\_num=500} and \texttt{max\_iter=100000} ($\sim$48 hours per variant). Note: the baseline \texttt{multiface/launch\_train\_job\_serial.py} ships with small defaults (\texttt{val\_num=50}, \texttt{max\_iter=100}); all other variants default to full reproduction values.

\textbf{Step 1: Training.} \hfill \\
Train all six model variants by running \texttt{launch\_train\_job\_serial.py} in each variant directory. Each produces a \texttt{best\_model.pth} checkpoint in \texttt{/work/training\_results/}. Configurable parameters include: quantization bitwidths (\texttt{bitwidth\_list}: 8--16), sparsity ratios (\texttt{sparsity\_list}: 0.2--0.8), and number of offloaded frequency components (\texttt{num\_freq\_comp\_offloaded\_list}: 2--14).

\textbf{Step 2: Testing.} \hfill \\
Run \texttt{launch\_test\_job\_serial.py} in each variant directory. This evaluates model quality (MSE, LPIPS) and saves latent codes (\texttt{z\_<id>.pth} and \texttt{z\_offload\_<id>.pth}) to \texttt{testing\_results/<project\_name>/latent\_code/}, which are required for noise calculation in Step~4.

\textbf{Step 3: Latency Profiling.} \hfill \\
Run \texttt{latency\_profiling\_script.py} in each variant directory. The baseline measures CPU latency (modeling VR headset); all others measure GPU latency (modeling cloud). For the partitioned variant, run both \texttt{latency\_profiling\_script\_local\_path.py} (local decoder) and \texttt{*\_offload\_path.py} (offloaded). Run \texttt{latency\_flops\_calculation.py} for FLOPs analysis across partition configurations.

\textbf{Step 4: Noise Calculation.} \hfill \\
Two noise mechanisms are supported: Differential Privacy (DP) based on L2 norm of latent codes, and PAC Privacy leveraging per-dimension covariance via SVD decomposition. Run DP noise scripts in \texttt{experiment\_scripts/dp\_analysis/} ({\small \texttt{dp\_noise\_generation\_for\_multiface.py} for baseline}; {\small \texttt{dp\_noise\_generation\_for\_partition\_multiface.py}} for partitioned configs covering both local and offloaded branches) and PAC noise script in \texttt{experiment\_scripts/pac\_analysis/} ({\small \texttt{pac\_noise\_generation\_for\_partition\_multiface.py}}). Both use mutual information bounds $[4, 3, 1, 0.1, 0.01]$ corresponding to posterior success rates $[98\%, 82.7\%, 40\%, 9\%, 3.5\%]$.

\textbf{Step 5: Noisy Inference.} \hfill \\
Run \texttt{launch\_noisy\_test\_job\_serial.py} in the partitioned frequency decomposition directory. Toggle \texttt{using\_pac\_noise} between \texttt{True} (PAC) and \texttt{False} (DP) to switch noise types. Results are saved to \texttt{/work/testing\_results/}.

\textbf{Step 6: Empirical Attack.} \hfill \\
Run \texttt{launch\_empirical\_attack.py} in the partitioned directory. The attacker guesses expressions by matching predicted high-frequency texture components to precomputed reference components. Supports both PAC and DP noise via \texttt{using\_pac\_noise} toggle.

\textbf{Step 7: NN-based Attack.} \hfill \\
Train a 3-layer fully-connected classifier (256$\to$128$\to$66) via \texttt{launch\_train\_nn\_attacker.py}, then evaluate under various noise levels via \texttt{launch\_test\_nn\_attacker.py}. Training data consists of one sample per expression from \texttt{selected\_expression\_frame\_list.txt}.

\textbf{Step 8: Frequency Covariance Analysis.} \hfill \\
Run \texttt{launch\_l2norm\_freq\_cov\_analysis.py} in the frequency decomposition directory to analyze the covariance trace of each of the 16 BDCT frequency components.

\subsection{Evaluation and expected result}

\textbf{Step 1: Training (full reproduction should set \texttt{max\_iter=100000}).}
All six variants train successfully and produce \texttt{best\_model.pth} checkpoints. 

\textbf{Step 2: Testing (using trained models).}
Expected test metrics (minor run-to-run differences due to random sampling in validation). ``no-noise" in \figref{fig:reconstruction_loss_vs_num_offloaded_components} shows results.

\squishlist
\item Baseline: screen$\approx$0.076, LPIPS$\approx$0.610
\item Partition-14: screen$\approx$0.077, LPIPS$\approx$0.612
\squishend

\textbf{Step 3: Latency Profiling (RTX 5090).} for  \figref{fig:total_users_quest_pro}.
\squishlist
\item  Baseline decoder (CPU): 15.47\,ms/inference

\item  Quantized decoder (GPU, 8-bit, traced): 0.685\,ms

\item  Sparse decoder (GPU): 0.87\,ms (10\% pruned) to 0.42\,ms (90\% pruned)

\item  Partitioned local path: 0.24--0.32\,ms across configs

\item  Partitioned offload path: 0.19--0.28\,ms across 7 configs

\item  FLOPs: 1.48\,G (14 offloaded) to 6.07\,G (2 offloaded), a 75.6\% reduction
\squishend

\textbf{Step 4: Noise Calculation.}
Noise trace values scale with mutual information budget as expected.
Both DP and PAC noise generation complete successfully. One randomly selected noise results are used to generate \figref{fig:noise_visualization}
and \figref{fig:privacy_vs_num_offloaded_components}.

\noindent $\bullet$ DP noise: 80 files total --- 5 (baseline complete offload) + 40 (partitioned offloaded branch) + 35 (partitioned local branch, 7 configs $\times$ 5 levels of mutual information). We use MI for mutual information in following contents.

\noindent $\bullet$ PAC noise: 75 files total --- both local and offloaded branches across 8 partition configs $\times$ 5 MI levels (minus local files for the direct-split config)

\textbf{Step 5: Noisy Inference.}
Noisy inference runs correctly with both PAC and DP noise. These contribute to ``noisy loss" of \figref{fig:reconstruction_loss_vs_num_offloaded_components}.

\textbf{Step 6: Empirical Attack.}
The empirical attacker achieves PSR=3.18\% or 1.54\% for different identities when taking partition-14 with MI=1 PAC noise, close to the prior rate of $1/65\approx1.54\%$, confirming that the noise effectively prevents expression identification attack. 

\textbf{Step 7: NN-based Attack.}
The NN attacker trains successfully (10 epochs) and achieves PSR=1.54\% on noisy latent codes, below the prior rate, confirming robustness against learned attacks. Results of both Step 6 and 7 are combined together to generate \tabref{tab:attack_cmp}.

\textbf{Step 8: Frequency Covariance Analysis.}
The covariance trace values for 16 frequency components match the expected output (e.g., component~0: $\sim$11308, component~15: $\sim$12.8). Low-frequency components carry $\sim$880$\times$ more variance than high-frequency components, confirming that high-frequency components are suitable for offloading.

\subsection{Experiment customization}

\squishlist
\item\textbf{Training duration:} Adjust \texttt{val\_num} and \texttt{max\_iter} in each \texttt{launch\_train\_job\_serial.py}. Functional test: \texttt{val\_num=30}, \texttt{max\_iter=1000}. Full reproduction: \texttt{val\_num=500}, \texttt{max\_iter=100000}.

\item\textbf{Partition configurations:} To test different numbers of offloaded frequency components (2, 4, 6, 8, 10, 12, 14), modify \texttt{num\_freq\_comp\_offloaded\_list}.

\item\textbf{Quantization bitwidth:} To use different precision, modify \texttt{bitwidth\_list} (8--16 bits) in the quantization \texttt{launch\_train\_job\_serial.py}.

\item\textbf{Sparsity ratio:} Modify \texttt{sparsity\_list} (0.2--0.8) in the sparsity \texttt{launch\_train\_job\_serial.py}.

\item\textbf{Noise type:} Toggle \texttt{using\_pac\_noise} between \texttt{True} (PAC) and \texttt{False} (DP) in noisy inference and attack launcher scripts.

\item\textbf{Mutual information budget:} Adjust \texttt{mi\_list} or \texttt{mutual\_info\_bound\_list} to test different privacy levels (default: $[4, 3, 1, 0.1, 0.01]$).
\squishend


\vspace{-1em}
\section{DAMP Noise Calculation}
\vspace{-0.5em}
\label{sec:theory_step_noise_calculation}
\label{sec:theory_pac}

\subsection{Computing Minimal Distribution-Aware Noise}
\label{sec:min_noise}
When we select noise $\bm{e}$ to be a multivariate Gaussian, i.e. $\bm{e} \sim \mathcal{N}\left(0,\Sigma_e\right)$, it suffices to optimize the noise covariance $\Sigma_e$ for required mutual information bound $v$, i.e. $\mathsf{MI}\left( X; \mathcal{F} \left(X\right) \+\bm{e}\right)\leq v$, as shown on the right hand side in (\ref{main_inequality}), which consequently bounds our objective $\left(1-\delta_{\rho}\right)$. Following Algorithm 1 in \cite{pac_alg}, optimizations over covariance $\Sigma_e$ are summarized below. 
\paragraph{Step 1: Compute Covariance Matrix} We first compute the covariance matrix $\Sigma_{\mathcal{F}\left(X\right)}$ of the profiled offloaded latent code $\mathcal{F}\left(X\right) \in \mathbb{R}^d$, a $d$-dimensional real vector ($d=256$), where we assume the input $X$ is uniformly selected from a data pool of expressions. And then we conduct the Singular Value Decomposition (SVD) of the covariance, denoted as
\begin{equation}
\label{equ:svd}
U \cdot s \cdot U^T = \Sigma_{\mathcal{F}\left(X\right)}
\end{equation}
where $s = \text{Diag}\left\{\lambda_{0:d-1}\right\}$ represents the diagonal matrix of eigenvalues $\lambda_i, i \in \left[0, d\right)$ while the unitary matrix $U$ corresponds to the eigenvectors. 
\paragraph{Step 2: Obtain Minimal Covariance of Noise for a Given Posterior Successful Rate} Given a posterior successful rate $\rho$ requirement, Eq. \ref{main_inequality} obtains the required mutual information $v$. Then, the $i$-th eigenvalue of minimal noise covariance, $\mathbf{\sigma}=\text{Diag}\{\sigma_{0:d-1} \}$, is obtained through
\begin{equation}
\label{equ:sigma}
\sigma_i = \frac{\sqrt{\lambda_i} \Sigma_{j=0}^{d-1}\sqrt{\lambda_j}}{2v}
\end{equation}
\paragraph{Step 3: Inject Random Noise per Offloaded Latent Code}
\framework samples noise $\bm{e}$ from the Gaussian distribution $\mathcal{N}\left(0, U \cdot \mathbf{\sigma} \cdot U^T \right)$ and release the noisy latent code $\mathcal{F}\left(X\right)\+\bm{e}$. The noise intensity satisfies $\mathbb{E}[\|\bm{e}\|^2_2]=\sum_{i=0}^{d-1} \sigma_i$.

By tailoring noise to statistical distribution, DAMP significantly reduces the noise magnitude compared to conventional DP approaches, thereby mitigating reconstruction accuracy degradation.

\section{Privacy Analysis of \framework Setup}
\subsection{Formal Privacy Guarantee (t-PSR)}
\label{sec:formal_noise_calculation}

For the Expression Identification Attack (EIA) in \secref{sec:attack_case_introduction}, we use $(\delta_{\rho},\rho,\mathsf{D})$–PAC Privacy to certify a \emph{posterior success rate} (PSR) bound, following \secref{sec:theory_pac}.

For the identity (user) in the multiface dataset with 65 categories of expressions, the prior success rate will be $1-\delta_{\rho,o}=1/65$ (uniform best guess), hence $\delta_{\rho,o}=64/65$. PAC Privacy (Eq.~\ref{main_inequality}) relates the target posterior successful rate $\delta_{\rho}$ to a mutual information (MI) budget $v$.
\[
\delta_{\rho}\ln \frac{\delta_{\rho}}{\delta_{\rho,o}}
+\left(1-\delta_{\rho}\right)\ln \frac{1-\delta_{\rho}}{1-\delta_{\rho,o}}
\;\le\; \mathsf{MI}\!\left(X;\mathcal{F}(X)+\bm{e}\right)\;\triangleq\; v.
\]
This mapping is \emph{monotone}: tighter privacy (lower bound for the t-PSR, $\delta_{\rho}$) requires a smaller $\mathsf{MI}$ ($v$). For our setting ($\delta_{\rho,o}=64/65$), the $v$ for different t-PSR bound are:
\[
\text{t-PSR upper bound } \{0.40,\ 0.09,\ 0.035\}\ \Rightarrow\ v \approx \{1,\ 0.1,\ 0.01\}
\]
We then choose the \emph{minimal} Gaussian noise covariance $\Sigma_e$ that achieves $\mathsf{MI}\!\left(X;\mathcal{F}(X)+\bm{e}\right)\le v$ following DAMP in \secref{sec:min_noise}. This per-user, distribution-aware calibration yields the minimum noise (\figref{fig:privacy_vs_num_offloaded_components}) necessary to certify the desired PSR against arbitrary attack under given t-PSR upper bound.

\subsection{e-PSR for Expression Identification Attack}
\label{sec:empirical_attack}
Under two empirical attack, \framework with loss noise level reduces its PSR from 86.2\% down to random guess, 1.54\%, emphasizing the efficacy of the provided horizontal partitioning and distribution based noise minimization.

\vspace{-1em}
\section{Detailed Attacker Setup}
\label{sec:formal_guarantee_empirical_attack}

\paragraph{Empirical Attacker} It randomly samples 1 angle from each expression in training dataset, and then partition it into frequency components to serve as frequency reference of a given expression. Then it will directly compare noisy reconstructed frequency components against all reference frequency components, making a guess of the expression whose reference frequency component has minimal difference to the reconstructed noisy frequency components. This attacker achieves 86.5\% e-PSR when offloading all frequency components, emphasizing its efficacy.

\vspace{-1em}
\paragraph{NN-based attacker} The attacker comprises three fully connected layers that map the intercepted noisy latent code to an estimate of expression directly. We pre-collect two samples per expressions to serve as training dataset of the NN-based attacker. And then mount attack by directly fed offloaded noisy latent code into it in attacking phase.


\vspace{-1em}
\section{TEE results on Quest Pro}
\begin{figure}[h]
    \centering
    \vspace{-4mm}
    \includegraphics[width=0.4\textwidth]{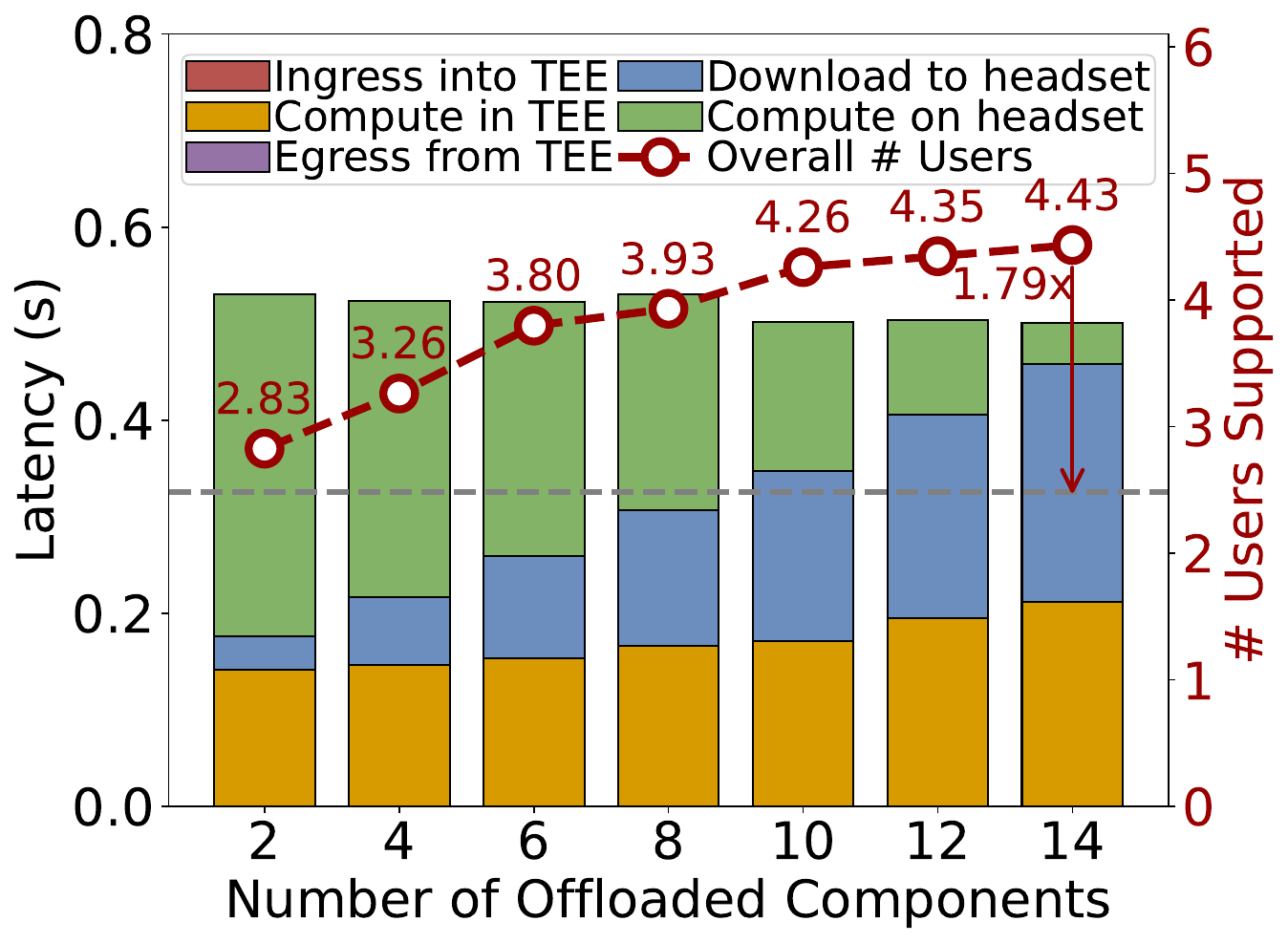}
    \vspace{-5mm}
    \caption{\textbf{Latency breakdown and overall throughput for CPU with secure memory encryption based secure partial offloading on \underline{Quest Pro}.} Bars decompose end-to-end per-frame latency into \emph{Ingress} (encrypt \& copy into enclave), \emph{Compute} (CPU inference inside the TEE), \emph{Egress} (copy out from TEE), \emph{Download} (return to headset), and \emph{Compute on the headset} (compute latency of reconstructing local frequency components). The red curve shows the overall throughput (the number of concurrently supported users at 60~FPS). Despite strong confidentiality, limited CPU inference latency restricts it throughput improvement to $\sim$1.79$\times$, which is less than 2.37$\times$ achieved by \framework.}
    \vspace{-6mm}
    \label{fig:total_users_tee_quest_pro}
\end{figure}

We use the latency profiled on AMD 9950 X3D with Secure Memory Encryption (SME) to serve as the latency for CPU TEE. In this setup, offloaded components without noises are directly processed in CPU with SME with stronger privacy guarantee. However, given the CPU SME being slower than the corresponding GPU without TEE setup, the maximal throughput improving for offloading $m=\{2,4,6,8,10,12\}$ frequency components is 1.79$\times$, which is 30\% less than \framework, as shown in \figref{fig:total_users_tee_quest_pro}. Among the collaborative local headset - CPU reconstruction path, the computation on both CPU and headset remain bottleneck. \textbf{\textit{Takeaway: }} CPU with TEE provides stronger privacy guarantee than \framework but being slower.

\section{Additional Results for Quest 3}
\label{sec:quest3_result}
\begin{figure}[t]
    \centering
    {\includegraphics[width=0.395\textwidth]{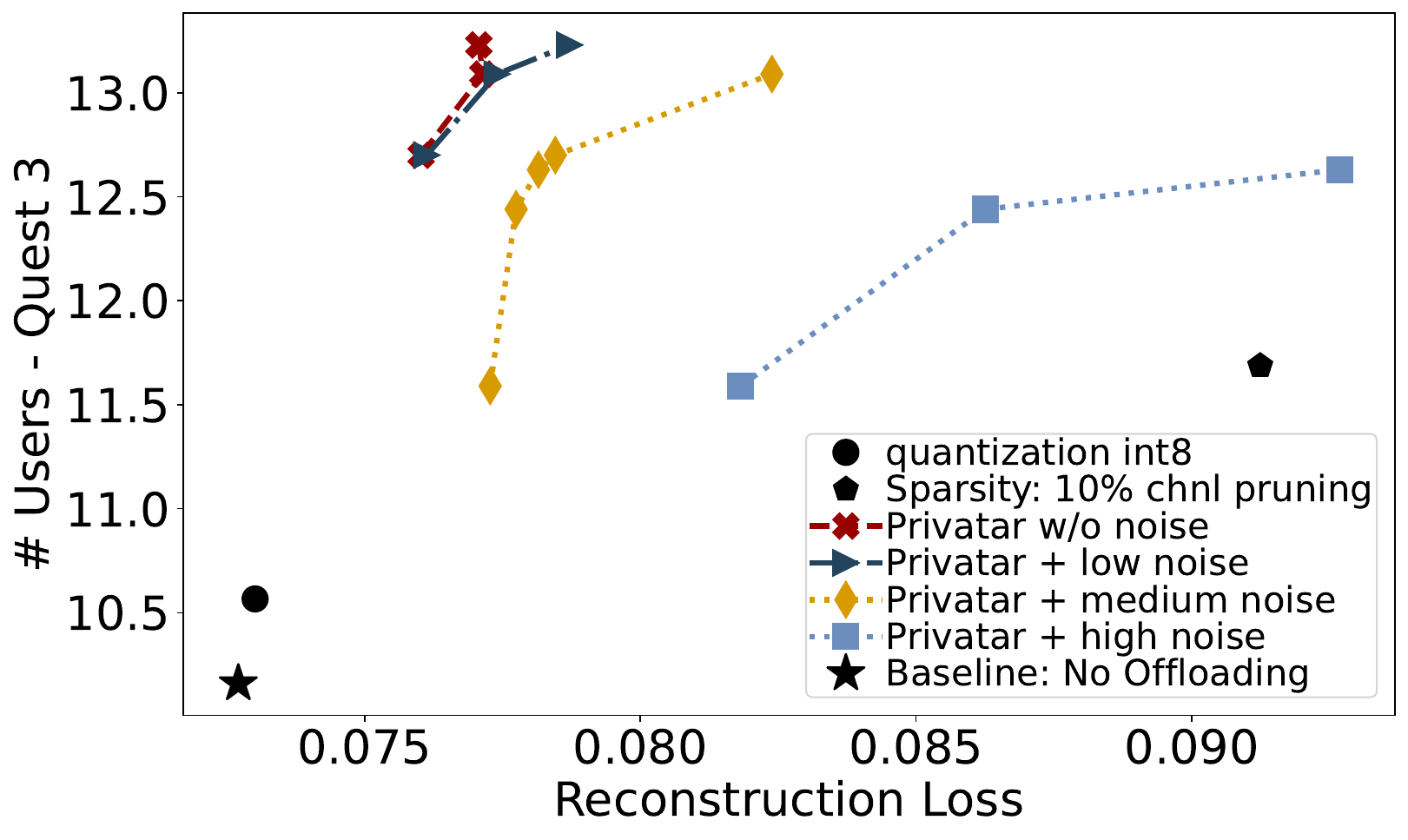}} 
    \vspace{-3mm}
    \caption{\textbf{Throughput-Loss comparison and ablation study with different number of offloaded frequency components for \underline{Quest 3}.} \framework enables Quest 3 to achieve 1.34$\times$ more users with 4.6\% loss, offering better pareto-frontier in throughput-loss curve when being compared against SotAs. }
    \vspace{-3mm}
    \label{fig:lat_acc_tradeoff_all_quest3}
\end{figure}

\begin{figure}[h]
    \centering
    \vspace{-6mm}
    \includegraphics[width=0.4\textwidth]{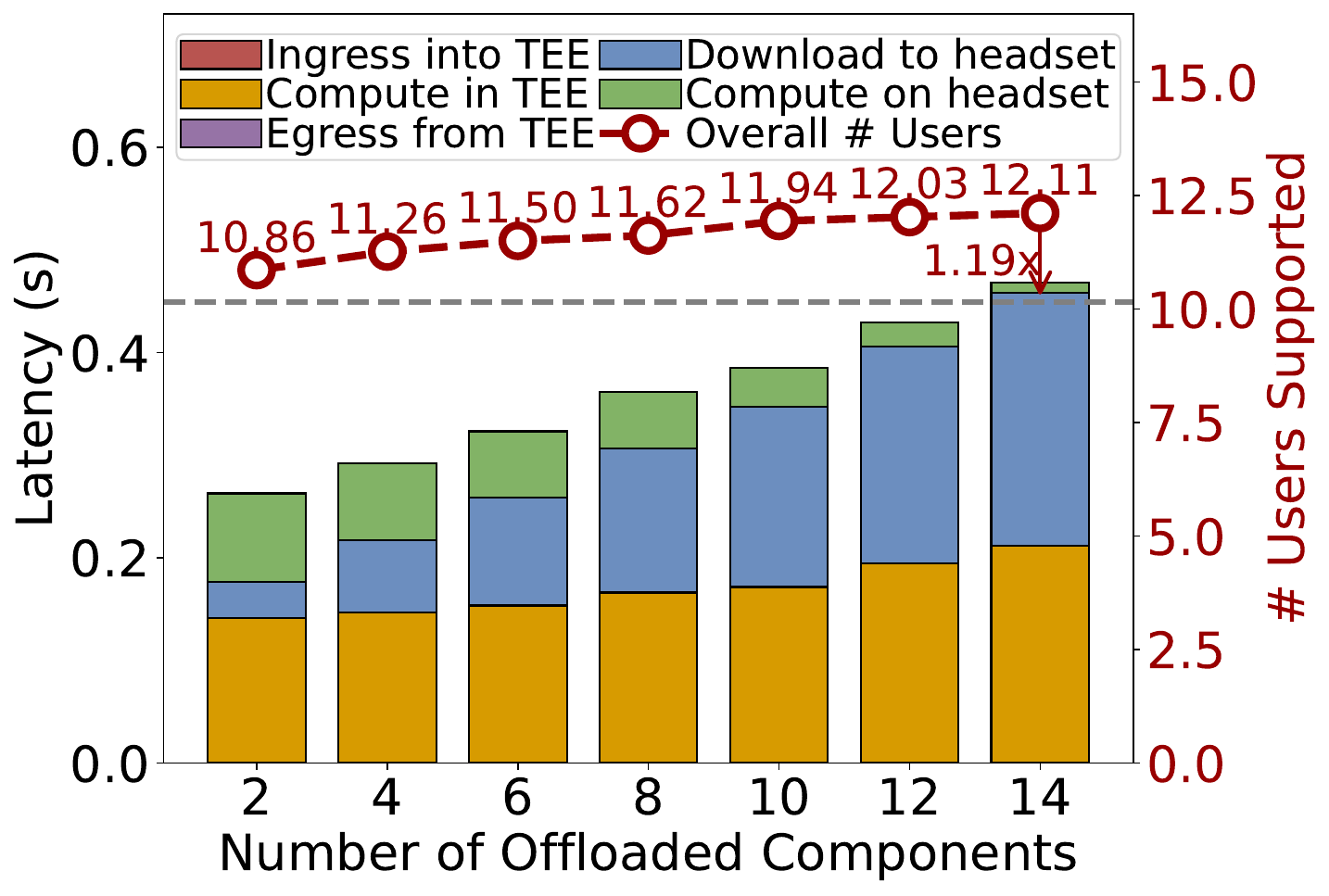}
    \vspace{-5mm}
    \caption{\textbf{Latency breakdown and overall throughput for CPU with secure memory encryption based secure partial offloading on \underline{Quest 3}.} Quest 3 provides higher local compute hence a lower green bar, and stronger baseline with less throughput improvement of $1.19\times$ under TEE. This aligns with the conclusion for Quest Pro: TEE provides stronger privacy guarantee but runs slower.}
    \vspace{-4mm}
    \label{fig:total_users_tee_quest_3}
\end{figure}

Quest 3 provides 4.1$\times$ (3709 GOPS) higher overall local compute capability than Quest Pro, enabling it to reconstruct more users (10.6 \# users in baseline for local reconstruction). We also provides \textbf{\framework vs SotAs} (\figref{fig:lat_acc_tradeoff_all_quest3}), \textbf{\framework Latency Breakdown} (\figref{fig:lat_acc_tradeoff_all_quest3}), and \textbf{TEE Latency Breakdown} (\figref{fig:total_users_tee_quest_3}) of Quest 3. \textbf{\textit{Takeaway: }} \framework provides better Pareto-frontier in throughput-loss curve on device with higher local computation capability, and communication is the system bottleneck for supporting more users from the collaborative local headset - GPU reconstruction.

Different partition choices induce different workload shapes, which map to the target hardware with different efficiencies. We therefore choose these choices of offloaded frequency components to maintain high on-device compute utilization for speedup. On reconfigurable hardware \cite{tong2026MINISA, tong2024FEATHER}, finer-grained partitioning exposes a larger design space between local compute and communication.




\begin{figure}[t]
    \centering
    {{\includegraphics[width=0.35\textwidth]{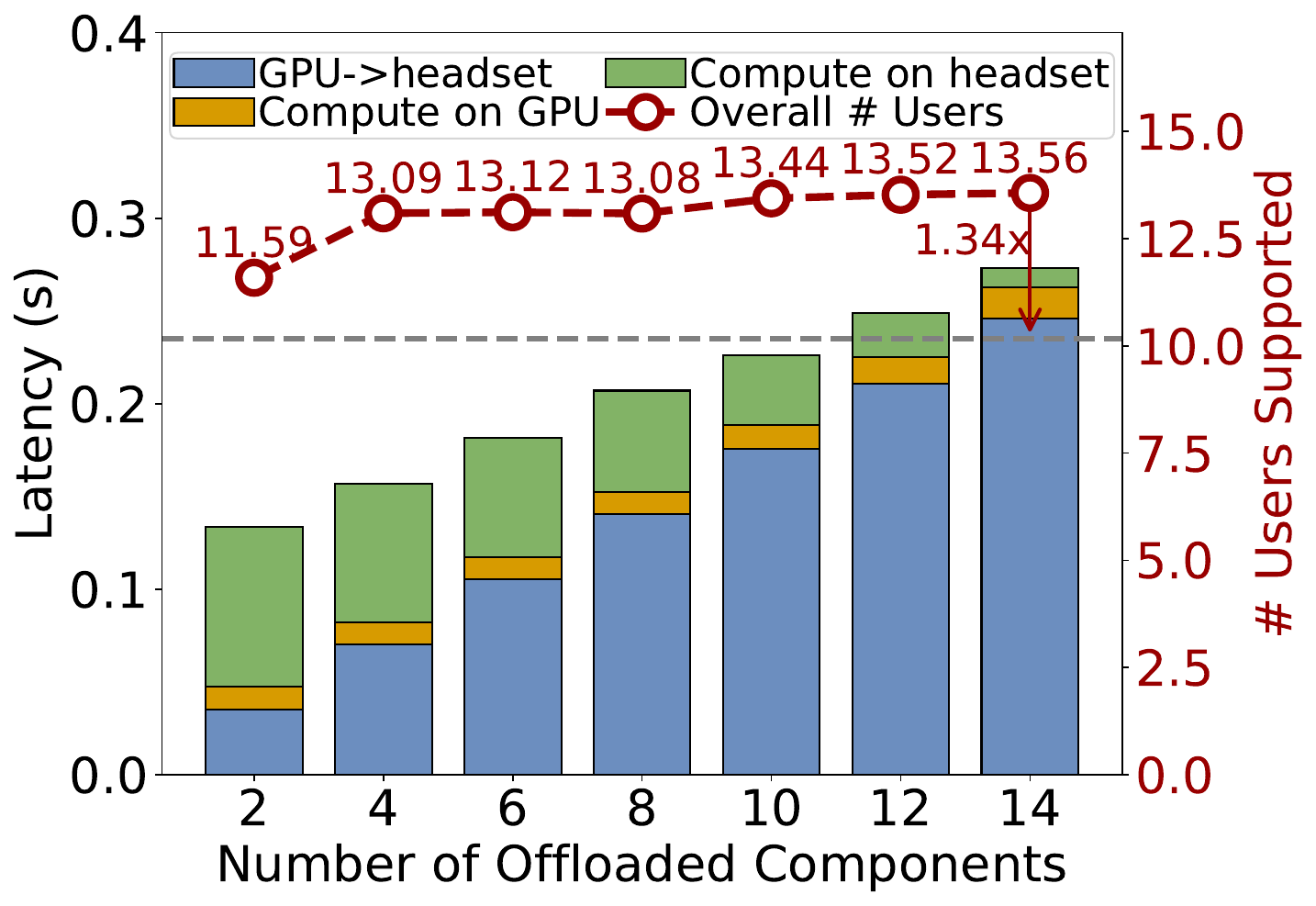}}}
    \vspace{-3mm}
    \caption{\textbf{Throughput Ablation Study on \underline{Quest 3}}. \framework achieves 1.34$\times$ throughput improvement for Quest 3 because it offers 3.7 TFLOPs peak throughput which is 4.1$\times$ larger than Quest Pro. Communication bandwidth between PC and headsets is the bottleneck of throughput improvement from partial offloading.}
    \label{fig:total_users_quest_3}
    \vspace{-3mm}
\end{figure}

\end{document}